\documentclass{aa}  

\usepackage{graphicx}
\usepackage{txfonts}

\def\kms{\hbox{km\,s$^{-1}$}}

\def\msun{M$_\odot$}
\def\lsun{L$_\odot$}

\def\msunyr{M$_\odot$\,yr$^{-1}$}
\def\cmdos{cm$^{-2}$}
\def\cmcub{cm$^{-3}$}
\def\x{$\times$}
\def\arcsec{$^{''}$}
\def\deg{$^{\circ}$}
\def\radec{RA, Dec.(J2000)}
\def\mic{$\mu$m}
\def\co{$^{12}$CO}
\def\tco{$^{13}$CO}
\def\cdo{C$^{18}$O}

\def\nhhp{N$_2$H$^+$}
\def\nddp{N$_2$D$^+$}
\def\msunyr{M$_\odot$ yr$^{-1}$}
\def\aztec{AzTEC-lup1-2}
\def\aztecb{AzTEC-lup3-5}
\def\iras{IRAS\,15398-3359}
\def\irasb{IRAS\,16059-3857}
\def\j{J160115-41523}
\def\sz{Sz\,102}
\def\merin{Merin\,28}

\begin{document}

   \title{Outflows, envelopes, and disks as evolutionary indicators in Lupus YSOs}


   \author{M. M. Vazzano
          \inst{1}
          \and
          M. Fernández-López\inst{1}
          \and
          A. Plunkett\inst{2}
          \and
          I. de Gregorio-Monsalvo\inst{3}
          \and
          A. Santamar\'ia-Miranda\inst{3}
          \and
          S. Takahashi\inst{4,5,6}
          \and
          C. Lopez\inst{4}
          }

   \institute{Instituto Argentino de Radioastronom\'ia, (CCT-La Plata, CONICET; CICPBA), C.C. No.  5, 1894, Villa Elisa,Buenos Aires, Argentina\\
              \email{mvazzano@iar.unlp.edu.ar}
         \and
         National Radio Astronomy Observatory, 520 Edgemont Rd, Charlottesville, VA 22903, U.S.A
         \and
         European Southern Observatory, Av. Alonso de C\'ordova 3107, Casilla 19001, Santiago, Chile
        \and
        Joint ALMA Observatory, Alonso de Córdova 3107, Vitacura, Santiago, Chile
        \and
        NAOJ Chile Observatory, Alonso de Córdova 3788, Oficina 61B, Vitacura, Santiago, Chile
        \and
        Department of Astronomical Science, School of Physical Sciences, The Graduate University for Advanced Studies, SOKENDAI, Mitaka, Tokyo 181-8588, Japan
        \\
             }

   \date{Received Agoust, 2020; accepted January 2021}

 
  \abstract
   {The Lupus star-forming complex includes some of the closest low-mass star-forming regions, and together they house objects that span evolutionary stages from pre-stellar to pre-MS.}
   {By studying 7 objects in the Lupus clouds from prestellar to protostellar stages, we aim to test if a coherence exists between commonly used evolutionary tracers.}
   {We present ALMA observations of the 1.3\,mm continuum and molecular line emission that probe the dense gas and dust of cores (continuum, \cdo, \nddp) and their associated molecular outflows (\co). Our selection of sources in a common environment, with identical observing strategy, allows for a consistent comparison across different evolutionary stages. We complement our study with continuum and line emission from the ALMA archive in different bands.}
   {The quality of the ALMA molecular data allows us to reveal the nature of the molecular outflows in the sample by studying their morphology and kinematics, through interferometric mosaics covering their full extent. The interferometric images in \iras\ appear to show that it drives a precessing episodic jet-driven outflow with at least 4 ejections separated by periods of time between 50 and 80 years, while data in \irasb\ show similarities with a wide-angle wind model also showing signs of being episodic. The outflow of \j\ could be better explain with the wide-angle wind model as well, but new observations are needed to further explore its nature.
   We find that the most common evolutionary tracers in the literature are useful for broad evolutionary classifications, but are not consistent with each other to provide enough granularity to disentangle different evolutionary stage of sources that belong to the same Class (0, I, II, or III). The evolutionary classification revealed by our analysis coincides with those determined by previous studies for all our sources except \j. Outflow properties used as protostellar age tracers, such as mass, momentum, energy and opening angle, may suffer from differences in the nature of each outflow, and therefore detailed observations are needed to refine evolutionary classifications. We found both \aztec\ and \aztecb\ to be in the pre-stellar stage, with the possibility that the latter is a more evolved source. \iras, \irasb\ and \j,  which have clearly detected outflows, are Class 0 sources, although we are not able to determine which is younger and which is older. Finally \sz\ and \merin\ are the most evolved sources in our sample and show signs of having associated flows, not as well traced by CO as for the younger sources.}
   {}


   \maketitle
%

\section{Introduction}
In the canonical picture of star formation, a dense core accretes material from a surrounding envelope, and, because of the conservation of angular momentum due to the accretion, a bipolar jet-outflow emerges. 
Prior to Class 0, a prestellar core provides the initial conditions for accretion and outflow-driving, and hence it is an important yet elusive initial component to study \citep{Dunhametal2014,Padoanetal2014}.
The accretion and jet-launching are the active processes during the `protostellar' (Class 0-I) phase \citep{ArceandSargent2006}, as a protostellar disk also begins to form. Once the envelope dissipates and the outflow subsides (Class II), the disk is exposed around the Young Stellar Object (YSO).  

Several indicators capable of providing information about the evolutionary state of pre- and protostellar sources exist in the literature. The most used are the bolometric temperature $L_{bol}$, and the ratios between the masses of the disk and the envelope ($M_{disk}$/$M_{env}$) and between the bolometric and submillimeter luminosity ($L_{bol}$/$L_{submm}$). 
Furthermore, for sources with associated outflows, it is possible to assess their evolution through parameters such as momentum, energy, mass-loss rate and momentum flux. 

The Lupus complex, located near the Scorpious OB2 association, is known for its nearby ($\sim$150 pc) clusters of low-mass, pre-main sequence stars \citep{Comeron2008,Knude2009}.  
Within 335\deg\ $<$ $l$ $<$ 348\deg\ and 0\deg\ $<$ $b$ $<$ 25\deg\ area, the Lupus complex was first found to have four sub-groups named as Lupus\,I, II, III, and IV. 
Around 70 T-Tauri stars are found around the Lupus complex with estimated ages of about 1-3\,Myr \cite[e.g.][]{Comeron2008, Ansdelletal2016, Ansdelletal2018, Alcalaetal2015}.
Lupus has been studied in \co, \tco, and \cdo\ in order to map the molecular gas (\citealt{Haraetal1999}; \citealt{Vilas-Boasetal2000}; \citealt{Tachiharaetal2001}; \citealt{Tothilletal2009}). These maps have shown that Lupus is actually divided into nine subgroups, where only Lupus\,I, III, and IV present evidence of ongoing star formation. The \cdo\ data were used to identify about 10 cores in the most massive one, Lupus\,I (with a mass of $\sim$ 1200 \msun), with column densities  N(\cdo) = (5-10)$\times$10$^{14}$ \cmdos\ indicating potential sites of star formation.
Lupus\,III has a mass of about 300 \msun\, and it hosts a rich cluster of T-Tauri stars, but not much star formation activity. Lupus\,IV is the third cloud of the complex to show evidence of star formation activity, including nine \cdo\ dense cores with column densities N(\cdo) = (4-10)$\times$10$^{14}$ \cmdos\ and three H$^{13}$CO$^+$ cores \citep{Tachiharaetal2007}.

The SOLA (`The Soul of Lupus with ALMA', \citealt{Saito15}) collaboration is dedicated to a multi-wavelength study of Lupus, including a catalog of all prestellar objects and YSOs (hereafter, `SOLA catalog') which will be published elsewhere. Previously, using the SOLA catalog, 37 protostellar sources were targeted to search for binaries in Lupus in Cycle 2 (PI:  Saito,  2013.1.00474.S).  Among  these  sources,  we  have  quantitative  information  about  their  sub-mm properties.  The spectral setup of those Cycle 2 observations covered continuum, and (simultaneously) the \co(3-2)and HCO$^+$(3-2) lines for evidence of outflow and infall at a single pointing for each source.  The data presented in this work, observed during the ALMA Cycle 4, complements and extends the Cycle 2 study: we now aim to study the cores, envelopes, disks, and the full extent of outflows from sources at different evolutionary phases from prestellar through protostellar in Lupus. In this way, we want to characterize all these phenomena associated with the star formation process in a set of isolated sources at different stages of evolution in a similar nascent environment, and infer their physical and kinematic properties in relation to their stage of evolution. 
Several variables that can influence the evolutionary tracers challenging this task. To the intrinsic difficulty in measuring some indicators (such as choosing the distance for measuring the outflows opening angles), we must add the possible effects of precession and/or outflow interaction with the environment, and the uncertainty in the interpretation of the results depending on the outflow model adopted.
However, we believe that the possibility of studying several sources in the same environment gives us the chance to evaluate how consistent the different indicators are. In addition, we have high-quality molecular outflow data that could shed light on the evolutionary state of the particular sources. 
In this way, we intend to evaluate the sources and test whether there is consistency between the evolutionary tracers commonly used.

Here we present the results of the study of 7 sources. Six of them were selected from the SOLA catalog and one more was included since it was catalogued as a young Lupus protostar in previous studies.
Although our sample is not statistically significant, we attempt to trace an evolutionary sequence in the sources of our sample by evaluating various evolutionary indicators from the literature. 
Spectral line and continuum observations from ALMA enable investigation of the molecular dust and gas environment surrounding the selected sources, on scales of 10$^3$ AU.
In Section \ref{sec:sample} we describe the sample selection, and in Section \ref{sec:data} we describe the observations and explain the calibration and imaging procedure. 
We organized the Results section in three subsections: continuum emission (Section \ref{subsec:continuum}), single dish line emission (Section \ref{subsec:line}), and outflow emission (Section \ref{subsec:outflows}), where we describe the outflow morphology, physical parameters and the kinematics. 
In Section \ref{sec:discussion} we split the discussion in two subsections: in \ref{subsec:individual_outflows} we refer to the individual outflows and in \ref{subsec:charac_source_sample} we discuss the characteristics of the source sample (SED, chemical and outflow evolution).
Finally we include an Appendix, where we present a brief overview of each source, mention the particularities of some sources, and show the complete velocity cubes.

\section{The sample} \label{sec:sample}

The SOLA catalog enacts a clump-finding algorithm and SED fitting to classify all sources in Lupus as prestellar or Class 0-III.  
The basis of this catalog is a set of millimeter and sub-millimeter images. Lupus\,I, II, III and IV clouds were observed at 1.1\,mm with the AzTEC bolometer mounted on the ASTE telescope \citep{Tsukagoshietal2011}, and the Lupus\,III cloud was also observed at 870 \mic\ with the LABOCA bolometer mounted on the APEX telescope. In total these observations cover 7.5 deg$^2$ on the sky.
A total of 251 sources were found in AzTEC and LABOCA images, utilizing CLUMPFIND as the structure search algorithm \citep{Williamsetal1994}\footnote{\url{http://www.ifa.hawaii.edu/users/jpw/clumpfind.shtml}}.
To classify these sources, we searched for optical to centimeter counterparts, estimating their spectral energy distributions (SEDs). We used catalogues and/or images from WFI, DENIS, 2MASS, AKARI, ATCA, SMA, SEST, WISE, IRAS, Spitzer and HERSCHEL telescopes, 
and correlated the positions at different wavelengths considering pointing errors from all telescopes. The AzTEC and LABOCA objects are classified as 189 starless, 4 Class 0, 7 Class I, 49 Class II, and 2 Class III objects. 

We chose 7 individual sources throughout the Lupus\,I, III, IV and VI molecular clusters that span protostellar evolutionary phases in similar nascent environments: two each from the prestellar, Class 0, and Class I evolutionary phases, and an additional late-Class I source (see Table \ref{tab:sources}). 
These sources were chosen based on the following source criteria: 
(1) We selected sources based on SED shape. 
(2) For protostar candidate sources, we selected sources that are considered as single (non-multiple) sources. This was confirmed with the ALMA Cycle 2 Band 7 continuum observations with the resolution of 0\farcs2, or 31 AU (PI: Saito, 2013.1.00474.S). This criterion results in 5 targets (2 in each of the protostar categories Class 0 and Class I, and one classified as late-Class I), a complete sample of non-binary protostars in Lupus.
(3) Finally, we chose the two pre-stellar candidates that have a single, strong peak based on the concentration factor, C = F$_{peak}$/F$_{3\sigma}$, using AzTEC/ASTE 1.1 mm observations. 
These have a density greater than the critical density, to imply they are not transient clumps.

Observing sources within Lupus allows us to assume similar environmental conditions (i.e. interstellar radiation fields, gas temperatures, chemical composition) and observing consistencies (i.e. beamsize, distance, spatial resolution, sensitivity).
Recent calculations of Lupus I and III distances based on {\it Gaia} DR2 data have revealed that these clouds are located at 153.35$\pm$4.64 and 154.75$\pm$9.59\,pc, respectively (Santamar\'ia-Miranda, accepted.), values which are in agreement with the distances derived by \citet{Sanchisetal2020}. 
This, along with the distances reported by \citet{Zucker20}, leads us to adopt a mean distance of 150\,pc for all the clouds in the complex.
In any case, our conclusions will not be dramatically affected by this choice of distance, since the consistency will always be better than choosing sources from entirely different regions.

A brief description of each source is provided in Appendix A.

\begin{table*}
    \centering
        \caption{Observed sources. ($\ast$): Classification from SOLA catalog (PC=Prestellar core, 0=Class\,0, I=Class\,I, LI=Late-Class\,I).}
     \resizebox{\textwidth}{!}{
    \begin{tabular}{lccccccccc}
    \hline
   Source name    &     Obs coordinates  &   Obs mode   & Map size & Time on source & Flux       & Phase      & Bandpass   & Lupus  & Clasif.$^{*}$ \\
                  &      R.A,Dec(J2000)  &              &          &   (min)  & calibrator & calibrator & calibrator &   cloud    &   \\
    \hline
AzTEC-lup1-2      & 15:44:59, --34:17:07 & Single-field & 60\arcsec & 2.0 & J1427-4206 & J1610-3958 & J1427-4206 &   I   & PC \\
AzTEC-lup3-5      & 16:09:00, --39:07:38 & Single-field & 60\arcsec & 2.0 & J1427-4206 & J1610-3958 & J1427-4206 &  III  & PC \\
IRAS\,15398-3359  & 15:43:02, --34:09:07 & Mosaic & 110\arcsec$\times$90\arcsec  & 10.2 &  Ganymede  & J1626-2951 & J1517-2422 &   I   &  0 \\
IRAS\,16059-3857  & 16:09:18, --39:04:53 & Mosaic & 280\arcsec$\times$110\arcsec & 18.8 & J1256-0547 & J1610-3958 & J1924-2914 &  III  &0 \\
Merin\,28         & 16:24:51, --39:56:32 & Mosaic & 280\arcsec$\times$110\arcsec & 18.8 & Callisto  & J1610-3958 & J1924-2914 &   VI  & I \\
Sz\,102           & 16:08:29, --39:03:11 & Single-field & 60\arcsec & 2.0 & J1427-4206 & J1610-3958 & J1427-4206 &  III &  I\\
J160115-41523     & 16:01:16, --41:52:36 & Mosaic & 280\arcsec$\times$110\arcsec & 18.8 & Ganymede  & J1604-4441 & J1337-1257 &   IV & LI\\
    \hline
    \end{tabular}
    }
    \label{tab:sources}
\end{table*}

\section{Data} \label{sec:data}
This work is mainly based on dedicated observations using the total power and 7m array of ALMA, including single-field and mosaic modes. We have complemented these observations with data from the ALMA Science Archive; in order to study compact sources, we have supported these dedicated observations with archival data of continuum emission from the ALMA 12m array, and high-resolution molecular line data for those sources that drive molecular outflows. We have also included infrared data associated with the compact sources over a wide range of wavelengths.

In Sections \ref{sec:SOLA_obs}, \ref{sec:cont-archive}, \ref{sec:archive} and \ref{sec:IR} we describe these four groups of data, respectively (our dedicated ALMA observations, archival continuum data, archival molecular data, and infrared data).  In section \ref{sec:imaging} we describe how the imaging and combination processes were carried out for each source.

\subsection{SOLA observations} \label{sec:SOLA_obs}

Using the ALMA 7m array and TP antennas, we observed the seven selected sources in Band 6 (1.3\,mm observations. Project code: 2016.1.01141.S, P.I: Takahashi, Satoko) . Three of the sources were observed with a single field, and four using small mosaics (see Table \ref{tab:sources}). 
The data were taken on 12-20 Dec 2016; 15-30 Apr 2017; and 19-21 May 2017.
The observations were made 
with baselines between 9 and 45 meters, from which we achieved angular resolutions of $\sim$8\arcsec (1240 AU).

The correlator configuration included four spectral windows with 4096 channels each. 
The two spectral windows dedicated to observe $^{12}$CO(J=2-1) ($\nu_{rest}$=230.538\,GHz) and SiO(5-4) ($\nu_{rest}$=217.10498\,GHz) covered 500\,MHz each at a frequency resolution of 122.1 kHz (velocity resolution 0.16\,\kms); the two spectral windows dedicated to observe N$_2$D$^+$(3-2) ($\nu_{rest}$=231.321828\,GHz) and C$^{18}$O(J=2-1) ($\nu_{rest}$=219.560358\,GHz) covered 125 MHz each at frequency resolution of 30.5\,kHz (velocity resolution 0.04\,\kms).
With respect to the weather conditions, the PWV values ranged between 0.25 and 2.16\,mm during the total power observations, and approximately 1.8\,mm for the interferometric observations. 
Calibration of the raw visibility data was performed using the standard reduction script for the Cycle 4 data provided by the ALMA Observatory. This pipeline ran within the Common Astronomical Software Application (CASA 5.1.0, \citealt{McMullinetal2007}) environment. The integration time and the calibrators used to correct for instrumental and atmospheric disturbances (flux, phase and bandpass) are listed in Table \ref{tab:sources}.

 In the total power observations, the frequency setup was exactly the same as in the case of the interferometric observations. In this work we directly use the total power images delivered by ALMA. 

The calibrated interferometric data were cleaned in CASA to produce continuum images. The spectral line cubes were produced by subtracting the continuum and applying a standard cleaning with primary beam correction. Imaging and data combination methods for the data are explained in greater detail in Section \ref{sec:imaging}.

\subsection{ALMA archive continuum data}\label{sec:cont-archive}
Aiming for a better understanding of the finer structure of the continuum sources, we complemented the observations with data products from the ALMA archive. Specifically, we searched the archive for all continuum data available for our sample, and retrieved high-angular resolution (taken with the so-called ALMA 12m array) continuum images of five of the sources: \iras,  \irasb, \merin, \sz\ and \j. The observational parameters, such us frequency band, central frequency, number of antennas, shortest and longest baseline, time on-source and flux, phase and bandpass calibrators are compiled in Table \ref{tab:obs_parameters}.

For some of these sources, we self-calibrated and cleaned the data, and we found that the quality of the continuum images (signal-to-noise ratio) did not improve considerably with respect to the archival product images (since the continuum sources are quite compact and the on-source exposure times are fairly short). Moreover, the small flux variations from non-self-calibrated and self-calibrated continuum emission was always within the 10\% uncertainty of the flux measurements. Hence in this case we decided to use the pipeline delivered images.

\begin{table*}
 \caption{Observational parameters of the high resolution continuum emission obtained from ALMA archive. Data described in Section \ref{sec:cont-archive}}
    \centering
        \resizebox{\textwidth}{!}{
    \begin{tabular}{cccccccccc}
    \hline
    Project & Band & Ctl freq &    Resolution     &\#\,Ant & B$_{min}$-B$_{max}$ & Time on source  & \multicolumn3{c}{Calibrator} \\
    \cline{8-10}
            &      &   (GHz)  & \arcsec\x\arcsec  &        &       (m)           &   (seg)   &  Flux  &  Phase  &  Bandpass \\
    \hline
     \multicolumn{10}{l}{{\bf IRAS\,15398-3359}}\\
    \hline   
    2013.1.00879.S	  &  6	& 225.646	& 0.44\x0.41 & 32	& 19- 650	& 1229	&   Titan 	 & J1534-3526 &	J1427-4206 \\ 
    2011.0.00628.S	  &  7	& 343.420	& 0.55\x0.36 & 32	& 19-558	& 180	&   Titan	 & J1534-3526 &	J15427-406 \\
    2013.1.00244.S	  &  8	& 402.004	& 0.35\x0.26 & 32	& 21-538	& 1511	&   Titan	 & J1517-2422 &	J1427-4206 \\
    \hline
     \multicolumn{10}{l}{{\bf IRAS\,16059-3857}}\\
    \hline   
    2015.1.00306.S	&  3	& 107.679	& 1.86\x1.54 & 36	& 15-452	& 2104	& J1517-2422	& J1610-3958	& J1517-2422 \\
    2013.1.00879.S	&  6	& 225.645	& 0.45\x0.40 & 32	& 19-650	& 1241	& Titan	        & J1534-3526	& J1427-4206 \\
    2013.1.00474.S	&  7	& 357.981	& 0.25\x0,14 & 32	& 15-1466	& 132	& Ceres	        & J1610-3958	& J1517-2422 \\
    \hline
    \multicolumn{10}{l}{{\bf J160155-41523}}\\
    \hline
    2016.1.00571.S	&  3	&  97.548	& 0.39\x0.29 & 40	& 18-3143	& 241	& J1427-4206	& J1610-3958	& J1517-2422 \\
    2016.1.00459.S	&  6	& 254.706	& 0.20\x0.14 & 40	& 16-2692	& 393	& J1517-2422	& J1610-3958	& J1517-2422 \\
    2013.1.00474.S	&  7	& 352.343	& 0.22\x0.13 & 32	& 15- 1574	& 143	& Ceres	        & J1610-3958	& J1517-2422 \\
    \hline
         \multicolumn{10}{l}{{\bf Sz102}}\\
    \hline   
    2016.1.01239.S	&  6	& 225.282	& 0.24\x0.18 & 40	& 16-2647	& 180	& J1427-4206	& J1610-3958	& J1517-2422 \\
    2016.1.01239.S	&  7	& 335.141	& 0.19\x0.18 & 40	& 15-1124	& 270	& J1517-2422	& J1610-3958	& J1517-2422 \\
    \hline
         \multicolumn{10}{l}{{\bf Merin\,28}}\\
    \hline   
    2013.1.00474.S	&  7	& 352.343	& 0.19\x0.13 & 32	& 15-1574	& 132	& Ceres	    & J1610-3958	& J1517-2422 \\
    \hline
    \end{tabular}
    }    
    \label{tab:obs_parameters}
\end{table*}

\subsection{ALMA archive molecular data}\label{sec:archive}

We also retrieved high-angular resolution data cubes from the ALMA archive for the sources presenting outflow activity (\iras, \irasb\ and \j).

The 12m array observations toward IRAS\,15398-3359 (2013.1.00879.S) were carried out during ALMA Cycle 2 on 2014 April 30 with 34 antennas and on 2014 May 19 and June 6 with 36 antennas. Since the extent of the outflow is small, these observations were only a single pointing, enough to observe the entire outflow. The spectral lines observed were \co(2-1), \cdo(2-1) and SO(J$_N$=6$_5$, 5$_4$).

Additional data from IRAS\,16059-3857 (also named Lupus\,3\,MMS) were taken using the 7m array, the 12m array and the total power (TP) from ALMA over the whole outflow extent (2017.1.00019.S). The total power data were observed during the period of 2018 April 7 -- 2018 June 9. The 7m data were observed during the period 2018 April 2 -- 2018 May 11, and 31 pointings were used to make a mosaic map. 12m data were observed on 2018 June 4 and 5 and 105 pointings were used to make the mosaic map. 

The high-angular resolution observations associated with J160115-41523 (2015.1.01510.S) consist of \co(2-1) observations with a single pointing towards the central source, and were taken on 2016 April 2. Note that the Band 6 continuum data used in this work does not correspond to this project, since we found in the archive this source observed in Band 6 with an angular resolution similar to that found in Band 3 and 7 (see Section \ref{sec:cont-archive}).

The observational parameters, such as configuration, achieved angular resolution, number of antennas, shortest and longest baseline, time on-source and flux, phase and bandpass calibrators are compiled in Table \ref{tab:obs_parameters2}.

\begin{table*}
    \caption{Observational parameters of the e high resolution ALMA archive's molecular data. Data described in Section \ref{sec:archive}
    }
    \centering
    \begin{tabular}{ccccccccc}
    \hline
     Project &  Band &  Resolution   &\#\,Ant & B$_{min}$-B$_{max}$ & Obs time  & \multicolumn3{c}{Calibrator} \\
    \cline{7-9}
       &   &  \arcsec\x\arcsec &    &       (m)              &   (min)   &  Flux  &  Phase  &  Bandpass \\
    \hline
     \multicolumn{9}{l}{{\bf IRAS\,15398-3359}}\\
    \hline   
      2013.1.00879.S & 6 & 0.57\x0.49 &  34-36	& 20-650 	& 88	& Titan  & J1534-3526  & J1427-4206  \\
         \hline
     \multicolumn{9}{l}{{\bf IRAS\,16059-3857}}\\
    \hline   
      2017.1.00019.S	& 6 & 1.73\x1.10  &  48-50 & 15-360 & 84.6	& J1517-2422 & J1610-3958	& J1517-2422 \\
    \hline
    \multicolumn{9}{l}{{\bf J160155-41523}}\\
    \hline
       2015.1.01510.S & 6 & 0.92\x0.85 &  43	&15-453	&	16 & J1617-5848 & J1604-4228 & J1427-420  \\
     \hline
    \end{tabular}
    \label{tab:obs_parameters2}
\end{table*}

\subsection{IR data}\label{sec:IR}

In order to fit the spectral energy distribution (SED) of the sample's sources we use data from IR to optical wavelengths (Lopez, SOLA collaboration et al. 2020 in prep.), along with the fluxes we measure in ALMA Band 6. 
A table detailing the data (band, wavelengths, FWHM and fluxes) has been included in Appendix B.

The association between ALMA and optical/IR counterparts was first based on a visual inspection. In the case that the association was not clear for any source, we compare the astrometric accuracy/pointing error of each telescope with the distance to the source position. Finally, we checked the angular resolution (FWHM) of each instrument if there was any doubt about the association. We followed the same procedure by Santamaría-Miranda et al. (accepted), where a more extensive list of telescopes/instruments and errors can be found.

\subsection{Imaging and data combination}\label{sec:imaging}

In this section we describe the imaging and data combination procedure we carried out for the different data sets, which includes observations from the SOLA project (hereafter 7m and total power observations, see Section \ref{sec:SOLA_obs}), and, for some sources, the data obtained from the ALMA archive (hereafter 12m observations) described in Section \ref{sec:archive}. 

Due to the physical size of the dishes and the minimum separation (shortest baseline) between any two, interferometers never sample the central region of the (u,v)-plane. The incompleteness of the (u,v)-coverage at low spatial frequencies (know as the short-spacing problem) makes interferometers insensitive to the emission on large angular scales. 
The effect of missing short-spacings is negligible for objects that are small in comparison to the extent of the primary beam, but for objects with large extended structures, the lack of sensitivity towards low spatial frequencies is a severe shortcoming. To overcome the short-spacing problem the data need to be combined with those of a single-dish telescope, which are capable of measuring the total power. For this reason we have combined interferometric and single-dish observations from those data sets where we encounter large-scale extended emission (outflows) in single-dish observations. 

Below we describe in detail the procedure that was carried out for each of the sources:

\begin{itemize}
    \item \aztec, \aztecb, \sz\ and \merin:
For these sources with only SOLA project data, the 7m calibrated visibilities were Fourier transformed and cleaned with the CASA task \texttt{tclean}. We set the Briggs weighting parameter robust$=0.5$ for the continuum and \co(J=2-1) images as a compromise between angular resolution and SNR (typical beam of 7\farcs0$\times$4\farcs5, PA=86\deg), and we set robust$=2$ -- which ensures higher signal-to-noise ratio -- for the \cdo(J=2-1), \nddp(3-2) and SiO(4-3) images, where the emission is fainter (typical beam of 8\farcs0$\times$4\farcs8, PA=82\deg). The rms noise level in the continuum images is around 2-4\,mJy/beam, and $\sim$ 100-150\,mJy/beam in the 0.17\,\kms\ channels of the line velocity cubes. The images produced from the total power single dish have an angular resolution of 29\arcsec. For these sources the total power data only shows cloud emission, therefore we did not perform the combination with the 7m array data.

\item \iras:
In the case of \iras the 7m array visibilities (\co\ and \cdo) were concatenated with the 12m visibilities from the archive data and cleaned together using the CASA task \texttt{tclean}, setting a Briggs weighting parameter robust$=0.5$ for the  \co(J=2-1) and robust$=2$ for the \cdo(2-1).
The FWHM of the clean beam after imaging the concatenated 7m+12m data is 0$\farcs$6$\times$0$\farcs$5, with a position angle 54\fdg0 and a rms noise level of 7\,mJy/beam in a spectral channel (velocity resolution 0.16\,\kms). The interferometric (7m and 12m) and total power images were combined using the CASA task \texttt{feather}. This algorithm converts each image to the gridded visibility plane, combines them, and reconverts them into a combined image. The combined image preserves the detailed structures revealed by the interferometric data while also recovering the extended emission filtered by them. 
We also note that the high-resolution archival data of this source includes the SO line, which was not observed under the SOLA project. We re-imaged these observations and obtained data cubes with an angular resolution of 0\farcs6\x0\farcs5 with a position angle 36\fdg5 and a velocity resolution of 0.16\,\kms, reaching an rms of 6.7 mJy/beam.

\item \irasb:
Combining the interferometric data (7m and 12m) available for \irasb\ we have obtained, after imaging with a Briggs weighting parameter robust$=0.5$, a \co(2-1) data cube with an angular resolution of 1\farcs7$\times$1\farcs1, a beam position angle 86\fdg5 and a signal to noise ratio of 13\,mJy/beam at a spectral channel width of 0.32\,\kms. 
Since we have total power data of the full outflow associated with this source, we combined it with the interferometric data via the CASA task \texttt{feather}, obtaining images that recover the extended emission flux while preserving the high angular resolution.

\item \j:
In the case of \j\ we do not have the complete outflow mapped by the 12m array, but only one single pointing \co(2-1) observation towards the central source. The angular resolution of this high resolution data is 0\farcs9$\times$0\farcs8 with a beam position angle 69\fdg1 and a rms noise level of 15\,mJy/beam at a velocity resolution of 0.16\kms, obtained using a value of Briggs weighting parameter robust$=0.5$.
The 7m observations of the full outflow were cleaned, setting the Briggs weighting parameter robust$=0.5$ for the continuum and \co(J=2-1) images, and combined with the total power data by using the CASA task \texttt{feather}, recovering the full outflow emission. The angular resolution of this combined images is 7\farcs5\x4\farcs7 (PA=-80\deg) and the velocity resolution 0.17\,\kms, achieving an rms 75 mJy/beam. 

\end{itemize}

\section{Results} \label{sec:results}

\begin{table*}
    \caption{Emission detected towards the 7 observed sources. Continuum emission, detected towards five sources, is identified as point-like in the dedicated 7m array observations (Band 6), while 4 of them are resolved in observations of 12m in Band 7 from the ALMA archive.
    ($\ast$): Only cloud emission is detected (no compact emission). In the last column we indicate in which sources we have detected \co(2-1) outflows associated ($\ast\ast$): only signs of the presence of outflows are detected.}
    \centering
    \resizebox{0.9\textwidth}{!}{
    \begin{tabular}{lccccccccc}
    \hline
   Source name    & \multicolumn{2}{c}{Continuum} &  \co\   & \nddp\   &   \cdo\   & SiO   &  DCN  & CH$_3$OH & Outflow \\
   \cline{2-3}    
                  &       7m    &    12m          &         &       &              &       &       &          &         \\
    \hline
AzTEC-lup1-2      &   no  	   &  no & TP,7m*     &	TP	   & TP,7m*    & no    & TP    &   no     & no  \\
AzTEC-lup3-5      &   no       &  no & TP,7m*     &	no     & TP,7m*    & no	   & no    &   no     & no  \\
IRAS\,15398-3359  & point-like & point-like & TP,7m,12m  &	TP,7m  & TP,7m,12m & TP,7m & TP,7m &   TP   & yes \\
IRAS\,16059-3857  & point-like & resolved & TP,7m,12m  &	TP     & TP,7m	   & no	   & TP    &   no   & yes \\
J160115-41523     & point-like & resolved   & TP,7m*,12m &   no     & TP      & no    & no    &   no    & yes  \\    
Merin\,28         & point-like & resolved & TP,7m      &	no	   & TP        & no	   & no    &   no   & yes** \\
Sz\,102           & point like & resolved & TP,7m &   no	   & TP,7m*	   & no    & no    &   no       & yes**  \\   
    \hline
    \end{tabular}
    }
    \label{tab:results}
\end{table*}

\subsection{Continuum emission} \label{subsec:continuum}

Millimeter continuum thermal dust emission was detected with ALMA toward 5 sources in our sample. 
As shown in Table \ref{tab:results}, the five sources detected as point-like in the dedicated 7m array data are resolved in high resolution ALMA archive data, except \iras\ which is detected as a point-like source within low-level extended emission, even in 12m ALMA data.

\begin{figure*}
 \centering
 \rotatebox{90}{
\begin{minipage}{\textheight}
 \includegraphics[width=1.0\textwidth]{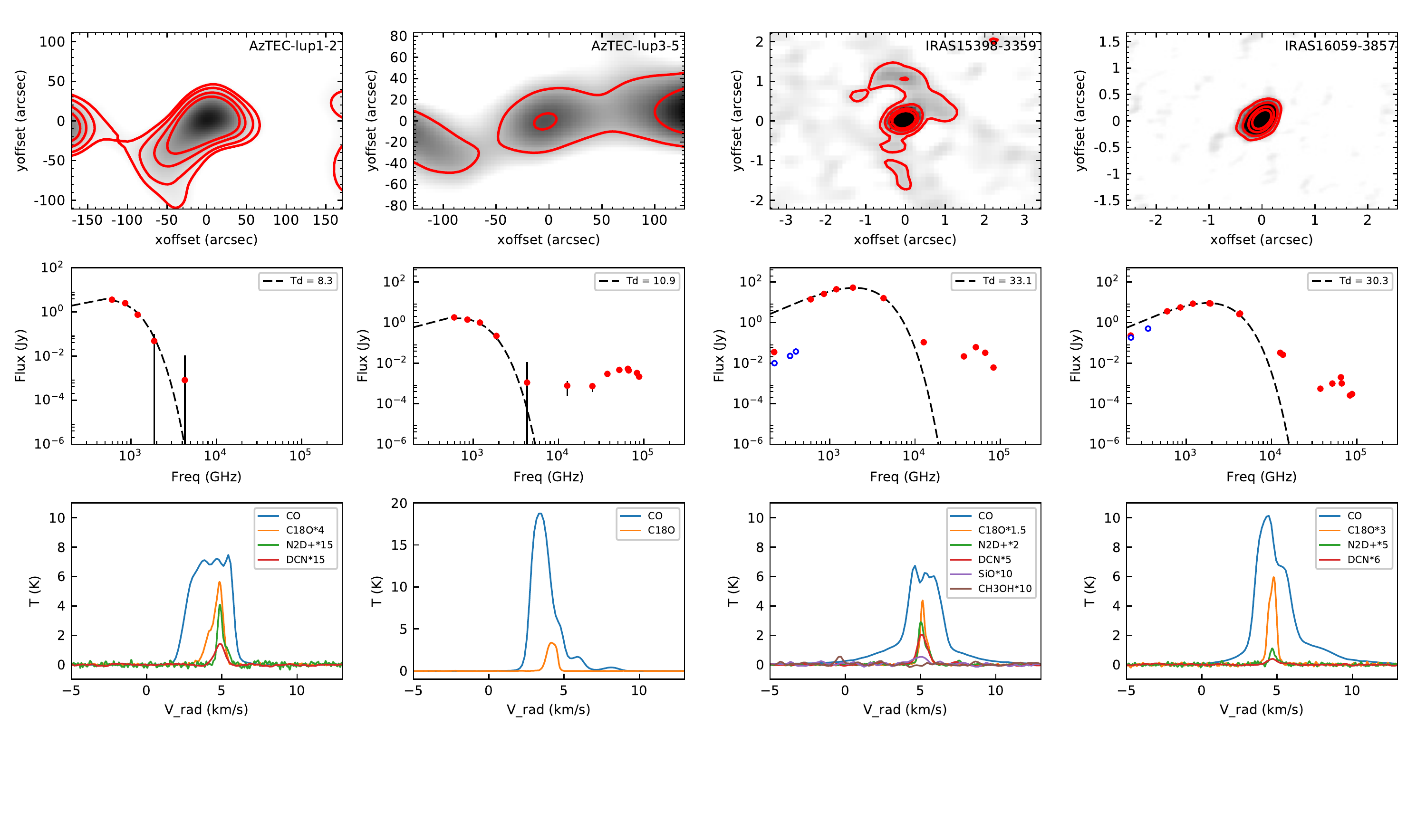}
 \caption {{\it Upper panels}: Continuum emission towards \aztec, \aztecb\ (ASTE, 1.1\,mm), \iras\ and \irasb\ (ALMA, 0.8\,mm).  Contours: 4, 10, 20 and 30 $\sigma$. {\it Middle row panels}:  Spectral energy distribution. A black-body could be fitted considering only the FIR and mm emission (black dashed line). The open blue dots corresponding to emission detected in the high resolution mm images are not included in the fit. The dust temperature derived from the fitting is shown in the top right corner. {\it Bottom panels}: Line emission detected towards the sources at the position of the protostar.}
   \label{fig:figure1}
  \end{minipage}
  }
\end{figure*}
\renewcommand{\thefigure}{\arabic{figure} (Cont.)}
\addtocounter{figure}{-1}
\begin{figure*}
 \rotatebox{90}{
\begin{minipage}{\textheight}
 \includegraphics[width=1.0\textwidth]{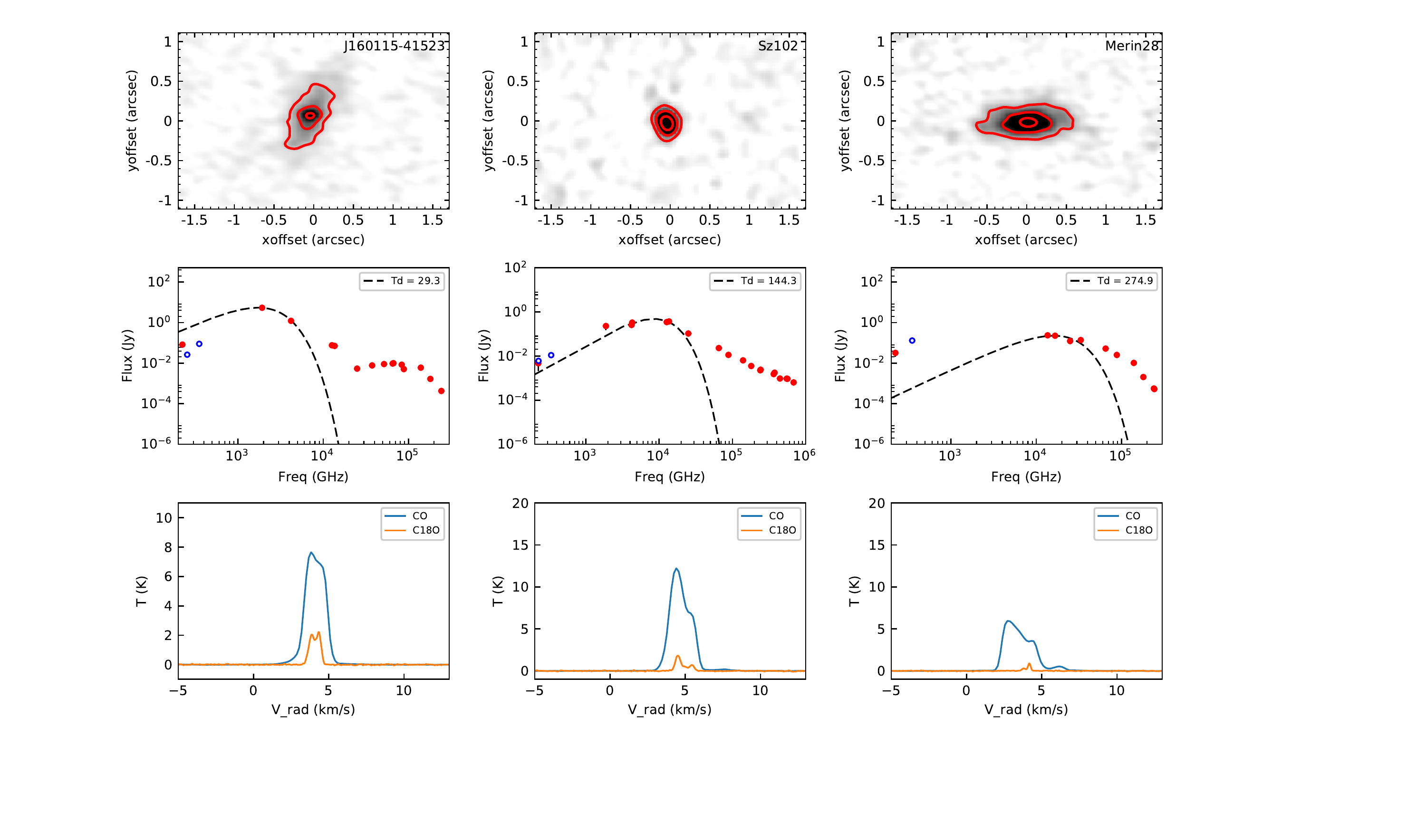}
 \caption {Same as Figure \ref{fig:figure1} for \j, \sz\ and \merin.}
   \label{fig:figure2}
  \end{minipage}
  }
\end{figure*}

\renewcommand{\thefigure}{\arabic{figure}}

In the upper panels of Figure \ref{fig:figure1} we show the continuum emission associated with each source. Compact emission from \aztec\ and \aztecb\ is not detected by ALMA,  so we display the AzTEC/ASTE 1.1\,mm continuum emission which traces the clumps where the sources are embedded. In \iras, \irasb, \j, \sz\ and \merin, the ALMA Band 7 continuum emission traces the disks, while extended emission is also detected associated with \iras.
These high resolution images, show that the sizes of the sources range between 0\farcs2 and 0\farcs8 ($\sim$40 to 120\,AU). 

In order to estimate the continuum emission parameters (major and minor axis sizes, position angles, integrated fluxes and peak fluxes) and their uncertainties, we fit 2D-Gaussians by using the CASA task \texttt{imfit}. Although this model may not be the best one for all our sources, we found that the maximum values of the residual maps are below 4$\sigma$, so they are a good approximation.
Assuming that the sources inherently have a circular shape, we estimate that the inclination angles of the disks are between 50$\degr$ and 70$\degr$\footnote{The disk inclination angle is defined as the angle between the plane of the disk and the plane of the sky, so that an edge-on disk would be inclined by $90\degr$.}.
Table \ref{tab:2gauss} lists for each source the Gaussian fitting parameters: position, deconvolved FWHM of the major and minor axes, position angle, size of the source and inclination. In the case of \iras, since the source is marginally resolved we can not be sure if the inclination calculated from the Gaussian fit is the real one, however it is in good agreement with that calculated previously by other authors based on H$_2$CO(5$_{15}$-4$_{14}$) ALMA observations, assuming that the outflow cavity has a parabolic shape and its velocity is proportional to the distance to the protostar \citep{Oyaetal2014}.  
We propose that the continuum emission detected in high resolution images corresponds mostly to the disks (typical disk sizes $<$ 150 pc), and the emission detected in the ALMA 7m observations is emitted by the dust envelope+disk, as the angular resolution of these observations ($\sim$7\arcsec) at the Lupus distance corresponds to $\sim$ 1000 AU.

In the central panel of Figure \ref{fig:figure1} we show the Spectral Energy Distributions (SEDs) of the seven sources. Filled red circles correspond to the IR fluxes described in Section \ref{sec:IR}, and to mm measurements taken from the ALMA 7m continuum images. Open blue circles represent the fluxes measured in the high-resolution ALMA images. The fluxes were fitted by a modified black-body model, shown with a black dashed line.

We estimate the dust temperature $T_d$ from fitting the IR SED of the sources (see Figure\,\ref{fig:figure1}). To obtain the temperature of the cold dust, we fit a black-body model to the emission at millimeter and FIR wavelengths. 
The fit was optimized using the Python scipy \citep{2020SciPy} task \texttt{curve\_fit}, which uses a non-linear least squares algorithm to fit a function to data. The errors come from the uncertainty of the black body fit, and are obtained from the covariance matrix provided by the task. Usually this emission is fitted by a grey-body model with a variable submillimeter slope, however due to the lack of information in the submillimeter spectrum and the low resolution of some infrared data we opted to fit a simple black body model.

In fitting, we do not include high angular resolution millimeter emission (open blue circles) as the fluxes could be affected by emission filtering, and in most cases are likely lower limits. 
A black body fits well with the data in the FIR for \aztec, \aztecb, \iras, \irasb\ and \j, and the derived temperatures show the presence of cold dust (between 8.3 and 32.0 K), while \sz\ and \merin\ fit with a black body at a higher temperature (144.3 and 275.0 K, respectively). However, for the latter source, the lack of FIR data makes this value unreliable.
The estimated $T_d$ values for each source are listed in Table \ref{tab:fluxes}. 

The SED corresponding to \aztec, which only includes FIR data, is typical of prestellar cores, and emission appears to pertain to isothermal dust with $\sim 10$K, peaking at $\lambda$ $>$ 100$\mu$m. The protostar seems to be completely covered by gas and dust and is obscured with a large optical depth by the dust envelope. No conclusion can be reached from the stellar-black body radiation. The source has no compact emission and only extended millimetric emission is detected, so we are probably observing the moment when the dust cloud starts to collapse and the disk has not yet formed.  

In the case of \aztecb, we do not have a mm detection either; however, this source emits over a wide wavelength range and the SED has the appearance of a less embedded object. On the other hand, the black body fit yields a dust temperature value close to 10\,K, so it would be one of the coldest sources in our sample. As we cannot see the disk because there is no compact mm emission we consider two possible scenarios: (1) that we are actually seeing two objects in the line of sight (a prestellar core and an extinct reddened star), or (2) indeed it is a protostar that has already collapsed and has not yet formed the disk around it, so we don't see the compact emission (this would be rare).
 
 
The SEDs we obtained for \iras, \irasb\ and \j\ correspond to Class 0/I objects, which are dominated by infalling material from the parent molecular cloud to the central object with the eventual presence of material flowing from the object towards the surroundings (outflows). When part of the envelope is dissipated by the outflows, the object becomes detectable in the near infrared. This scenario is in accordance with \iras, \irasb\ and \j, which are sources that have outflows associated. The FIR emission in these sources can be fit by black bodies with temperatures $\sim$\,30K.
 
In contrast to any other source of our sample, the SEDs of \sz\ and \merin\ show peak flux at a wavelength near to 22 \mic, rather than in the FIR. In both cases, also in contrast to the other sources, the mm points are above the black body curve, so the fitting is probably not appropriate, especially in the case of \merin, where no IR observations are available at wavelengths larger than 22 \mic.

Furthermore, we estimate the mass associated with the sub-millimeter continuum emission. Assuming isothermal dust emission, well-mixed gas and dust, and optically thin emission, the dust masses are given by,

\begin{equation}
M_d = \frac{S_\nu D^2}{B_\nu(T_d) \kappa_\nu}
\label{eq:m_dust}
\end{equation}

\noindent where $S_\nu$ is the continuum flux density, $D$ is the distance to the source, $B_\nu$ is the Planck function, $T_d$ is the dust temperature and $\kappa_\nu$ is the mass dust opacity coefficient \citep{Hildebrand1983}. The dust opacity was estimated from $\kappa_\nu$ = 0.1($\nu$/10$^{12}$Hz)$^\beta$ \cmdos\ gr$^{-1}$ (e.g. \citealt{BeckwithandSargent1991}). The value of $\beta$ was estimated by fitting the flux densities at different millimetric wavelengths with a power law, $F_\nu \propto \nu^{2+\beta}$, based on measures of the present observations and the literature. 
Since we are assuming optically thin dust emission and not taking into account possible scattering effects \citep{Zhu2019}, the derived masses should be considered as lower limits.  We derive the dust temperature by fitting the SED of every source.
For this purpose we also use ALMA archival data from the sources in different spectral bands (see Table \ref{tab:obs_parameters}). The measured fluxes are listed in columns 2 to 6 of the Table \ref{tab:fluxes}.
The error in $\beta$ was derived from the fit assuming a 10\% uncertainty in the ALMA observations due to calibration errors.
For sources with insufficient measurements to make the fit, we adopted a value of $\beta$ = 1.8$\pm$0.2 a typical opacity spectral index value for the ISM at submillimetric wavelengths \citep{Draine2006}. 
In all cases, the $\beta$ values obtained from the fitting (see Table \ref{tab:fluxes}) range between 0 and 1, according to the values of $\beta$ reported by other authors \citep{Draine2006,BeckwithandSargent1991,Ribasetal2017,Ansdelletal2018}.

The values of $M_d$, listed in the last three columns of the Table \ref{tab:fluxes} (expressed in jovian masses), were calculated from fluxes measured in the dedicated observations and in high resolution continuum archive images, showing similar results, which are within the errors. The errors in the dust masses were calculated by propagating the errors in fluxes, dust temperatures and $\beta$.
Since for both \aztec\ and \aztecb\ we cannot detect compact emission, the estimated dust mass values were calculated by using a flux equal to three times the $rms$ and therefore represent an upper limit.
For three of the sources where compact emission was detected (\iras, \irasb\ and \j) we note that the dust mass calculated from the high resolution observations is lower than that calculated from the 7m observations. 
This could have two explanations: the flux loss produced by the extended emission filtering, or to the fact that at high resolution we are measuring the mass of the dust associated to the disk and at low resolution the mass associated with the envelope and the disk.
This second hypothesis seems to work well for  relatively large samples \citep[e.g.,][]{Tobinetal2020}. We make the caveat that unresolved edge-on disks with large envelopes can be confused with disks with high inclination and the 12m observations may include emission from these envelopes. In the case the case of \sz\ and \merin\ (where the masses calculated from the measured fluxes with the 7m and 12m arrays are practically the same) we consider that the envelope mass would be negligible, and we are only measuring the disk mass. Particularly in the case of Sz102, other studies conducted with ALMA observations have shown that no traces of a massive envelope are observed \citep{Louvetetal2016}.
 
 \begin{table*}
     \caption{Deconvolved parameters of the continuum sources from a 2D-Gaussian fit obtained from the 12m observations in Band 7. To obtain the derived disk inclinations, we measured the angle between the plane of the disk and the plane of the sky, so that a face-on disk has $i=0\degr$ and an edge-on disk has $i=90\degr$.}
    \centering
         \resizebox{\textwidth}{!}{
    \begin{tabular}{lccccccccc}
    \hline
    Source           & Position & Wavelength &   MAxis   &   mAxis   & Pos ang &  Spatial size & Int Flux & Peak Flux & Inc\\
                     &  \radec  &  mm        & [marcsec] & [marcsec] &  [\deg] &   UA  &   mJy    & mJy/beam  & [\deg]\\
            \hline
    IRAS\,15398-3359 & 15:43:02.24, --34:09:06.72 & 0.87 & 306$\pm$43 & 124$\pm$54 & 139$\pm$12 & 45.9$\times$18.6 &  24$\pm$1   & 19$\pm$0.5   & 66$\pm$14 \\
    IRAS\,16059-3857 & 16:09:18.09, --39:04:53.30 & 0.84 & 431$\pm$9  & 262$\pm$7  & 142$\pm$2 &  64.6$\times$39.3 & 507$\pm$8   & 116$\pm$2    & 53$\pm$2 \\
    J160155-41523    & 16:01:15.53, --41:52:35.51 & 0.85 & 673$\pm$33 & 274$\pm$19 & 154$\pm$2 & 100.9$\times$41.1 &  92$\pm$4   & 11.8$\pm$0.5 & 66$\pm$3 \\
    Sz\,102          & 16:08:29.72, --39:03:11.42 & 0.90 & 190$\pm$12 & 100$\pm$14 & 13$\pm$6  &  28.5$\times$15.0 &  12$\pm$1 & 6.6$\pm$0.1  & 58$\pm$3 \\ 
    Merin\,28        & 16:24:51.77, --39:56:32.95 & 0.85 & 680$\pm$22 & 253$\pm$8  & 92$\pm$1  & 102.0$\times$37.9 &  124$\pm$4  & 15.3$\pm$0.4 & 68$\pm$1 \\
    \hline
    \end{tabular}
    }
    \label{tab:2gauss}
\end{table*}

\begin{table*}
    \caption{($\ast$): No compact emission is detected, calculated flux and mass values are upper limits assuming Flux=3$rms$ (M$_J$=9.55$\times$10$^{-4}$ \msun).}
    \centering
     \resizebox{0.97\textwidth}{!}{
    \begin{tabular}{lcccccccccc}
    \hline
     Source &   \multicolumn{5}{c}{Fluxes [mJy]}      & $\beta$ & $T_d$ [K] & \multicolumn{3}{c}{$M_d$ [M$_J$]} \\
     \cline{2-6} \cline{9-11}
            & Band\,6 (7m) &Band\,3 & Band\,6 & Band\,7 & Band\,8 &  &  &   7m  & \multicolumn{2}{c}{12m} \\
            \cline{10-11}
            &              &        &         &         &         &  &  &       &    Band\,6 & Band\,7  \\
    \hline
    AzTEC-lup1-2*    & $<$11.5 & ...  &  ...  &   ... &  ... &  1.8$\pm$0.2    &   8.3$\pm$0.2 & 61.0$\pm$13.9  &     ...        &     ...        \\
    AzTEC-lup3-5*    & $<$19.0 & ...  &  ...  &   ... &  ... &  1.8$\pm$0.2    &  10.9$\pm$0.3 & 64.8$\pm$13.5  &     ...        &     ...        \\
    IRAS\,15398-3359 &   35.0 &  ...  &  10.5 &  22.7 & 49.9 &  0.13$\pm$0.20  &  33.1$\pm$0.9 & 2.3$\pm$0.3    & 0.68$\pm$0.08  & 0.66$\pm$0.08  \\
    IRAS\,16059-3857 &  230.0 &  32.2 & 175.9 & 472.1 &  ... &  0.24$\pm$0.12  &  30.3$\pm$0.6 & 20.2$\pm$2.3   & 15.3$\pm$1.7   & 16.4$\pm$2.0   \\  
    J160115-41523    &   81.9 &  1.94 &  21.3 &  78.5 &  ... &  0.80$\pm$0.10  &  29.3$\pm$0.6 & 16.3$\pm$2.1   & 3.0$\pm$0.4    & 4.9$\pm$0.7    \\
    Sz102            &    4.7 &  ...  &   6.0 &  10.6 &  ... &  0.06$\pm$0.52  &   144$\pm$6   & 0.06$\pm$0.006 & 0.07$\pm$0.006 & 0.06$\pm$0.006 \\
    Merin 28         &   31.9 &  ...  &  ...  & 130.2 &  ... &  0.40$\pm$0.31  &   270$\pm$20  & 0.31$\pm$0.04  &     ...        & 0.44$\pm$0.07  \\
    \hline
    \end{tabular}
    }
    \label{tab:fluxes}
\end{table*}

\subsection{Line emission} \label{subsec:line}

The lower panels of Figure \ref{fig:figure1} show the line emission associated with the continuum sources described in Section \ref{subsec:continuum}. 
The spectra shown were obtained from the total power data. The interferometric data alone, with the achieved sensitivity, show the detection of only a few lines. Although it is not possible to spatially resolve the line sources in the total power data, we extract the spectra within a 2" box at the position where the continuum sources are located. This procedure is sufficient to achieve our goal, which is simply to identify the molecules present in the source envelopes. The Gaussian fit performed to the lines is very simple, and although it is not perfect for lines with more complex structure (double peaked or asymmetric profiles), it allows us to determine a systemic velocity and an overall line width. The fit was optimized using the python task \texttt{curve\_fit}, and the parameters are shown in Table \ref{tab:gauss_param} along with their uncertainties.

It is possible to see that the chemical complexity varies from one source to another.
All sources exhibit \co(2-1) and \cdo(2-1) emission at velocities ranging 4-5 \kms, but \aztec, \iras\ and \irasb\ are chemically more complex: all three show \nddp(3-2) and DCN(3-2) emission, and \iras\ presents weak SiO(5-4) and CH$_3$OH(6(1,5)-7(2,6)) lines, the latter one at a different velocity than the rest (at --0.4\kms).
In all sources the CO line width is many times wider than the other lines. The systemic velocities were derived from the peak velocities of the molecules which trace highest densities (\cdo, \nddp\ and DCN), considering an error equal to half the channel width. These values coincide with the velocity of the absorption dips in the optically thick lines.


Both \aztec\ and \aztecb\ spectra show multiple components. Since these sources have not been spatially resolved using infrared observations nor the millimeter emission from ASTE, and they do not show any compact emission in the ALMA data, we could not rule out the possibility that there could be more than one source projected in the plane of the sky toward these positions.
However, the \aztec\ spectrum is chemically more complex than that of \aztecb, showing \nddp\ and DCN emission. These molecules are high density tracers: \nddp\ shows the presence of cold gas and sometimes traces quiescent clumps, and DCN survives at low temperatures where other molecules freeze out (see discussion in section \ref{disc:chem}).

The \iras\ spectrum shows prominent \co\ wings, which are a typical feature of outflow sources. 
This source is the only one that shows SiO emission (with v$_{peak}$=4.9\,\kms\ and $\Delta$v=1.4\,\kms) that is usually associated with  young energetic outflows.
Furthermore, it presents a methanol line at a different velocity (v=--0.4 \kms), which does not trace the velocity of the cloud, and it is probably generated in a shock.

The spectrum in \irasb shows an asymmetric double peak of \co\ with large broadening wings, which are probably associated with its outflow.

The three remaining sources (\j, \sz\ and \merin) show only \co\ and \cdo\ emission, with multiple peaks and asymmetric lines.
We note that the two \cdo\ emission peaks in \j\ have similar amplitudes, while those detected in \sz\ are asymmetric. \merin\ has a single \cdo\ peak and the maximum of \co\ is blue-shifted, with v$_{peak}$ at 3.3\,\kms compared to the v$_{lsr}$ at 4.2\kms.




\begin{table}
    \caption{Gaussian fitting parameters of the total power emission. Note that the CH$_3$OH emission in \iras\ is the only with a different mean value.}
    \centering
    \resizebox{0.49\textwidth}{!}{
    \begin{tabular}{lcccc}
    \hline
    Source  & Molecule & Mean & FWHM & Amplitude \\
            &          & [\kms] & [\kms] & [K]   \\ 
    \hline
    AzTEC-1-2  & CO         & 4.24\,$\pm$\,0.005 & 2.83\,$\pm$\,0.013 &  8.03\,$\pm$\,0.033\\
               & C$^{18}$O  & 4.72\,$\pm$\,0.002 & 0.93\,$\pm$\,0.005 &  1.21\,$\pm$\,0.006 \\
               & N$_2$D$^+$ & 4.91\,$\pm$\,0.046 & 0.48\,$\pm$\,0.108 &  0.25\,$\pm$\,0.048 \\
               & DCN        & 4.87\,$\pm$\,0.008 & 0.74\,$\pm$\,0.019 &  0.09\,$\pm$\,0.002 \\
    AzTEC-3-5  & CO         & 3.50\,$\pm$\,0.002 & 1.50\,$\pm$\,0.005 & 19.05\,$\pm$\,0.059 \\
               & C$^{18}$O  & 4.19\,$\pm$\,0.001 & 0.73\,$\pm$\,0.002 &  3.61\,$\pm$\,0.008 \\
    IRAS15398  & CO         & 5.25\,$\pm$\,0.004 & 2.62\,$\pm$\,0.010 &  6.55\,$\pm$\,0.022 \\ 
               & C$^{18}$O  & 5.16\,$\pm$\,0.001 & 0.49\,$\pm$\,0.003 &  2.54\,$\pm$\,0.013 \\    
               & N$_2$D$^+$ & 5.10\,$\pm$\,0.003 & 0.55\,$\pm$\,0.008 &  1.32\,$\pm$\,0.016 \\
               & DCN        & 5.11\,$\pm$\,0.005 & 0.76\,$\pm$\,0.011 &  0.41\,$\pm$\,0.005 \\
               & SiO        & 4.90\,$\pm$\,0.098 & 1.40\,$\pm$\,0.228 &  0.05\,$\pm$\,0.007 \\
               & CH$_3$OH   & -0.37\,$\pm$\,0.051 & 0.52\,$\pm$\,0.121 & 0.06\,$\pm$\,0.011 \\
    IRAS16059  & CO         & 4.60\,$\pm$\,0.005 & 2.38\,$\pm$\,0.011 &  9.48\,$\pm$\,0.037 \\
               & C$^{18}$O  & 4.70\,$\pm$\,0.001 & 0.59\,$\pm$\,0.003 &  1.97\,$\pm$\,0.009 \\
               & N$_2$D$^+$ & 4.71\,$\pm$\,0.009 & 0.50\,$\pm$\,0.022 &  0.20\,$\pm$\,0.009 \\
               & DCN        & 4.69\,$\pm$\,0.025 & 0.83\,$\pm$\,0.006 &  0.64\,$\pm$\,0.004 \\
    J160115    & CO         & 4.10\,$\pm$\,0.009 & 1.46\,$\pm$\,0.004 &  8.12\,$\pm$\,0.021 \\
               & C$^{18}$O  & 4.09\,$\pm$\,0.002 & 0.83\,$\pm$\,0.005 &  2.18\,$\pm$\,0.012 \\
    Sz102      & CO         & 4.61\,$\pm$\,0.002 & 1.56\,$\pm$\,0.006 & 11.58\,$\pm$\,0.038 \\
               & C$^{18}$O  & 4.54\,$\pm$\,0.003 & 0.51\,$\pm$\,0.007 &  1.70\,$\pm$\,0.019 \\
    Merin28    & CO         & 3.33\,$\pm$\,0.005 & 2.05\,$\pm$\,0.012 &  5.65\,$\pm$\,0.028 \\
               & C$^{18}$O  & 4.16\,$\pm$\,0.002 & 0.24\,$\pm$\,0.006 &  8.86\,$\pm$\,0.017 \\
    \hline
    \end{tabular}
    }
    \label{tab:gauss_param}
\end{table}

\subsection{Molecular Outflows} \label{subsec:outflows}

In this section we introduce the outflows associated with three of the sources of our sample (IRAS\,15398-3359, IRAS\,16059-3857 and J160115-41523), by showing the CO(2-1) integrated intensity emission from their red and blue lobes (Figure \ref{fig:mom_outflows}) which delineate and describe well their overall morphology. We estimate the physical parameters of each outflow (Table \ref{tab:param_outflows}) and study their kinematics via the moment 0, 1 and 2 maps (Figures \ref{fig:IRAS15_moments}, \ref{fig:IRAS16_moments} and \ref{fig:J160115_moments}) and the position--velocity (PV) diagrams (Figures \ref{fig:IRAS15-pv}, \ref{fig:IRAS16-pv-red} and \ref{fig:J160115-pv}). We present the velocity cubes of all three outflows in Figures \ref{fig:IRAS15-ellipses}, \ref{fig:IRAS16-chan-red-ellipses} and \ref{fig:J160155-chan-high-resol}.  

We also report the emission associated with \sz\ and \merin, in which we can observe gas at different velocities than the systemic velocity. 

\subsubsection{First Glimpse at the CO Outflow Emission} 

\begin{figure*}
    \centering
    \includegraphics[width=1.0\textwidth]{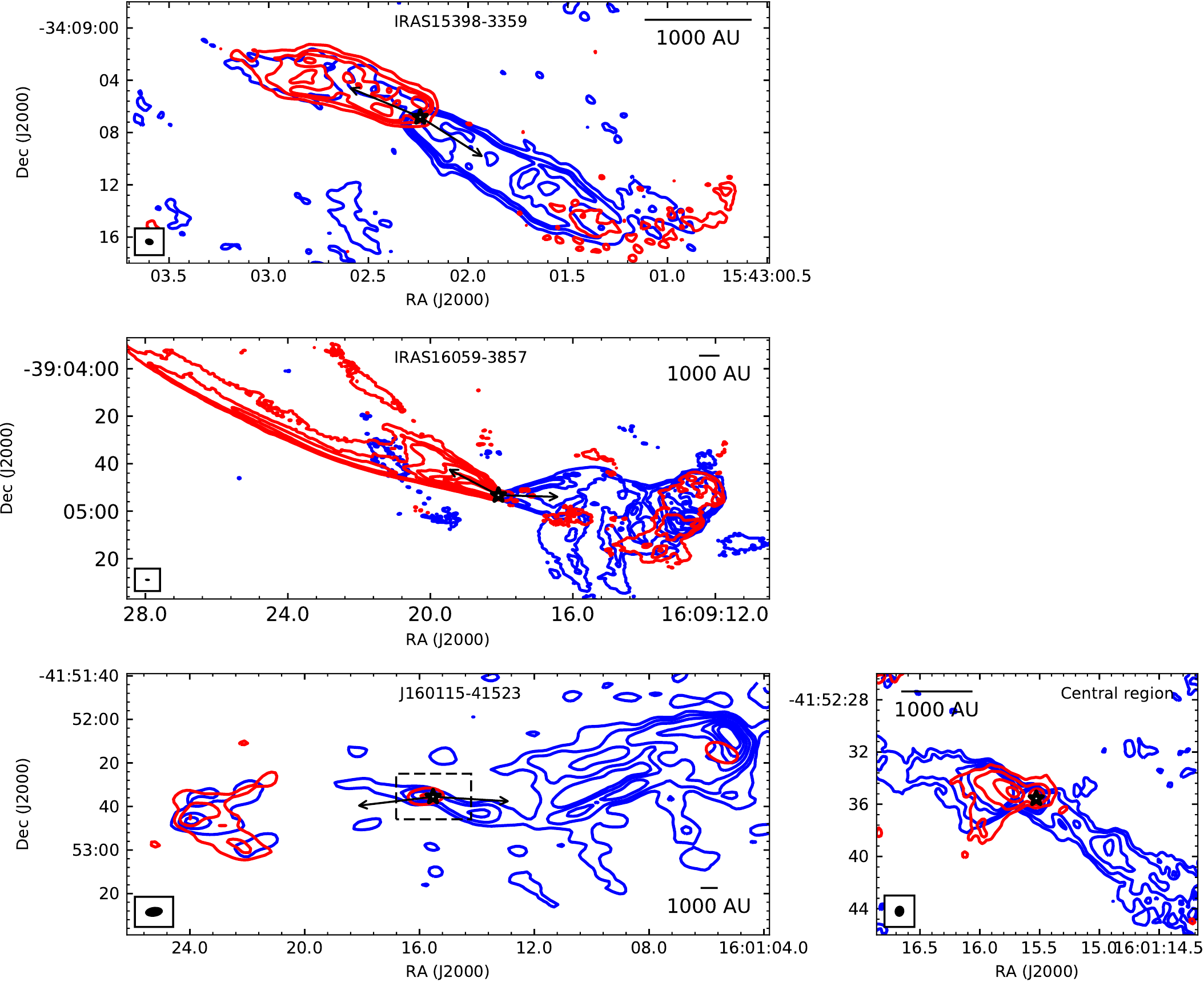}
    \caption{\co(2-1) integrated intensity images showing the blue-shifted and red-shifted emission for the outflows associated with \iras\ (top panel, 12m+7m data), \irasb\ (middle panel, 12m+7m data) and \j\ (bottom panels, 7m data at left and 12m data at right). Each panel is labeled with the name of the source. The minimum and maximum velocities over which the emission is integrated are chosen based on visual inspection of the velocity channels (Table \ref{tab:param_outflows}). For the left panels the contours correspond to 10, 20 and 40 times the $rms$ noise level (the $rms$ noise for the full outflow moment 0 blue/red maps are 0.03/0.04, 0.04/0.05 and 0.2/0.08 Jy/beam \kms\ for \iras, \irasb\ and \j, respectively). The bottom right panel shows the \co\ high-angular resolution integrated intensity emission of the blue- and red-shifted lobes toward the central region of \j, indicated with a box in the bottom left panel. The contours correspond to 3, 5 and 7 times the rms noise levels ($rms$ of the blue/red maps 0.04/0.1 Jy/beam \kms). 
    The locations of the driving sources match the corresponding millimeter continuum emission, and are marked with black stars in the images. The estimated P.A. of each outflow lobe is indicated by black arrows. The beam sizes are shown as black ellipses in the lower left of each panel.}
    \label{fig:mom_outflows}
\end{figure*}

Figure \ref{fig:mom_outflows} shows the integrated \co(2-1) intensity images associated with the three observed sources in which outflows have been detected. In Table \ref{tab:param_outflows} we list the systemic velocity, the integration ranges used to build the redshitfed and blueshifted images ($\Delta$v$_{rad}$, determined by channels with emission above $3\sigma$), the position angle (P.A.), the opening angle ($\theta_{op}$), the size, and the dynamical timescales ($t_{dyn}$) of the outflows, among other measurements obtained from a homogeneous analysis of the moment images for all three outflows. We derive these quantities for the red- and blue-shifted lobes of each outflow separately. $\Delta$v$_{rad}=\mid{\rm v}_{sys}-{\rm v}_{max}\mid$ is the outflow spread in radial velocity, where v$_{sys}$ is the systemic velocity and v$_{max}$ is the velocity up to which it is possible to detect outflow emission over the 3$\sigma$ threshold.
The outflow opening angles are derived from the width of the outflow at a projected distance of 1000\,AU from the originating source.
The outflow position angles are taken as the bisectors of the angles formed by the lines used to define the opening angles. The outflow inclinations are obtained by assuming that they are perpendicular to the disks (Table \ref{tab:2gauss}). 
The sizes are estimated in projection in the plane of the sky (considering the emission over 3\,$\sigma$ threshold), and then corrected by the inclination. The errors in the measured sizes are given by the angular resolution of each image ($\Delta$size = beam/2).  
We estimate the dynamical timescales using the sizes and the radial velocity spreads, $\Delta$v$_{rad}$, deprojected by each outflow inclination angle (which is the complementary angle of the disk inclination i): $t_{dyn}$= size/(v$_{rad}$ cos(90-i)).

A common characteristic of all three outflows is that the outflow opening angle at a given distance decreases as the radial velocity increases. 
Also, in general, the opening angles are smaller if measured at larger projected distances from the originating source. This speaks of the intrinsic difficulty in measuring the outflow opening angle and compare measurements from different authors in the literature. Here we have opted to take the angles from inspecting the integrated intensity images at a certain projected distance of 1000\,AU, since at that distance all three outflows show a structure in which it is possible to perform this measurement, while at greater distances they become irregular.

The outflow associated with the compact continuum source linked with \iras\ (top panel in Figure \ref{fig:mom_outflows}) has been very well studied (see section \ref{sec:IRAS15} and Appendix B). This outflow is very collimated ($\theta_{op}\sim$ 30\deg) and shows a distinct bipolar morphology. It shows brighter CO emission at the cavity walls of the blue-shifted southwest lobe with some arc bridges crossing it, and a more knotty or irregular morphology in the red-shifted lobe. Faint red-shifted emission is detected at the tip of the blue-shifted lobe.
This outflow spreads over 27\,\kms\ in radial velocity, with the blue and red lobes extending 15.3 and 11.6\,\kms\ from the systemic velocity, respectively.
Its size (deprojected using an inclination of 66\fdg14 with respect to the line of sight) is 4350$\pm$94\,AU. The dynamical time derived from the size and velocity $\Delta$v$_{rad}$ of each lobe, is of about 300$\pm$20\,yr, revealing that it would be the shortest of the outflow sources in our sample.

A compact continuum millimeter source is coincident with \irasb\ and we consider it the driving source of the outflow shown in the central panel of Figure \ref{fig:mom_outflows}. This outflow is less collimated than that associated with \iras\ ($\theta_{op}\sim$ 75\deg) and although it looks bipolar, its two lobes are not exactly opposed, but make a 148\deg\ angle. The CO outflow extends for 23\kms, with the blue and red lobes extending for 9.6 and 13.2\kms\, respectively.
The measured outflow size corrected by its inclination (53\fdg2) is 39250$\pm$320\,AU, with the red lobe being almost two times larger than the blue one. The red lobe size estimate represents a lower limit since it is possible that this outflow may extend beyond the observed field of view. We also estimate its average dynamical time to be 5500$\pm$460\,yr.

Close to the central source, a V-shaped structure in both lobes traces the walls of the outflow cavity. As pointed out before by \citet{Yenetal2017}, the blue-shifted lobe is oriented due west and the red-shifted towards the northeast. Away from the central source, the emission from the red-shifted lobe is mostly seen in its southern wall, which extends at least up to the limits of the field of view. 
The blue-shifted lobe presents two singular features: (i) about 3400\,AU west from the source there are several gas shreds that extend due south, 
(ii) about 7300 AU west from the source lies the center of a shell-like structure. This structure spatially coincides with gas moving at red-shifted velocities and a previously reported Herbig-Haro object (HH\,78). The possible causes of this peculiar structure will be discussed in Section \ref{sec:IRAS16}.

The bottom panels of Figure \ref{fig:mom_outflows} show the CO emission distribution of the molecular outflow associated with the millimeter continuum source linked to \j. The outflow velocity spread of the blue and red lobes are 5.6 and 2.6 \kms, respectively, with a total spread of 8.2\,\kms, and the total size is 40550$\pm$1214\,AU. Furthermore, its average dynamical time is $>$18000$\pm$700\,yr, making it the largest outflow in our sample.

Close to the central source (bottom right panel of Figure \ref{fig:mom_outflows}), the outflow shows a bipolar morphology with the red and blue-shifted lobes pointing in a southeast/west direction. It comprises two V-shaped structures associated with the outflow cavity walls of both lobes, and we measure an average opening angle of $\sim$60\deg. 

The presence of blue-shifted emission eastward from the central source (i.e., toward the red-shifted lobe) could be due to the closeness of the outflow main axis to the plane of the sky (note that it is inclined about $24\degr$ with respect to the plane of the sky), and that the wide-angle flow is expanding toward both blue- and red-shifted directions. Further from the central source, the CO emission from the outflow comes mainly from the walls of the cavity dragged by the wind and two bow-like structures at the tips of both lobes. Noteworthy, at a distance of about 5500\,AU west from the central source, the blue-shifted lobe deviates about 20\deg north (from a PA of 266\deg close to the young star to a PA of 282\deg at the tip of the blue-shifted lobe of the outflow), turning into a northwest direction. 

\begin{figure}
    \centering
    \includegraphics[width=0.49\textwidth]{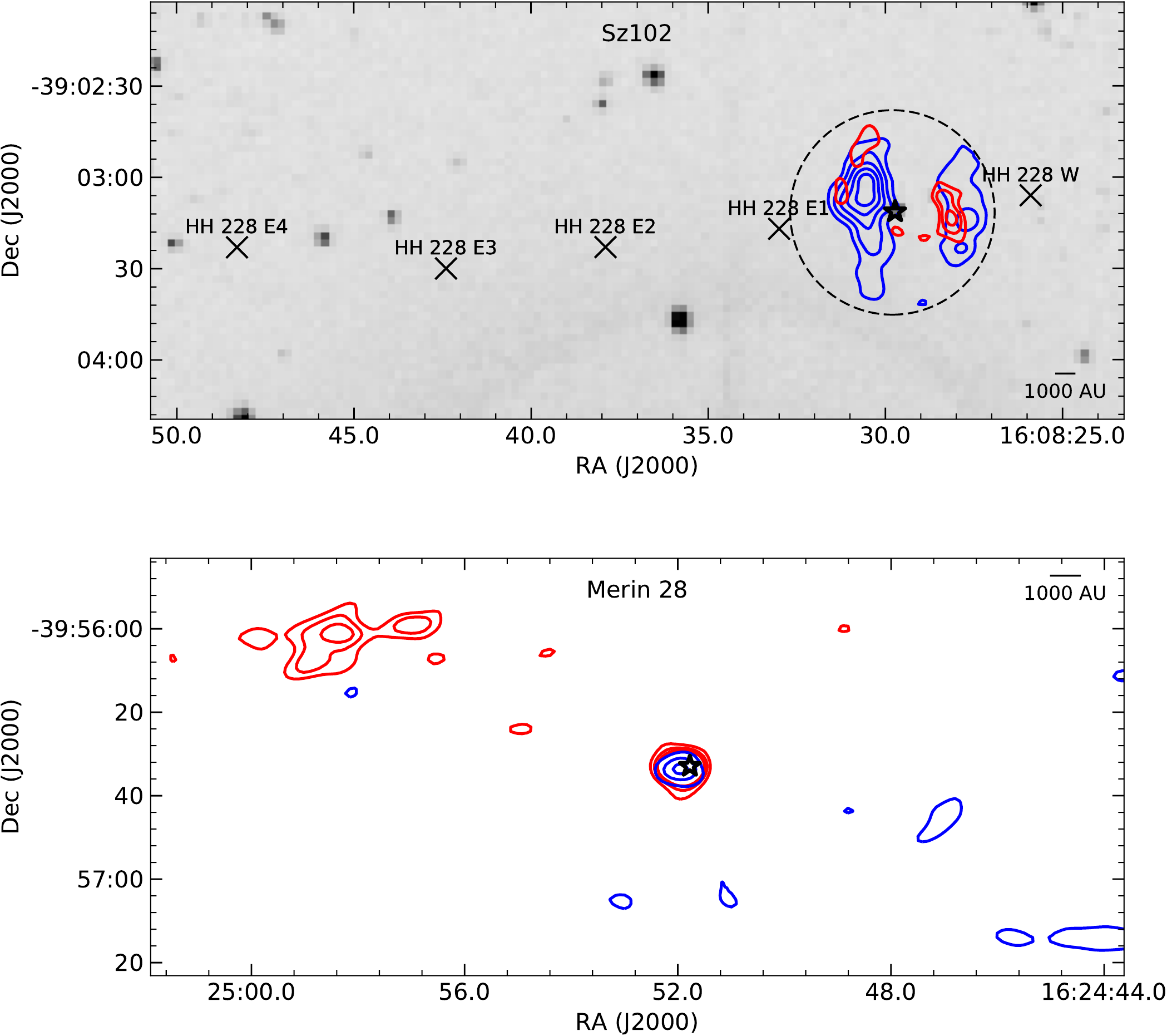}
    \caption{Blue- and red-shifted emission associated with \sz\ (blue contours: 2, 4, 6, 8 and 10\,rms; red contours: 2, 3, and 4\,rms) and \merin\ (contours: 4, 8, and 12\,rms). Plus signs in the top panel display the position of the Herbig-Haro objects HH\,228 (over the DSS2 IR image), that lie out of the region mapped by the dedicated 7m array  observations (black circle). The positions of the continuum emission are indicated with a black star.}
    \label{fig:outflows-sz+merin}
\end{figure}

In two other sources of our sample we have detected signs of the presence of outflows, although we have not been able to delineate them clearly. In Figure \ref{fig:outflows-sz+merin} we show moment 0 maps of the blue- and red-shifted emission associated with \sz\ and \merin.

\sz\ is associated with the Herbig-Haro objects HH\,228\,W and HH\,228\,E1, E2, E3 and E4, that extend far beyond the observed field. The dedicated 7m array observations are able to detect blue- and red-shifted emission in the east-west direction, which is in agreement with the position angle of the jet (first reported by \citealt{HeyerandGraham1989}) associated with \sz\ (see Appendix A). The velocity ranges in which the blue- and red-shifted emission is detected are 1.0 and 0.7\,\kms, respectively. Probably we are detecting the remnants of the molecular envelope in which the source formed. 

The red-shifted emission associated with \merin\ is an arc-shape structure located at 82\arcsec\ of the central source subtending an angle between PA 55\deg\ and 75\deg. This structure is only present in a few channels covering a velocity range of only 0.5\,\kms.
The blue-shifted emission is detected barely over the noise level and only in two velocity channels. We probably are not able to detect it since it is contaminated with cloud emission. Both maps show strong emission at the central source position.


\begin{table*}
    \caption{Outflow parameters.$^{(a)}$Aperture angles measured at 1000\,AU of the central source. $^{(b)}$Sizes measured in the plane of the sky and corrected by inclination. $^{(c)}t_{dyn}$ = (size cos(i))/v$_{rad}$, with i the disk inclination. $^{(*)}$Lower limits, as the outflow could extend beyond the FOV. The values of $M_{out}$, $P_{out}$, $E_{out}$, $\dot{M}_{wind}$ and $\dot{P}_{wind}$  listed were calculated assuming excitation temperatures of 25\,K and 100\,K.
    }
    \centering
    \resizebox{\textwidth}{!}{
    \begin{tabular}{lcccccccccc}
    \hline
      &$\Delta{\rm v}_{rad}$ &  P.A.  & $\theta_{op}$ $^{(a)}$ & Size$^{(b)}$ & $t_{dyn}$ $^{(c)}$ & $M_{out}$ & $P_{out}$ & $E_{out}$ & $\dot{M}_{out}$ & $\dot{P}_{out}$ \\
      &     (\kms)       &(\deg) &   (\deg)  & (AU) & (yr)  &($10^{-4}$\msun)  & ($10^{-3}$\msun\ \kms) & ($10^{40}$erg) & ($10^{-8}$\msun/yr) & ($10^{-7}$\msun \kms/yr) \\
       & & & & & & (25K/100K)& (25K/100K) & (25K/100K) & (25K/100K) & (25K/100K) \\
     \hline
     IRAS\,15398-3359 &&&&&&&&&&\\
     \multicolumn{11}{l}{v$_{sys}$= 5.1$\pm$0.1 \kms}\\
     \hline
      Blue & 15.3$\pm$0.08 & 232.0$\pm$0.2  & 27.8$\pm$0.3 &  2550$\pm$50 & 320$\pm$20 & 1.0/2.5 & 0.8/1.9 &  7.9/18.7 & 5.1/12.4 & 1.7/4.2 \\
      Red  & 11.6$\pm$0.08 &  64.9$\pm$0.2  & 31.8$\pm$0.3 &  1800$\pm$50 & 300$\pm$20 & 0.7/1.8 & 0.6/1.4 &  5.2/12.6 & 4.0/9.8  & 1.3/3.2 \\    
      Total &              &            &                  &  4350$\pm$95 &            & 1.8/4.3 & 1.4/3.3 & 12.9/31.3 &    ...    & ... \\
     \hline
     IRAS\,16059-3857 &&&&&&&&&&\\
     \multicolumn{11}{l}{v$_{sys}$=4.7$\pm$0.1 \kms} \\
     \hline 
     Blue  &  9.6$\pm$0.2 & 268.5$\pm$0.3 & 70.3$\pm$0.6 & 13900$\pm$200         & 4160$\pm$100         & 4.5/11.0  & 2.4/5.9  & 13.6/32.9 & 10.8/26.5 & 5.9/14.2 \\
     Red   & 13.2$\pm$0.2 & 57.1$\pm$0.3  & 79.0$\pm$0.6 & 25400$\pm$200$^{(*)}$ & 5500$\pm$500$^{(*)}$ & 6.3/15.3  & 3.5/8.6  & 23.5/57.4 & 11.4/27.8 & 6.4/15.5 \\ 
     Total &              &               &              & 39200$\pm$300         &                      & 10.9/26.4 & 6.0/14.5 & 37.0/90.3 & ... & ... \\
     \hline
    J160115-415235 &&&&&&&&&&\\
    \multicolumn{11}{l}{v$_{sys}$= 4.1$\pm$0.1 \kms} \\
    \hline
     Blue &  5.6$\pm$0.08 & 266.9$\pm$0.15 & 64.6$\pm$0.3 & 20900$\pm$600  & 18000$\pm$700  & 4.2/10.3 & 1.8/4.3 & 8.3/20.3 & 2.4/5.7 & 0.7/1.7 \\
     Red  &  2.6$\pm$0.08 & 103.2$\pm$0.2  & 78.5$\pm$0.3 & 19700$\pm$600  & 23000$\pm$1000 & 0.6/1.5  & 0.2/0.5 & 0.8/1.9  & 0.3/0.6 & 0.1/0.2 \\
     Total &              &                &              & 40600$\pm$1200 &                & 4.8/11.7 & 2.0/4.8 & 9.1/22.2 & ... & ... \\
    \hline
    \end{tabular}
        }
    \label{tab:param_outflows}
\end{table*}

\subsubsection{Mass, Momentum and Energy from the Outflows} 

In order to recover the whole emission from the outflows and therefore to better estimate their masses, we combined the interferometric and total power data of \iras, \irasb\ and \j, as mentioned in section \ref{sec:imaging} above. 

We use the expressions from \citet{MangumandShirley2015} adapted to the \co(2-1) transition to estimate the column densities. From these, we derive the masses, the momenta and kinetic energies of each lobe separately, and the corresponding values for the entire outflows. In particular, we use the following equation to derive the column density of the gas (note that this expression is valid in the optically thin case only):

\begin{equation}
    N_{tot}(CO) = \frac{3h}{8\pi^3\mu^2J} \frac{Q_{rot}e^{E_u/kT_{ex}}}{e^{h\nu/kT_{ex}}-1}\frac{\int T_b d{\rm v}}{[J_{\nu}(T_{ex})-J_{\nu}(T_{bg})]}
\end{equation}

\noindent where $J_{\nu}(T) = (h\nu/k)/(e^{h\nu/kT}-1)$ is the Rayleigh-Jeans equivalent temperature. For the \co(2-1) transition, it can be expressed as:

\begin{align}
    N_{tot}(CO) = 1.196\times10^{14} \frac{(T_{ex}+0.921)e^{16.596/T_{ex}}}{e^{11.065/T_{ex}-1}} \nonumber \\
     \times \frac{T_B \Delta_{\rm v}}{[J_{\nu}(T_{ex})-J_{\nu}(T_{bg})]}
\end{align}

To get this expression we use the line strength S = J/(2J + 1) = 2/5, the dipole moment $\mu$ = 1.1$\times$10$^{19}$ C cm, the partition function Q$_{rot}$ = $k$T$_{ex}$/(hB$_0$)+1/3 with a rigid rotor rotation constant B$_0$ = 57.635968 GHz, and the degeneracy $g$ = 2J+1 = 5. We put the T$_B$ in K and the $\Delta$v velocity interval in \kms. 

We measure the average brightness temperature for the outflow (T$_B$) for each velocity channel assuming a background temperature of 2.7\,K. We avoid the channels contaminated with cloud emission since it could not be disentangled from that of the outflow (the velocity ranges of cloud emission are 4.2-6.8, 3.0-6.8 and 3.14-5.0\,\kms\ for \iras, \irasb\ and \j, respectively).
In addition, we estimate the density for two different values of T$_{ex}$: 25\,K and 100\,K. This range of temperatures have been used to estimate masses of outflows associated with other young stars (\citealt{ArceandSargent2006}, \citealt{vanKempenetal2009}, \citealt{Arceetal2013}). Using these T$_{ex}$ values we find that some channels are optically thick at the emission peak, which sets a lower limit for the densities.   

Using all this, we estimate the mass as: 

\begin{equation}
M_{out} = \mu m h \Omega N_{tot}/X_{CO},    
\end{equation}

\noindent where $\mu$ is the mean molecular weight, which is assumed to be equal to 2.76 after allowing for a relative helium abundance of 25$\%$ by mass \citep{Yamaguchietal1999}, $m$ is the hydrogen atom mass ($\sim 1.67 \times 10^{-24}$g), $\Omega$ is the area, and a CO abundance of X$_{CO}$ = 10$^{-4}$ \citep[e.g.,][]{Lacyetal1994}.

We further calculate the momentum and the kinetic energy of the outflows correcting by each of the outflow inclinations and using:

\begin{equation}
    P_{out} = M_{out} \left( \frac{V_r - V_{LSR}}{cos(i)} \right)
\end{equation}

\begin{equation}    
    E_{out} = \frac{1}{2} M_{out}  \left( \frac{V_r - V_{LSR}}{cos(i)} \right)^2
\end{equation}    

The outflow masses, along with their momenta and energies, are listed in columns 7, 8 and 9 in Table \ref{tab:param_outflows}.
We note the caveat that these values should be treated as lower limits, since we do not correct for the missing flux at cloud velocities, nor the opacity of the emission. The masses corresponding to a T$_{ex}$ of 25\, are typically a factor of 2.5 smaller than those estimated using 100\,K, and this factor is a fair representation of the uncertainty in the mass measurements. In our outflow sample, the masses are in the range of $10^{-4}$ and $10^{-5}$\msun, the momenta are roughly a few times $10^{-3}$\msun\,\kms, and the energies are between 10$^{40}$ and 10$^{41}$\,erg.

In the case of \iras\ the derived mass, momentum and energy agree, in order of magnitude, with those derived by \citet{Bjerkelietal2016} and \citet{Dunhametal2014}, based on \co(2-1) data observed with the Submillimeter Array (SMA) and the JCMT, respectively.  

We estimate lower limits for the average mass loss rate ($\dot{M}_{out}$ = $M_{out}/t_{dyn}$) and the average rate of linear momentum injected by each outflow, also known as the flux force ($\dot{P}_{out}$ = $P_{out}/t_{dyn}$). These values are also listed in Table \ref{tab:param_outflows} (columns 10 and 11), resulting in mass loss rates ranging $10^{-9}-10^{-7}$\,\msunyr, and rates of linear momentum ranging $10^{-8}-10^{-6}$\,\msunyr\  for the outflow lobes. Finally, a rough estimate of the accretion rate toward each young star can be made by assuming it is 10\% of the outflow mass loss rate ($\dot{M}_{acc}$ = 0.1 $\dot{M}_{out}$, \citealt*{PudritzandBanerjee2005}, \citealt{Ellerbroeketal2013}). The average numbers for $\dot{M}_{acc}$ are on the order of $10^{-8}-10^{-7}$\,\msunyr.

\subsubsection{Kinematics}\label{sec:kinematics}

\begin{figure*}
    \centering
    \includegraphics[width=0.9\textwidth]{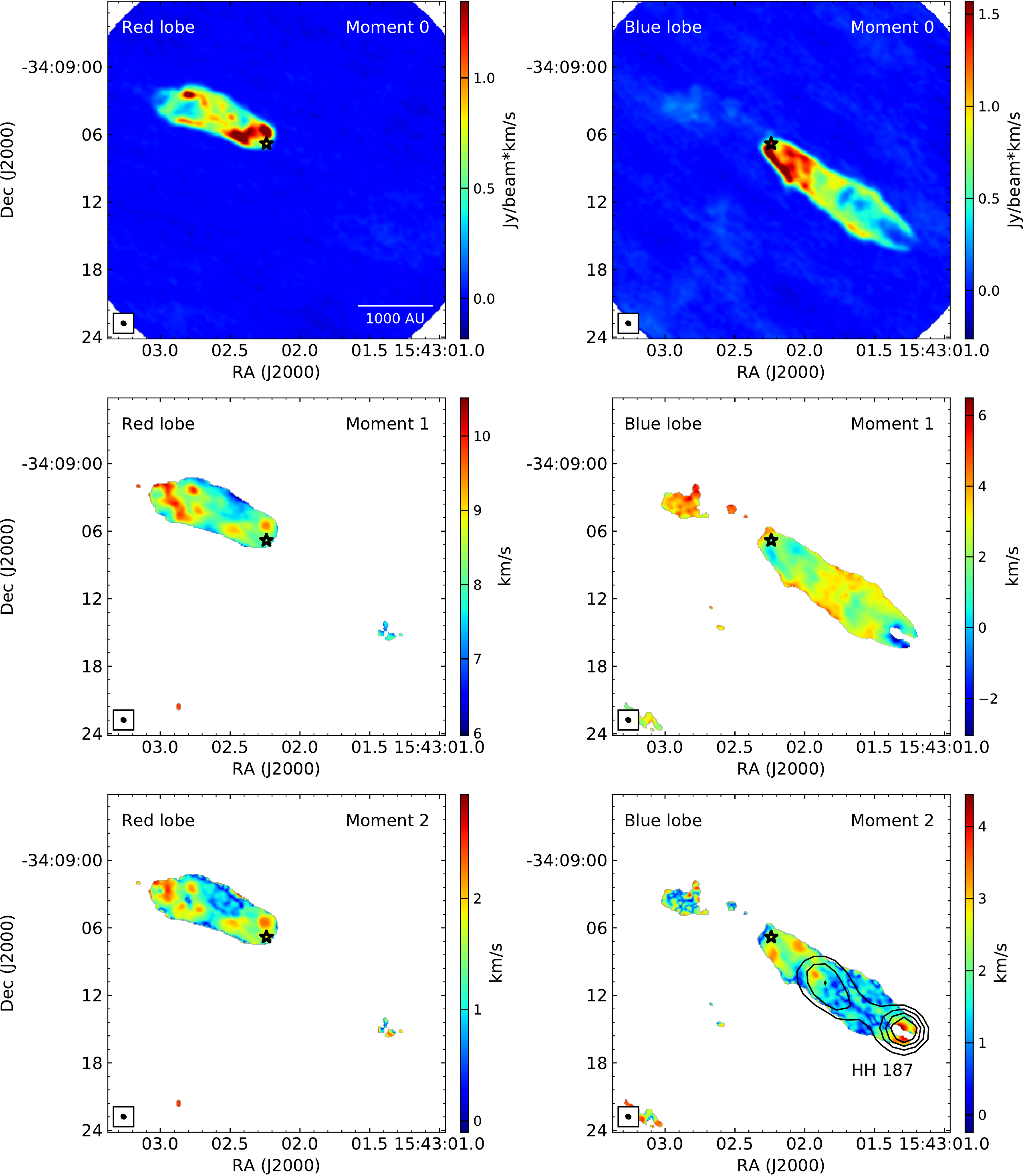}
    \caption{{\bf \iras.} {\it Left:} Red lobe moment maps (from 5.9 to 16.7\,\kms). {\it Right:} Blue lobe moment maps (from --10.2 to 4.8\,\kms). The systemic velocity is $\sim$5.1\,\kms. The position of HH\,185 matches with the 2MASS K$_s$ band peak (15:43:01.311, --34:09:14.81)  \citep[see also][]{HeyerandGraham1989} which is shown in black contours in the bottom right panel. Moment 1 and 2 maps only show the emission over 4$rms$.}
    \label{fig:IRAS15_moments}
\end{figure*}

\begin{figure*}
    \centering
    \includegraphics[width=0.9\textwidth]{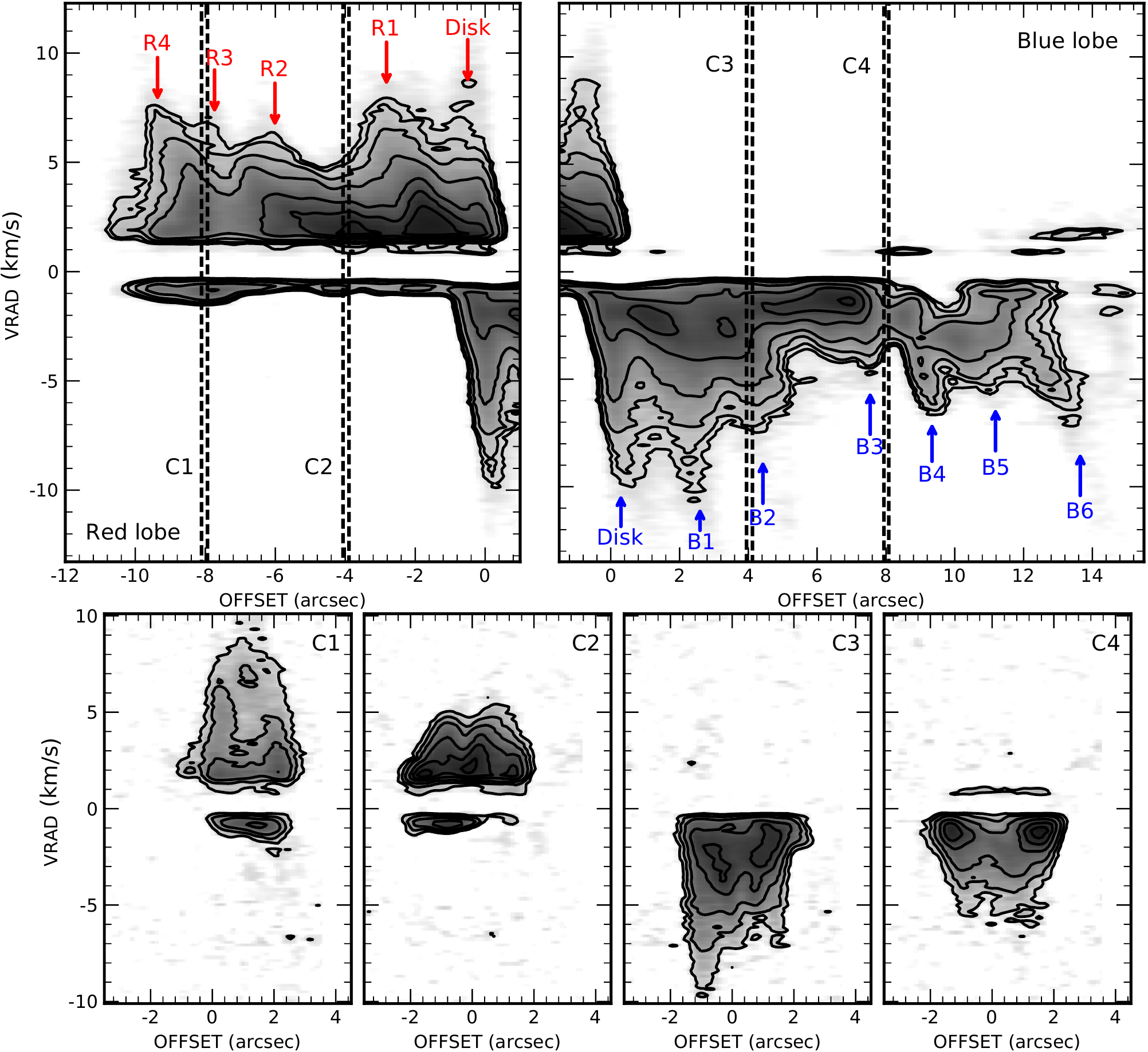}
    \caption{{\bf \iras}. {\it Upper panel:} Position-velocity diagram of the \co(2–1) emission along the outflow axis with a cut width of 1\arcsec. The arrows show the presence high velocity gas in both lobes. {\it Bottom panels:} Position-velocity diagrams of the \co(2–1) emission across to the outflow axis at an angular distance of 4\arcsec\ and 8\arcsec\ from the central source in each lobe.}
    \label{fig:IRAS15-pv}
\end{figure*}

\begin{figure*}
    \centering
    \includegraphics[width=1.0\textwidth]{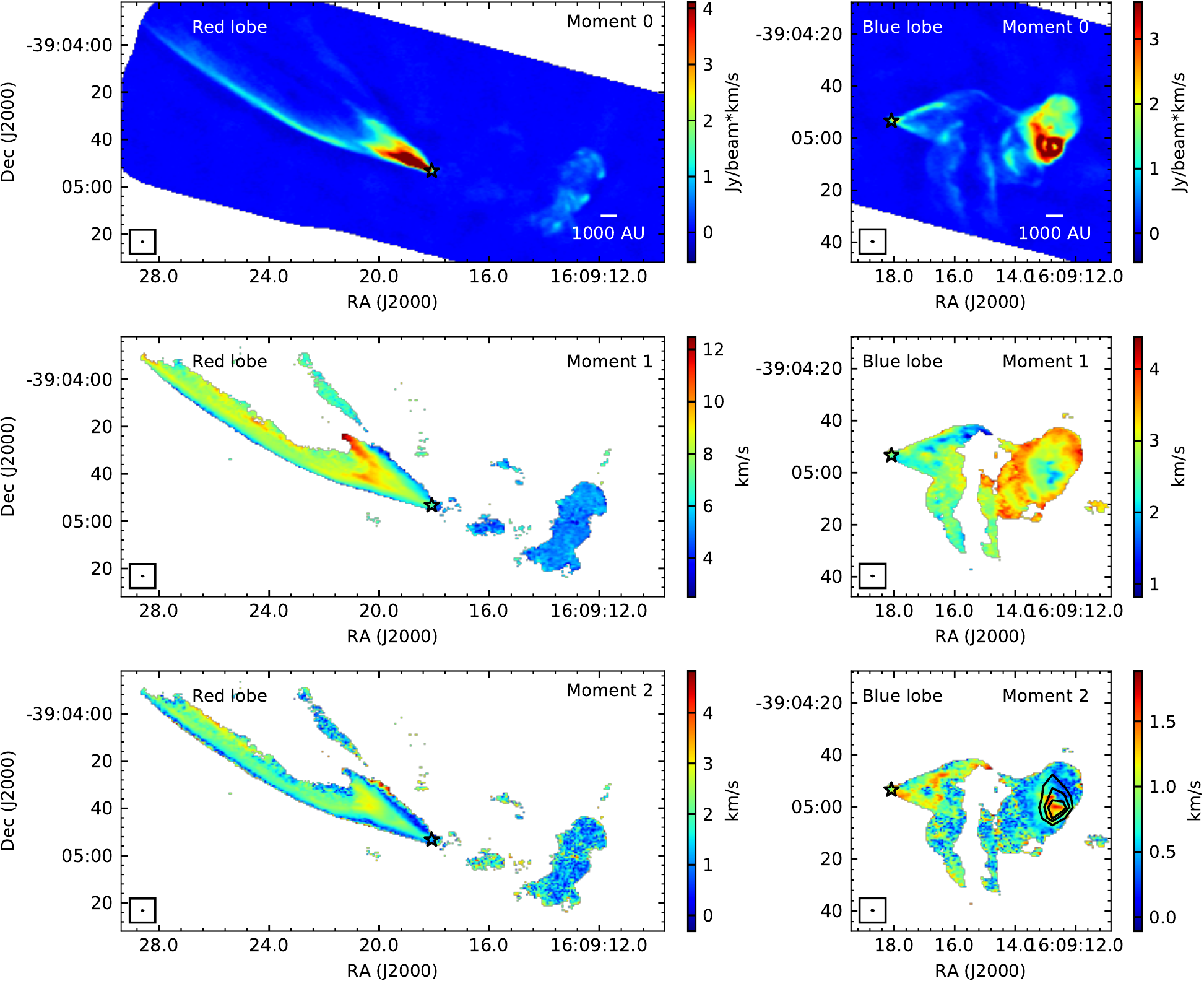}
    \caption{{\bf \irasb.} {\it Left:} Red lobe moment maps (from 4.9 to 17.9\,\kms). Note the red-shifted emission spatially coinciding with the blue lobe. {\it Right:} Blue lobe moment maps (from --4.9 to 4.28\,\kms). Note the emission that extends southward perpendicular to the lobe. The systemic velocity is $\sim$4.7\,\kms. At the end, a bubble-shaped structure is detected. Over the moment 2 map the black dashed contours show the DSS2 emission (6300-6900 \AA) that reveals the presence of the Herbig-Haro object HH\,87, matching a region of high velocity dispersion. Moment 1 and 2 maps only show the emission over  4$\sigma$.}
    \label{fig:IRAS16_moments}
\end{figure*}

\begin{figure*}
    \centering
    \includegraphics[width=0.9\textwidth]{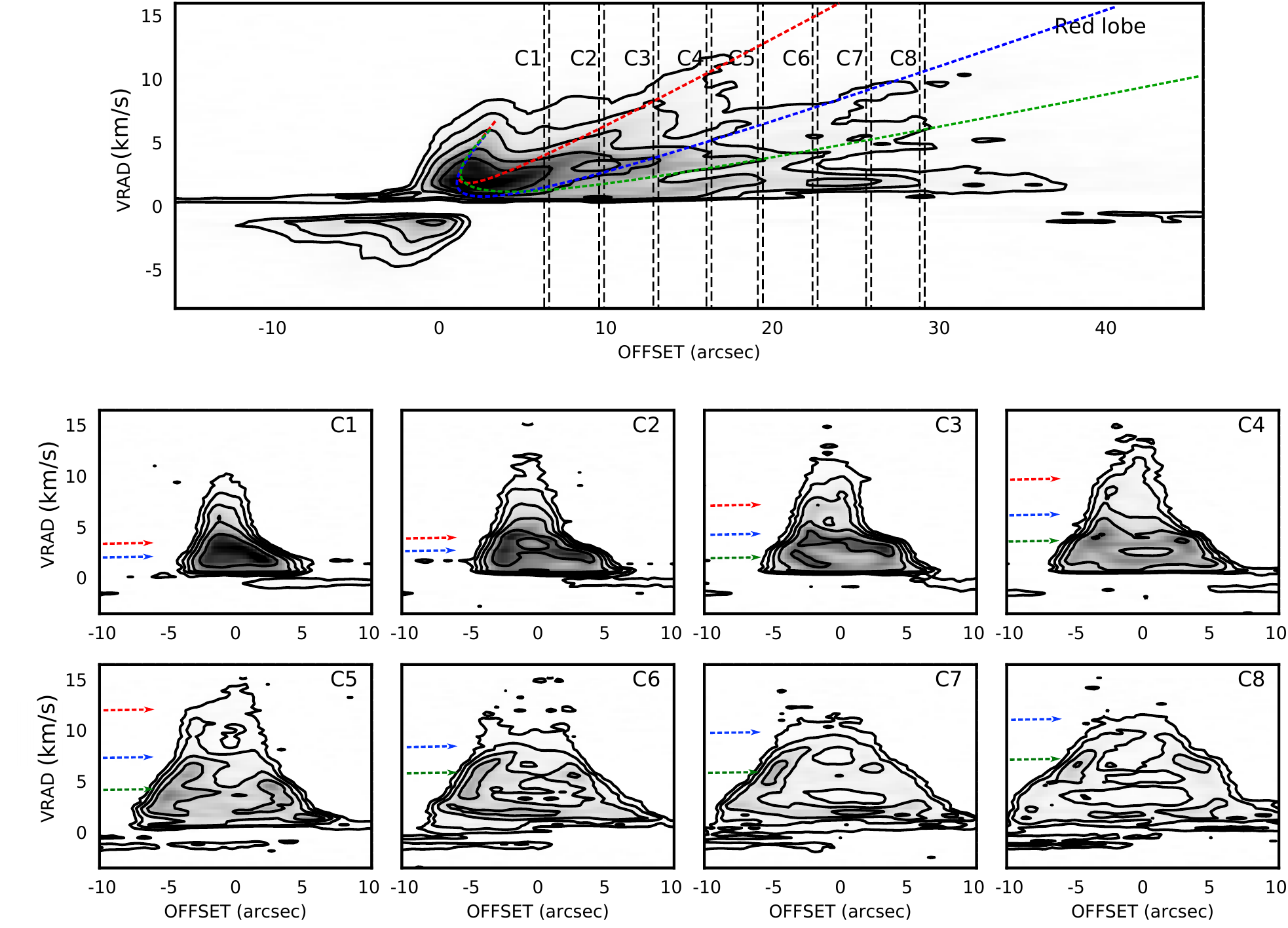}
    \caption{{\bf \irasb}. {\it Up}: Position-velocity diagram of the \co(2–1) emission along the red lobe outflow axis with a cut width of 1\arcsec. {\it Bottom}: Position-velocity diagrams of the \co(2–1) emission along 1\arcsec-wide cuts perpendicular to the outflow axis at distances of 1000\,AU (C1), 1500\,AU (C2), 2000\,AU (C3), 2500\,AU (C4), 3000\,AU (C5), 3500\,AU (C6), 4000\,AU (C7) and 4500\,AU (C8).}
    \label{fig:IRAS16-pv-red}
\end{figure*}

\begin{figure*}
    \centering
    \includegraphics[width=1.0\textwidth]{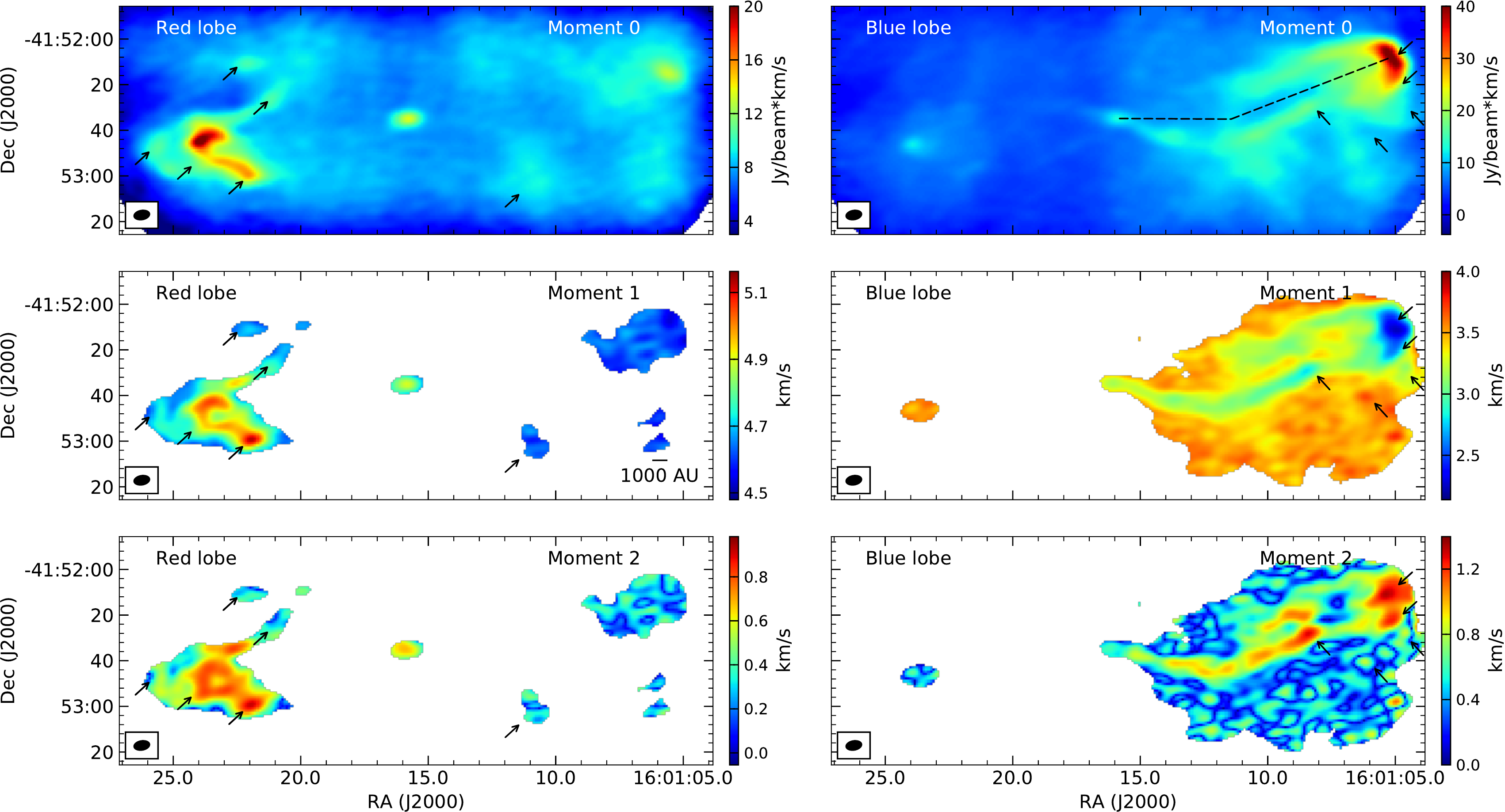}
    \caption{{\bf \j:} Moment maps of the red ({\it left}, from 4.1 to 6.7\,\kms) and blue ({\it right},from --1.5 to 3.6\,\kms) lobes. The systemic velocity is $\sim$4.1\,\kms. Moment 1 and 2 maps only show the emission over 4$rms$. The black dashed lines in the right upper panel show a 27\degr\ deflection in the blue lobe. The arrows point to the features described in the main text, where high velocity or dispersion emission is detected.}
    \label{fig:J160115_moments}
\end{figure*}

\begin{figure*}[h!]
    \centering
    \includegraphics[width=0.7\textwidth]{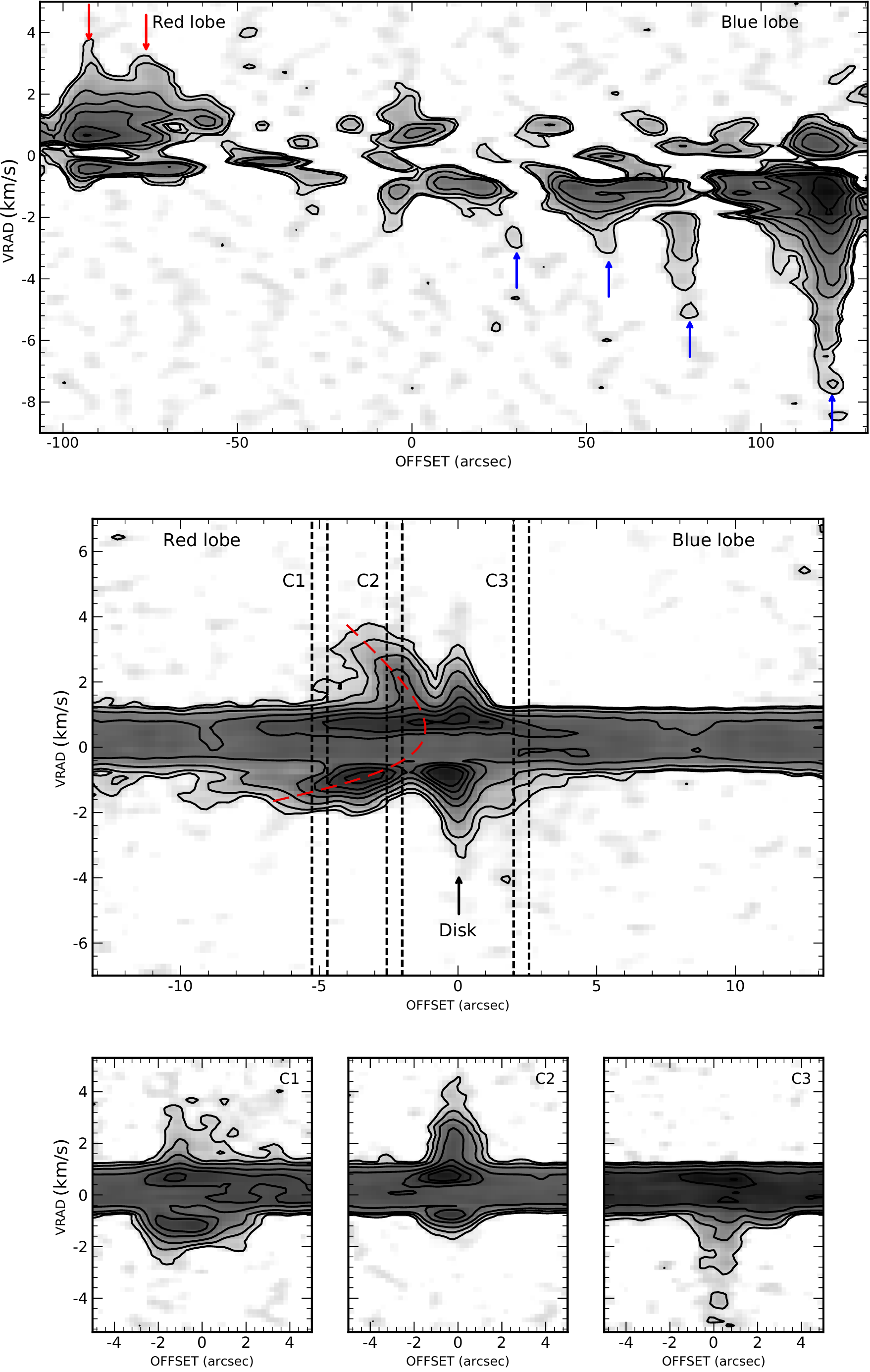}
    \caption{{\bf \j}. {\it Upper panel:} Position-velocity diagram of the \co(2-1) emission along the outflow axis with a cut width of 30\arcsec\ (low resolution, full outflow). Arrows point out high velocity emission. {\it Middle panel:} Position-velocity diagram of the \co(2-1) emission along the outflow axis with a cut width of 1\arcsec\ (high resolution, central region). We indicate the emission from the disk. A parabolic structure, indicated with a red dashed line, crossing the cloud velocity is detected in the red lobe. {\it Bottom panels:} Position-velocity diagrams of the \co(2–1) emission across to the outflow axis at an angular distance of 2\arcsec\ and 5\arcsec\ from the central source in the red lobe, and 2\arcsec\ in the blue lobe.}
    \label{fig:J160115-pv}
\end{figure*}

This section contains figures showing the moment maps of first and second order (integrated weighted radial velocity and integrated weighted velocity dispersion, respectively) separated by the cloud velocity, along with several position-velocity (PV) diagrams along and across the outflows of \iras, \irasb\ and \j\ (Figures \ref{fig:IRAS15_moments}, \ref{fig:IRAS15-pv}, \ref{fig:IRAS16_moments}, \ref{fig:IRAS16-pv-red}, \ref{fig:J160115_moments} and \ref{fig:J160115-pv}). We analyze the observations and comment on our findings for each individual outflow. 
We will discuss these findings and interpret them with respect to the nature of the outflows in the next section.

The moment 1 of the \co\ emission of \iras\ (Figure \ref{fig:IRAS15_moments}, middle panels) shows a non-uniform velocity pattern on each lobe of the outflow. For instance, in the redshifted lobe there are several high-velocity spots, while in the blueshifted lobe, the velocity changes with distance from the protostar from about +1\kms (cyan) to +3\kms (yellow) and then again to +1\kms ($v_{cloud}=5.1$\kms). A characteristic high-velocity spot is distinguished close to the tip of the blueshifted lobe. All these high-velocity spots are identified in the moment 2 images of \iras\ (lower panels), as well as three more spots close to the protostar position in the blueshifted side of the outflow. It is worth noting that some of these spots positionally coincide with the emission from the HH\,185 object, which is comprised by two sources in the 2MASS images: one unresolved, at the tip of the blueshifted lobe and the other, extending along the outflow body. The faint emission detected southeast of the outflow is discussed in Appendix B.

Figure \ref{fig:IRAS15-pv} shows a cut along the \iras\ outflow axis (upper panel) revealing the presence of four knots (which we define as sudden increases of radial velocity at an almost fixed position) in the red lobe and six additional knots in the blue lobe within 14\arcsec\ from the driving source. These are labeled as R1, R2, R3, R4, B1, B2, B2, B4, B5 and B6, and most coincide with the spots traced in the moment maps. Red and blue-shifted high-velocity emission is also detected close to the disk position. In addition, the cuts made at 4\arcsec\ and 8\arcsec\ from the protostar (600 and 1200\,AU at the considered distance in projection) across the outflow axis (see lower panels of Figure \ref{fig:IRAS15-pv}) reveal two or even three (see e.g., C2 cut) structures with large radial velocity dispersion extending from the cloud velocity to high velocities almost parallel. Most of these structures are probably related to the walls or the tips of the outflow.  

Figure \ref{fig:IRAS16_moments} shows the moment maps for \irasb\ (moment 0 map have already been shown in contours in Figure \ref{fig:mom_outflows}). A large difference between the two lobes is quickly noticed. Regarding the kinematics of the outflow, the redshifted lobe is smooth and shows a change of velocity from the outer parts of the outflow cavity walls (close to the cloud velocity of 4.7\kms) to the inner parts ($>10$\kms). Unlike in the case of \iras, the redshifted lobe of \irasb\ does not show any high-velocity or high-dispersion spot or knots. The blueshifted side of the outflow in the vicinity of the source ($<$3500\,AU) is v-shaped, but it loses regularity abruptly and continues further west in a blueshifted intricate emission at the position of the HH\,87 (16:09:12.8, --39:05:02), surrounded by gas nearly at the cloud velocity. Evidence of the presence of shocked gas can be found in the increase of velocity dispersion (moment\,2 images in Figure \ref{fig:IRAS16_moments}) and, of course, its perfect positional match with HH\,87. In addition, redshifted emission is detected matching spatially with the blue lobe. We discuss in the Appendix B the possible origin of this structure. 

Since the redshifted side of the \irasb\ outflow shows a more regular behaviour, we only perform PV analysis along this lobe (Figure \ref{fig:IRAS16-pv-red}), rather than including the blueshifted lobe. We identify three different parabolic structures marked with the red, blue and green dashed lines in the top panel of the figure. There may be more parabolas, but we think our identification may serve well as a proof of concept of the existence of several overlapping kinematic structures.
In the PV diagrams transversal to the outflow axis (middle and bottom panels of Figure \ref{fig:IRAS16-pv-red}) we identify various ring-like structures that could be associated with the parabolas found in the longitudinal PV cut. The top of these structures, approximately flat in radial velocity, are indicated by red, blue and green arrows. The rings move from lower to higher radial velocities as the cut is taken further from the protostellar position. In addition, these transversal PV cuts clearly show the smooth velocity gradient between the edges of the cavity (lower velocities) and the inner regions of the outflow.

The moment maps of the \j\ outflow apparently trace its cavity walls well in the blueshifted side (Figure \ref{fig:J160115_moments}). The cavity structure ends in a strong blueshifted arc of emission, possibly formed by shocked gas at about 116\arcsec and a PA of -78\degr\ from the protostar position (there are other high-velocity dispersion spots at 117\arcsec\ and PA=-83\degr, 121\arcsec\ and -87\degr, and at 104\arcsec\ and -93\degr). Another arc-like feature with higher velocity dispersion is detected in the path of this side of the outflow, at about 80\arcsec from the protostar with a PA of -85\degr. It is also worth noting that close to the protostar, the most prominent side of the cavity is the southernmost. 
 A faint cavity wall is also detected in the left side lobe, tracing an X shape centered on the protostar (see Figure \ref{fig:J160155-chan-high-resol}).
In the red-shifted side of the outflow (left panels of Figure \ref{fig:J160115_moments}) the observations show gas at different position angles from the protostar (from 80\degr\ to 110\degr). The most prominent features (higher velocity and dispersion velocity) are a point-like spot at about 77\arcsec\ and PA=108\degr, and an arc-like complex structure subtending an angle between PA 90\degr\ and 100\degr\ at about 92\arcsec. There is also another arc further from the protostar and closer to the cloud emission at 114\arcsec\ with a PA $\sim$ 96\degr. There are still another two CO cloudlets at PA (78\arcsec, 70\degr) and (63\arcsec, 80\degr). 
We also detected red-shifted emission southwest of the source (60\arcsec,$-117\degr$).
All these features are indicated with black arrows in the middle and bottom panels in Figure \ref{fig:J160115_moments}.

Figure \ref{fig:J160115-pv} shows the position-velocity diagrams obtained from the low- and high-angular resolution data (full outflow and close to the position of the source \j, respectively), following the outflow axis.
In the upper panel the PV diagram obtained from the low-resolution data (cut width of 30\arcsec, PA=98\deg) shows features related with high velocity gas suggesting the presence of shocks, even though (except in the terminal region of the outflow) the emission is not very clear.  
In the middle panel, the PV-diagram of the central region (cut width of 1\arcsec, PA=89\deg) shows high-velocity emission from the disk of the young star detected at the central position. Also, a parabolic structure crossing the cloud velocity is detected in the red lobe. This structure is related with the cavity walls of the redshifted side of the outflow, which should be oriented close to the plane of the sky in order to show such a crossing of the cloud velocity. It has no correspondence in the blueshifted side.

\section{Discussion} \label{sec:discussion}
We split this section in two parts. First we analyze the three individual outflows observed in our sample (IRAS\,15398-3359, J160115-415235 and IRAS\,16059-3857) to discern their nature. In second place we discuss the physical characteristics of the whole sample of young protostars, trying to classify them in an evolutionary series based on their properties. We also discuss the links between evolutionary stage, the molecular observations, and the outflow characteristics.

\subsection{Individual outflows} \label{subsec:individual_outflows}
The CO outflows are generally seen as low-velocity shell structures around jets that produce high-velocity shocks along their paths. The higher velocity emission is usually found farther out from the source \citep{Leeetal2000} and the position-velocity diagrams usually show one of two distinctive kinematic features: a parabolic structure originating at the driving source, or a convex spur structure with the high-velocity tip near known H$_2$ bow shocks or HH objects. The former parabolic PV structures could be produced by wide-angle winds \citep{RagaandCabrit1993}, while the latter are generally modeled under a jet-driven bow shock scenario \citep{Shuetal991}. 

Regarding the ejection, it is supposed to be intimately related with accretion \citep{1995Hartigan}. A variety of instabilities in the accretion disk can cause the accretion of material from a circumstellar disk of a forming star to be episodic (e.g., \citealt{DunhamandVorobyov2012}), and this would in turn correspond to an episodic outflow rate \citep{Vorobyoetal2018}. In Class 0 and I sources, which are still embedded in their parental cores, the most significant evidence of episodic accretion comes from jets that show a series of knots along their axis \citep{SantiagoGarciaetal2009,Hiranoetal2010,Plunkettetal2015}.

We adopt the wide-angle wind and the jet-driven wind models to discuss the nature of the outflows of our sample through interpreting the details of the ALMA observations. 
However, in this study we do not pretend a direct identification between the outflows and these models. We are aware that they are simplified models and that more sophisticated ones would be needed to explain in detail the ejection mechanisms.

\subsubsection{IRAS\,15398-3359} \label{sec:IRAS15}
\begin{figure*}
    \centering
    \includegraphics[width=0.9\textwidth]{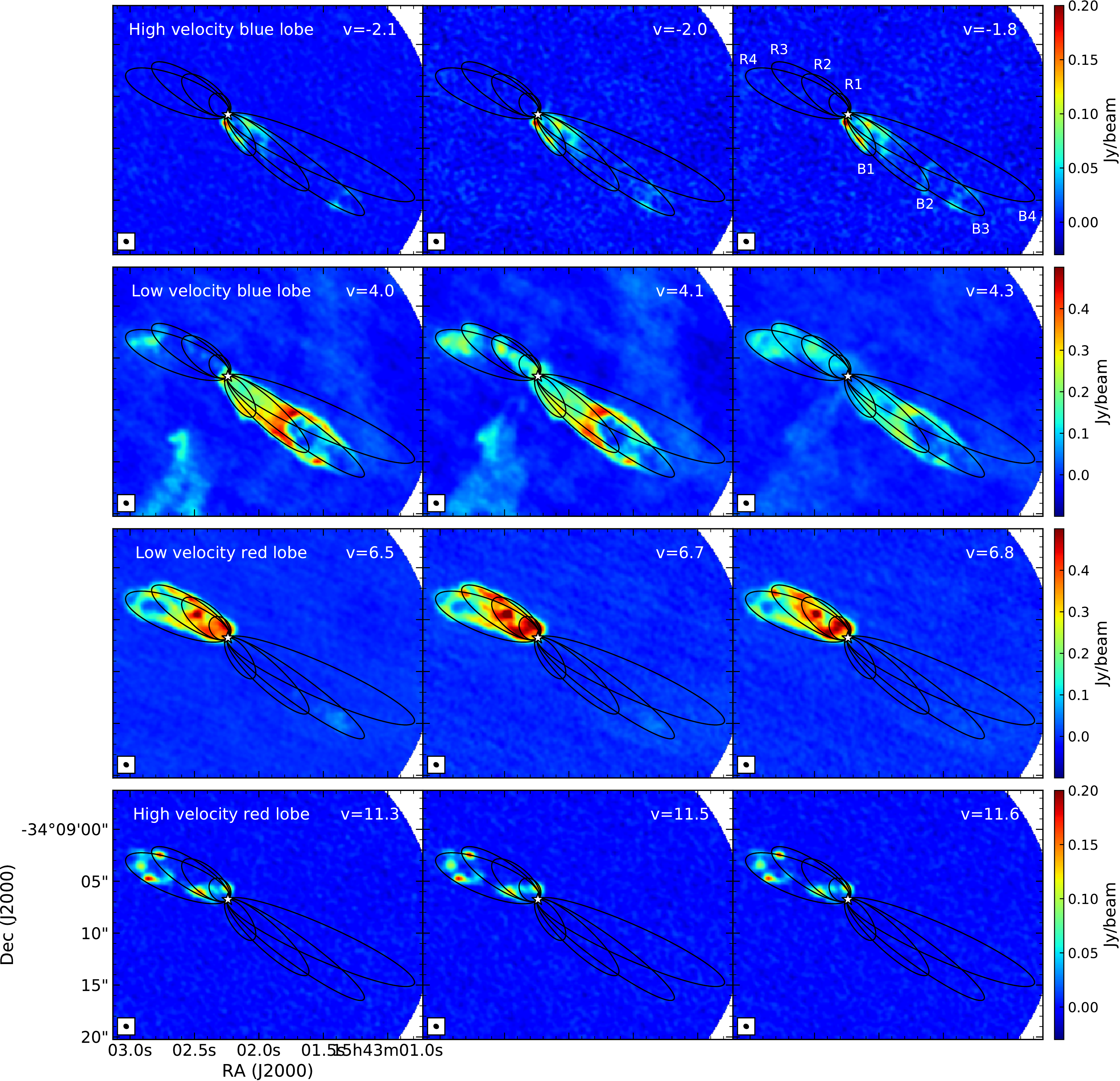}
    \caption{{\bf \iras:}\co(2-1) emission velocity channel maps. The star indicates the position of the compact continuum source. Black ellipses show the four bipolar structures.}  
    \label{fig:IRAS15-ellipses}
\end{figure*}

Figure \ref{fig:IRAS15-ellipses} displays some selected \co(2-1) velocity channels of the outflow associated with \iras\ at high and low velocities (see Appendix C for the complete velocity cubes of the three outflows).

A close inspection of the redshifted low-velocity channels ($v\sim$6.5-6.8\,\kms) led us to notice the presence of some CO loops or bow-shocks at different position angles measured from the protostar position. These outflow lobes seem to trace the expanding shell of separate ejections. Moreover, this possibility is strengthened by the fact that the position of the bullet-like shocks shown in the position-velocity diagram in Figure\,\ref{fig:IRAS15-pv}, are not associated with gas in the central lane of the outflow, but at different position angles from the protostar (see high-velocity channels in Figure\,\ref{fig:IRAS15-ellipses}). The loops and hot-spots seen at different velocities in the CO velocity cube agree well with an scenario with ejections in different directions from the central source. As a proof of concept, we try to identify these outflows using four highly eccentric ellipses 
(shown with black lines and labelled as R1, R2, R3 and R4 in the Figure \ref{fig:IRAS15-ellipses}). These four structures have different sizes and position angles, and we have drawn them so all four share the same position in one of their apices. The semi-major axes of the ellipses are larger the greater the position angle is.  
By looking carefully at the rest of the redshifted velocity channels we notice evidence of the presence of these structures along the whole velocity range reached by the outflow. The CO emission appears sometimes delineating the border of the ellipses and in other cases approximately at the intersections (hot-spots) of two or more of them, as happens in the higher velocity channels (see lower panels of the Figure\,\ref{fig:IRAS15-ellipses}).
We further extend this analysis to the more spread-out blueshifted lobe identifying four ellipses with semi-major axes longer than those of the redshifted lobe by a factor of 1.85 (labelled as B1, B2, B3 and B4 in the Figure\,\ref{fig:IRAS15-ellipses}). These ellipses are not perfectly counter-aligned with their redshifted counterparts by an angle of about 1\deg-3\deg. The geometric parameters of all these ellipses are listed in Table \ref{tab:elipse_i15398}. The two sets of ellipses describe qualitatively well the emission observed at both low and high velocities. The set of ellipses of one of the outflow's sides have various sizes. They seem to change their sizes and, perhaps more interestingly, their orientations in intervals of $\sim2000$\,AU and $\sim10\degr$, respectively, from smaller to larger.

Although these structures were identified manually, and the total number and position of the ellipses may not be exactly as presented here, this analysis allows us to have a good first approximation, and gives us the opportunity to discuss possible ejection mechanisms of the source.
Our analysis leads us to identify each pair of counter-aligning ellipses with an episodic bipolar ejection of material from \iras. In principle, if the features were ejected with a similar velocity and inclination with respect to the line of sight, the largest ellipses would be associated with the oldest ejections. We estimated the dynamical time for each ejection taking the inclination effect into account (Table \ref{tab:elipse_i15398}) and the younger ejection may appear to be more spaced in time than the older for both sides of the outflow (about 50 and 80 years from the oldest and the youngest in the western side). Due to their smaller size, the dynamical times of the redshifted (eastern) ejections are $\sim1.4$ times shorter. Since the ejections appear to be mostly bipolar in the plane of the sky, it would also be expected that each pair of counter-aligned ejecta share the same inclination with respect to the line of sight. Therefore, the differences in the sizes of two opposite ejecta should be caused by differences in the density of the environment or in the ejection velocity. The former scenario seems more plausible, given the high degree of alignment of the paired ejections and the symmetry of the whole set of ellipses. Nevertheless, it is true that the four pairs of ejections are not exactly bipolar by a few degrees. The deviations from strict bipolarity could be explained by an intrinsic difference in the ejection angle on both sides or, more possibly the displacement of the protostar in the plane of the sky (proper motions) or even the orbital motion of a multiple system. In fact, outflows with a similar appearance have been found associated with high-mass protostars; see for instance the cases of Cepheus\,A HW2 and S140 \citep{Cunninghametal2009, Zapataetal2013, Weigeltetal2002}, where the multiple ejections with different orientations have been explained as produced by a tilt of the ejecting system during the periastron passage of a companion in a very eccentric orbit.

\begin{figure*}
    \centering
    \includegraphics[width=0.6\textwidth]{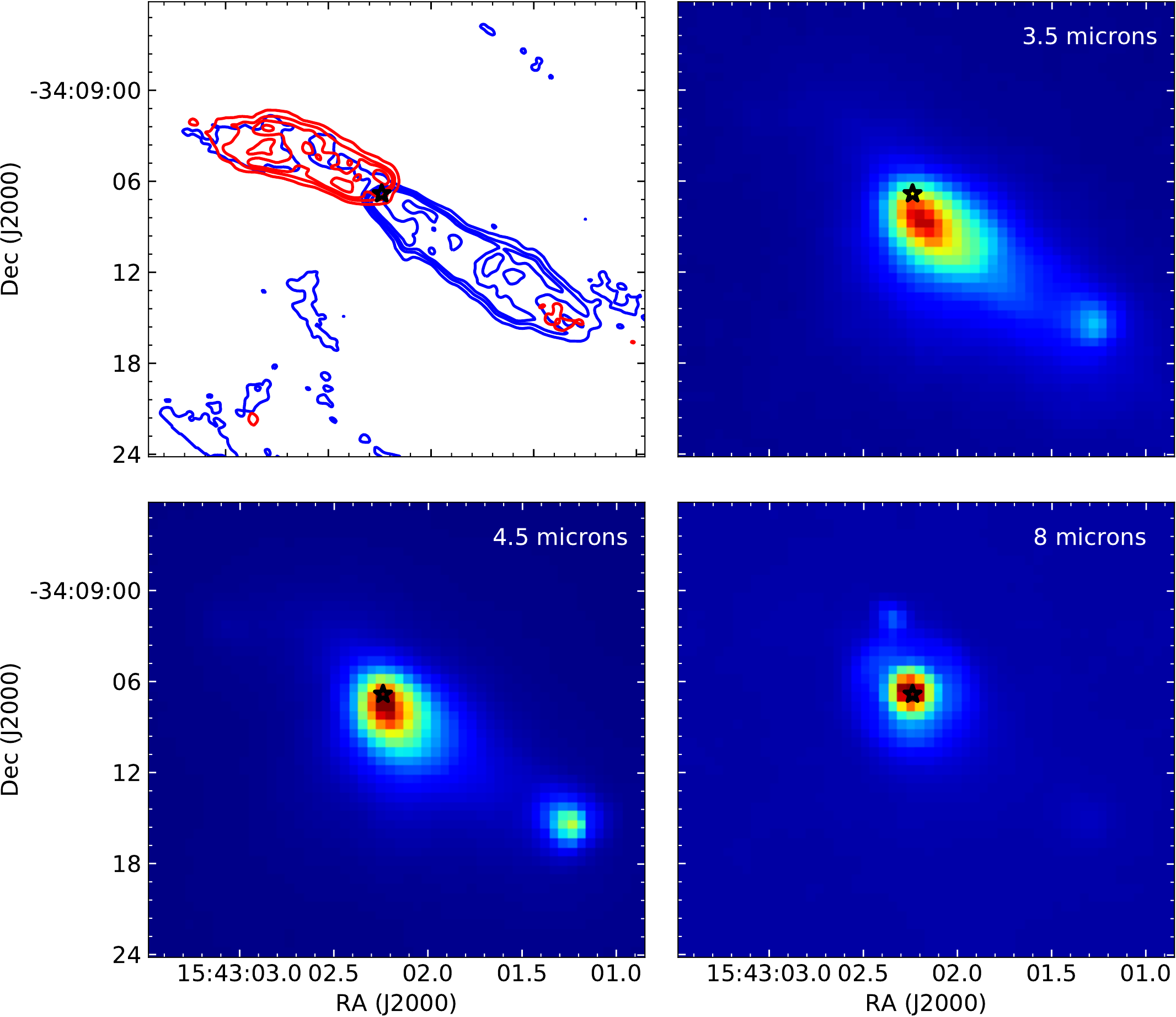}
    \caption{Spitzer IRAC emission at 3.5\,\mic\ (top right), 4.5\,\mic\ (bottom left) and 8.0\,\mic\ (bottom right)associated with \iras. Top left pannel display the blue-shifted and red-shifted \co(2-1) emission. The black star indicates the position of the compact source. The infrared emission extending southwestward may stem from shocks of the blueshifted outflow close to the star. The spot in \radec=(15:43:01.3; --34:09:15.4) seen at 3.4 and 4.5\,\mic\ coincides with the terminal part of the blue lobe}
    \label{fig:IRAS15_IR}
\end{figure*}

The difference in the position angles of the ejections in \iras\ suggests that the outflow is ejecting material episodically and that the originating source is perhaps precessing, as already proposed by \citet{Bjerkelietal2016}. These authors proposed that the different PA between the two outflow sides could possibly be due to precession of the ejection axis, originated by the tidal interaction of a binary companion. However, an inspection of the 3.6, 4.5 and 8.0 $\mu$m Spitzer images show that the infrared emission is not directly associated with \iras\ but may probably stem from shocks of the blueshifted outflow close to the star (see Figure \ref{fig:IRAS15_IR}). Regardless of the cause for the tilt on the system, \cite{Jorgensenetal2013} found what seems to be the chemical footprint of an accretion burst in the past 100 to 1000 years (a ring of H$_{13}$CO$^{+}$ which has been destroyed at the center by the water vaporized during the luminosity bump created by the accretion burst). This time estimate coincides quite well with the dynamical time of the youngest of the ejections (400-600 years for the redshifted and the blueshifted side respectively).

Another clue that makes us think that the outflow could have an episodic behavior is the pattern detected in the position-velocity diagram (Figure \ref{fig:IRAS15-pv}). Inspecting this figure we detect four high-velocity spur-like features (jumps in radial velocity) in the red lobe and six in the blue lobe. These spur-like features in a PV diagram are reminiscent of the jet-driven wind model described in \cite[e.g.][]{Leeetal2000}. In addition they match several large-density knots positions on the velocity cubes. The spurs are spaced quite evenly, suggesting the existence of an episodic outflow (e.g, \citealt{Plunkettetal2015}). They probably trace bow-shocks formed by variations in the mass loss rate or jet velocity, which likewise can be caused by variations in the accretion rate, or side-shocks that are produced when a new outflow ejection is launched in a different orientation and sweeps partially the trail of dragged material left by a previous ejection. As seen in the upper and lower panels of the Figure \ref{fig:IRAS15-ellipses}, some high-velocity spurs could be originated in the intersections of two or more ellipses (i.e., the collision of different ejections). 
The shape of each of the high-velocity signatures in the PV matches with a jet-driven model, showing a spur structure in the PV diagram along the jet axis, as described by \citet{Leeetal2000}.

Finally, if we consider the fact that \iras\ may be somehow undergoing bursts of accretion and driving a new outflow ejection in different orientations, it is likely that the estimate of our previous outflow's dynamical time is wrongly reported. The different inclinations with respect to the plane of the sky of each ejection would have implications in the calculated dynamical times. It is likewise worth to note that knowledge about the nature of an outflow is crucial for correlating this dynamical time estimate against any other relevant measurement of the protostar.

\begin{table}
    \centering
    \caption{{\bf \iras:} Parameters of the ellipses identified to the episodic ejections of \iras. }
    \resizebox{0.5\textwidth}{!}{
    \begin{tabular}{ccccccc}
    \hline
        \multicolumn{6}{c}{ Episodic ejections of \iras}  \\ 
        \hline
        & Position & M\_axis & m\_axis & Angle & Length & $t_{dyn}$\\
        & \radec  & \arcsec & \arcsec & \deg & AU & yr \\
    \hline
    B4 & 15:43:01.519, -34:09:10.86 & 9.8 & 1.7 & -114 & 1609 & 268 \\
    B3 & 15:43:01.714, -34:09:11.60 & 8.1 & 1.3 & -124 & 1330 & 221 \\
    B2 & 15:43:01.930, -34:09:10.38 & 5.3 & 1.2 & -133 & 870 & 144  \\
    B1 & 15:43:02.145, -34:09:08.74 & 2.3 & 0.9 & -146 & 377 & 62 \\
    R4 & 15:43:02.632, -34:09:04.71 & 5.3 & 1.7 & 69 & 870 & 144 \\
    R3 & 15:43:02.531, -34:09:04.31 & 4.4 & 1.3 & 57 & 722 & 120 \\
    R2 & 15:43:02.412, -34:09:04.91 & 2.9 & 1.2 & 50 & 476 & 79 \\
    R1 & 15:43:02.302, -34:09:05.83 & 1.2 & 0.9 & 37 & 197& 33 \\
    \hline
    \end{tabular}
    }
        \label{tab:elipse_i15398}
\end{table}

\subsubsection{IRAS\,16059-3857} \label{sec:IRAS16}

The outflow associated with \irasb\ presents two lobes with very different characteristics: the red lobe, previously analyzed in Section \ref{sec:kinematics}, shows the typical kinematic features of an episodic wide-angle outflow, while the blue lobe shows a peculiar emission distribution. The presence of more than one parabolic structure apparently sharing a similar origin in the PV diagram of Figure \ref{fig:IRAS16-pv-red} could indicate an episodic outflow that has had multiple ejections (e.g., \citealt{Zhangetal2019}). Interestingly, unlike in the case of \iras\, these parabolic structures in the PV diagrams correspond better with the wide-angle wind model \citep{Leeetal2000}. The episodic ejections of \irasb\ do not show abrupt spurs in velocity (i.e., no strong velocity shocks), but a continuous velocity increase as they move farther from the driving system. In this case, the different ejections seem to be launched with a similar orientation, but small perturbations may be hindered due to the fact that they seem to be launched almost isotropically, but with different thrusts depending on the polar angle (see wide-angle wind model details in the literature, e.g. \citealt{Leeetal2000}).

\begin{figure*}
    \centering
    \includegraphics[width=1.0\textwidth]{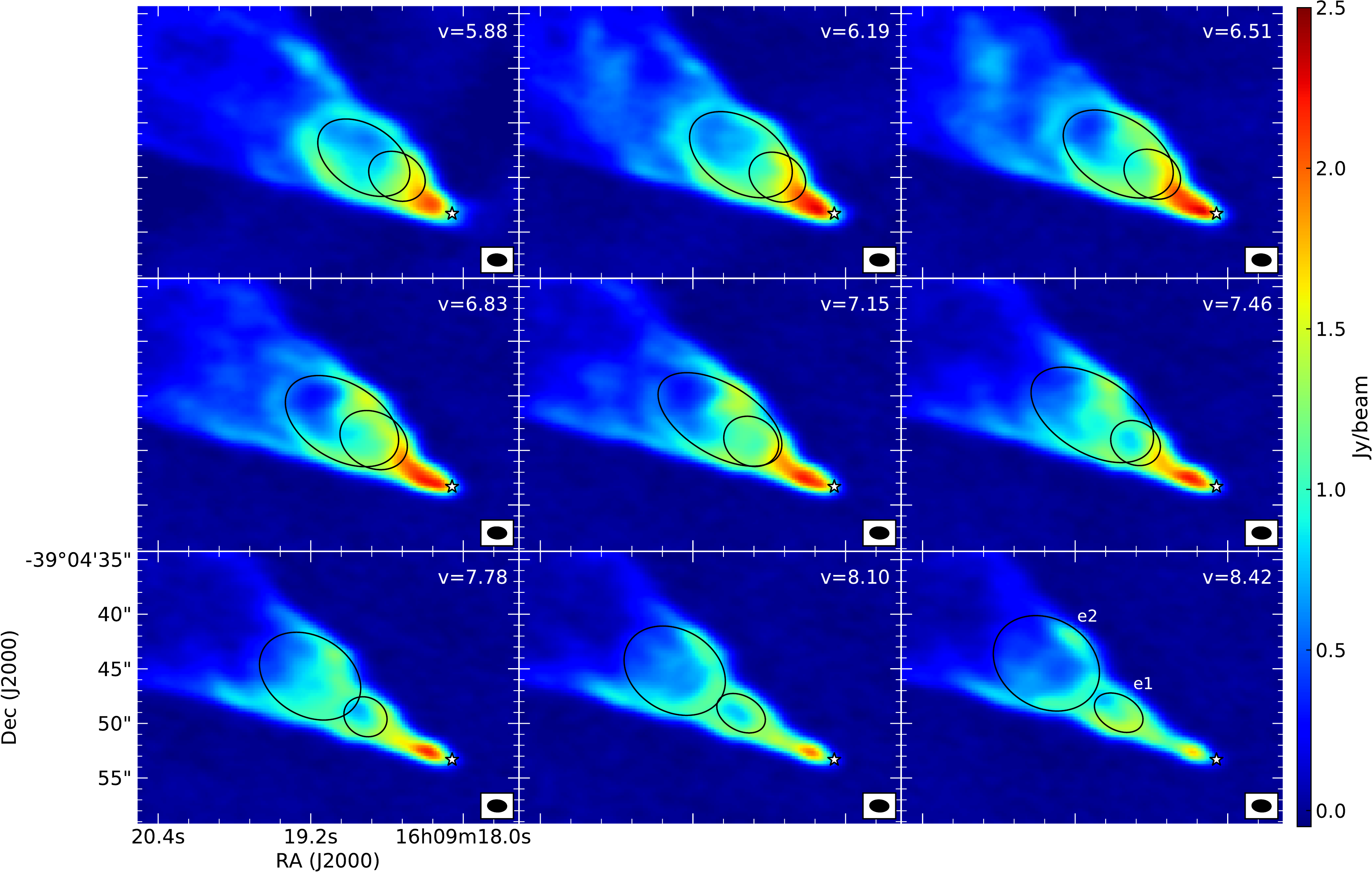}
    \caption{{\bf \irasb:} \co(2-1) emission velocity channel maps. The star indicates the position of the compact continuum source. We identified elliptical structures labeled as {\bf e1} and {\bf e2}.}
    \label{fig:IRAS16-chan-red-ellipses}
\end{figure*}

Figure \ref{fig:IRAS16-chan-red-ellipses} shows the channel maps of the redshifted lobe from 5.88 to 8.42 \kms. After analyzing the images, we could identify several elliptical structures. The parameters of these ellipses (central position, major and minor axis and position angle) are listed in Table \ref{tab:elipse_param}. We realized that there are at least two elliptical features in each velocity channel, making two distinct sets that apparently evolve with velocity. By moving away from the systemic velocity (4.7 \kms), the central position of the ellipses moves along the axis of the outflow, away from the continuum source. This Hubble-law-like behaviour is again expected when a wide-angle wind is launched into an environment with a density distribution sinusoidally stratified along the polar axis (more dense in the equatorial plane than in the polar caps). Hence we hypothesize that in the case of \irasb\ a wide-angle wind is episodically ejected. This different nature compared with that of the jet-driven wind in \iras\ may be due to environmental conditions (different density distributions) or to intrinsically different ejection mechanisms (isotropic vs polar wind), but present observations do not allow us to discern between both scenarios.

\begin{table*}
    \centering
    \caption{{\bf \irasb:} Parameters of the ellipses fitted to the red lobe outflow emission. See Figure \ref{fig:IRAS16-chan-red-ellipses}.}
    \resizebox{0.8\textwidth}{!}{
    \begin{tabular}{l|cccc|cccc}
    \hline
        & \multicolumn{4}{c}{{\bf e1}}       & \multicolumn{4}{c}{{\bf e2}} \\ 
        \hline
    Vel & Position & M\_axis & m\_axis & Angle & Position & M\_axis & m\_axis & Angle \\
    \kms & \radec  & \arcsec & \arcsec & \deg & \radec & \arcsec & \arcsec & \deg \\
    \hline
    5.88 & 16:09:18.5, -39:04:49.8 & 2.708 & 2.144 & 61 & 16:09:18.8, -39:04:48.2 & 4.642 & 2.966 & 57 \\
    6.19 & 16:09:18.5, -39:04:49.9 & 2.708 & 2.144 & 61 & 16:09:18.8, -39:04:47.9 & 5.185 & 3.284 & 57 \\
    6.51 & 16:09:18.7, -39:04:49.1 & 3.237 & 2.524 & 61 & 16:09:18.9, -39:04:47.9 & 5.521 & 3.339 & 59 \\
    6.83 & 16:09:18.7, -39:04:49.1 & 3.237 & 2.524 & 61 & 16:09:18.9, -39:04:47.3 & 5.679 & 3.472 & 59 \\
    7.15 & 16:09:18.7, -39:04:49.2 & 2.600 & 2.203 & 61 & 16:09:18.9, -39:04:47.2 & 6.321 & 3.242 & 59 \\
    7.46 & 16:09:18.7, -39:04:49.3 & 2.365 & 1.960 & 61 & 16:09:19.1, -39:04:46.7 & 6.198 & 3.487 & 59 \\
    7.78 & 16:09:18.8, -39:04:49.4 & 2.037 & 1.763 & 61 & 16:09:19.2, -39:04:45.7 & 4.949 & 3.625 & 59 \\
    8.10 & 16:09:18.8, -39:04:49.1 & 2.386 & 1.587 & 61 & 16:09:19.3, -39:04:45.2 & 4.919 & 3.746 & 59 \\
    8.42 & 16:09:18.9, -39:04:49.0 & 2.386 & 1.587 & 61 & 16:09:19.4, -39:04:44.5 & 5.116 & 4.067 & 59 \\
    \hline
    \end{tabular}
    }
    \label{tab:elipse_param}
\end{table*}

\subsubsection{\j}
\begin{figure*}
    \centering
    \includegraphics[width=0.9\textwidth]{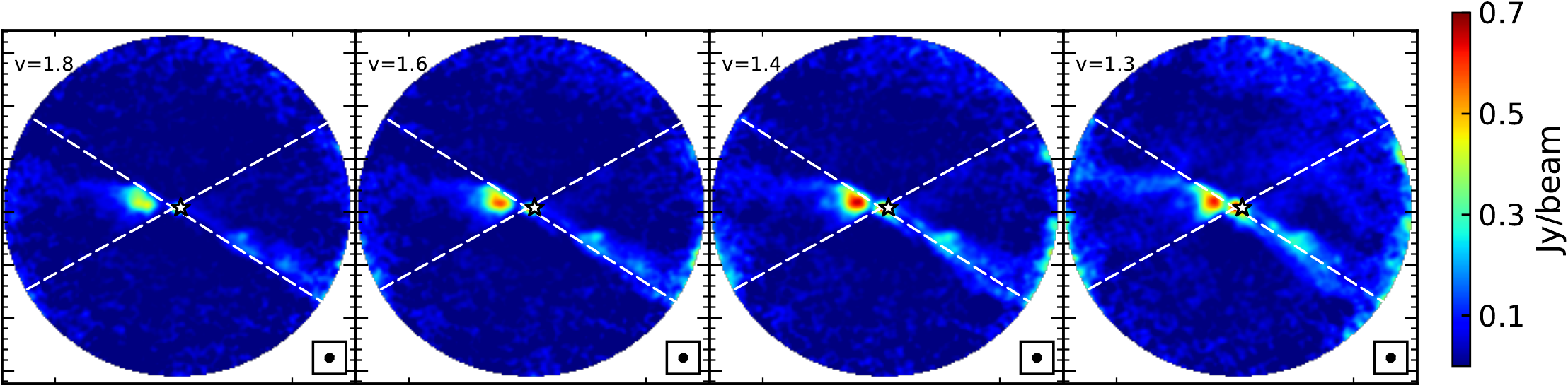}
    \includegraphics[width=0.9\textwidth]{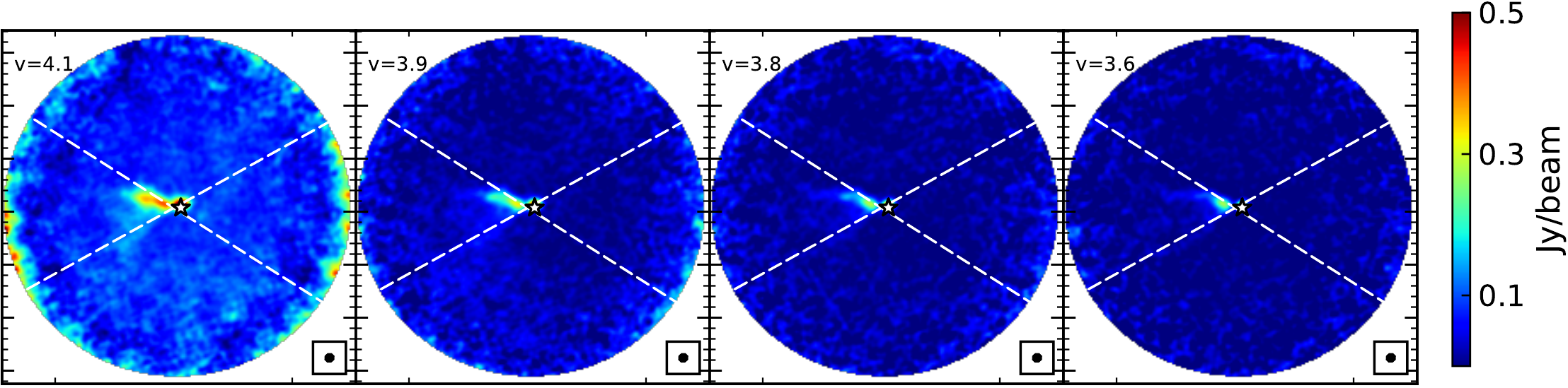}
    \caption{{\bf \j}: Blue (\textit{up}) and red (\textit{down}) high resolution velocity channel maps.}
    \label{fig:J160155-chan-high-resol}
\end{figure*}

The outflow from \j\ is apparently not as complicated as the other two in our sample. It shows at first sight more or less classical cavity walls, probably excavated by a high-velocity jet. Close to the protostar, the high-angular resolution ALMA data apparently shows the typical biconical shape of outflows (see Figure \ref{fig:J160155-chan-high-resol}). 
Although most of the evidence indicates that the outflow coincides with a jet-driven model, there is other evidence that agrees with a wide-angle wind model, as the parabolic structure in the CO emission of the PV diagram of its redshifted lobe (Figure \ref{fig:J160115-pv}). The eastern and western tips of the outflow end up both in bow-shock structures that show larger velocities and dispersions.  

There is however the possibility that this outflow is not so, let us say, well-behaved. The upper panels of Figure \ref{fig:J160115_moments} show the moment 0 maps of the outflow associated with \j. We already noted that the morphology of each lobe is different , and a change of direction of the blueshifted lobe is noticeable at about 6000\,AU from the central source (see dashed lines in Figure \ref{fig:J160115_moments}). In this same blueshifted lobe, between the outflow deviation and the outflow end there are various quasi-parallel stripes of gas oriented close to north-south and seen at the cloud's velocity. Also, as noted in previous sections, the redshifted lobe shows CO emission at position angles north of the main outflow lobe. Are these features at both sides of the protostar position related? Is it possible that \j's outflow drove several ejections with different orientations like \iras's? If this is the case, the north-south stripes of gas in the blueshifted side of the outflow, could be related with the interaction of different outflow ejections and the cloud surface (which emission is more prominent in the western side of the FoV). These hypotheses are right now mere speculations, but the data show hints that the quiet and typical scenario may not be the final answer for this source. New high-angular resolution and better sensitivity observations along the outflow may help to interpret the nature of the outflow.       

\subsection{Characteristics of the source sample} \label{subsec:charac_source_sample}

In the following section we discuss the results obtained from the SEDs, chemistry and outflow parameters with the purpose of determining the evolutionary state of the sources in our sample. 

\subsubsection{SED evolution} \label{subsub:SED}
As pre-main sequence stars evolve, their envelope mass is expected to decrease due to accretion onto the disk-protostar system and/or dispersion by outflowing winds. Thereby, the ratio of the disk/envelope mass can be regarded as an age indicator of a protostar. The envelope material also affects the characteristic spectrum of the escaping light. It absorbs light from lower wavelengths, and reprocesses it into IR and (sub)mm wavelengths, providing the characteristic two-bump Spectral Energy Distribution of many protostars: the submm bump is explained by a $T_{eff}<100$\,K blackbody which corresponds to the cold dust of the envelope emission ($T_d$ in Table \ref{tab:fluxes}, hereafter $T_{cold}$), and the second near-infrared bump, with larger effective temperature, corresponds to the warmer dust emitting in the inner disk, as well as the scattered and reprocessed light from the protostar (we characterize it as $T_{warm}$). The SEDs of pre-main sequence stars and the ratio between the bolometric luminosity and the luminosity from radio up to submillimeter wavelengths only ($L_{bol}$/$L_{submm}$), have been thus utilized as age indicators, based on the relative importance of these two bumps. Hence the spectral index between 2 and $\sim$25\,\mic\ \citep{Adams1987,Greene1994,Young2005}, and an average color measurement translated into a blackbody temperature: the bolometric temperature, $T_{bol}$ \citep[][]{Myers1993}, are commonly invoked to classify the pre-main sequence stars into evolutionary stages or classes \citep{Andre1993}. However, there are some known problems with such indicators due to the non-isotropic nature of the accretion/outflow processes and the different inclinations of the protostellar objects \citep[e.g.][and references therein]{Crapsi2008,Evansetal2009}. Sometimes various indicators are needed to obtain a more realistic evolutionary situation, although the bolometric temperature has been the most trusted age proxy, and it is the base upon which some age estimates are constructed \citep[e.g.,][]{Young2005}.

In this work we collect some commonly used protostellar age indicators for our sampled objects. We summarize some of them in Table \ref{tab:bol} along with certain physical characteristics. 
A quick inspection of the SEDs (Figure \ref{fig:figure1}) shows that only one of the sources does not have emission bluewards of $\sim$25\,\mic\ (\aztec) suggesting this source is probably in the prestellar phase, only surrounded by cold dust. \aztec\ has a one-bump SED while the rest have emission at near infrared wavelengths and most of them show a second bump peaking bluewards of 25$\mu$m as well. We estimate the black-body temperatures of these two bumps for every SED ($T_{cold}$ and $T_{warm}$, see Table \ref{tab:bol}). At the warmer end of the SEDs we have \sz\ and \merin. For these two sources, $T_{warm}$ was estimated by fitting the fluxes at the shortest wavelengths, since the separation between the two bumps is not so evident, suggesting that they are probably in a more evolved stage in which the surroundings of the young stars is becoming more transparent. We should keep in mind that stars with face-on or low-inclined disks can also show a SED with two ill-separated bumps \citep[e.g.][]{Crapsi2008,Enoch2009}, but in this case, the disks of \sz\ and \merin\ are closer to edge-on. The rest of the sources display SEDs with two clear bumps. We will see in the following that fine-tuning the age classification of our young stellar sample is not straightforward, even after using several of the age indicators at hand.

We use the SEDs to estimate $T_{bol}$ as: 
$$T_{bol}= 1.25 \times 10^{-11} \langle\nu\rangle\,{\rm K}\quad,$$ 
where the mean frequency is 
$$\langle\nu\rangle = \frac{\int \nu S{_\nu} {\rm d}\nu}{\int S{_\nu} {\rm d}\nu}\quad.$$ 
We also estimate the bolometric luminosity, $L_{bol}$:
$$L_{bol}=4\pi d^2 \int S_\nu {\rm d}\nu\quad,$$  
where $d$ is the distance and $S_\nu$ is the flux at the frequency $\nu$. We define the ratio $L_{bol}$/$L_{submm}$, in which $L_{submm}$ is obtained integrating the SED at $\lambda>350\mu$m. Their values and uncertainties are in Table \ref{tab:bol} and were obtained integrating by two different methods (Simpson and trapezoidal rules), before further averaging (final estimate) and subtracting (uncertainty) them both. The infrared and (sub)millimeter spectral indices ($\alpha_{IR}$\ and $\beta_{mm}$, respectively) of the source sample are another typical evolutionary indicators shown in Table \ref{tab:bol}.  The table contains the ratio between the disk and envelope mass, $M_{disk}$/$M_{env}$ as well, calculated from the masses estimated by the continuum emission observed using the 12m-only ($M_{disk}$) and the substracting between the 7m-only (M$_{env+disk}$) and the 12m-only array. This ratio cannot be calculated for \sz\ and \merin, since, as mentioned in Section \ref{subsec:continuum}, the envelope masses would be zero.

\aztec\ and \aztecb\ are the two sources with no disk detection ($M_{disk}$/$M_{env}$=0) as well as a  $L_{bol}$/$L_{submm}$ ratio lower than 2; therefore, we consider that they are probably in the prestellar phase. While \aztec's $T_{bol}$ is, as expected, very low (10\,K), \aztecb's is over 100\,K due to the second bump in the SED with T$_{warm}=1000$\,K. Higher spatial resolution infrared observations are needed to decide whether this second warm dust component is truly associated with the cold dust component or it is just spatially overlapped. In the meantime, it seems that for \aztecb\, $T_{bol}$ is not a robust tracer of its evolutionary stage. 

On the upper end of the $M_{disk}$/$M_{env}$ ratio (which diverges), \sz\ and \merin\ have a similar luminosity (L$_{bol}\sim0.1$\lsun) and show the two largest $T_{bol}$ values ($>450$\,K) and  $L_{bol}$/$L_{submm}$ ratios ($>130$). Following Table 1 of \cite{Enoch2009}, these two sources may be classified as late Class\,I protostars, since 300\,K$<$ T$_{bol}<650$\,K. Their infrared spectral indices however are not flat or negative as expected for Class\,I stars, indicating the problems of $\alpha_{IR}$ as a reliable age proxy in these early evolutionary stages. We notice too that \sz\ has a warm temperature, larger than 4000\,K, which may indicate that the protostar has already taken off the envelope veil, maybe aided by a more face-on disk which is in addition much smaller and less massive than \merin's disk. \sz\ has a lower $T_{cold}$ than \merin\ though. These two sources are likely transiting a similar evolutionary stage, but the $T_{bol}$ and the ratio  $L_{bol}$/$L_{submm}$ do not agree about which is at an earlier stage. We note at this point that these SEDs are not corrected by extinction, which may produce differences in the results. 

The remaining three sources (\iras, \irasb and \j) show prominent CO emission associated with their outflows, suggesting that they entered into a prominent accretion phase recently (outflow dynamical times less than a few 10$^4$ years). Following again the classification scheme of \cite{Enoch2009} based on $T_{bol}$, \irasb\ may be the youngest of the three (early Class\,0), while \iras\ and \j\ may be more evolved (both in the late Class\,0 or early Class\,I stage) (see also \citealt{Chenetal1995}). We obtain the exact same order when taking the ratio $L_{bol}$/$L_{submm}$ into account. Regarding the disk to envelope mass ratio, the three sources are arranged in the opposite order as set by the bolometric luminosity. If we take as valid the trend based on the $T_{bol}$, then the ratio between disk and envelope mass should be taken with care when dating objects in a similar evolutionary stage. We should make the caveat that \iras\ is one order of magnitude brighter than the other two, though.
It should be noted that since the disks are near edge on, they may actually be older than their SED appears \citep{Crapsi2008}.

Overall, the most used evolutionary indicators ($T_{bol}$, and the ratios $L_{bol}$/$L_{submm}$ and $M_{disk}$/$M_{env}$) seem to agree in diagnosing which are generally more or less evolved sources, but they do not agree completely, and sometimes the different indicators lead to contradictions when arranging sources within a similar stage. The infrared or submillimeter spectral indices $\alpha$ and $\beta$ are not robust tools for this sample.

\begin{table*}
    \centering
    \caption{$T_{cold}$ and $T_{warm}$ obtained from fitting a black-body model to FIR+mm and NIR peak emission, respectively. $T_{bol}$ and $L_{bol}$ values were obtained applying two different integration methods and averaging between them (Simpson and Trapezoidal rules). The $M_{disk}$/$M_{env}$ values were calculated as the ratio between the emission measured with the 12m and the 7m-12m array (this ratio diverges for \sz\ and \merin). 
    ($\ast$)The last two columns include values for IR spectral index (measured between 2 and 24 $\mu$m) and $T_{bol}$ extracted from \citet{Dunhametal2015} (uncorrected/corrected by extinction). \iras\ was not reported in that paper. \iras\ $T_{bol}$ from \citet{Evansetal2009}.}
    \resizebox{\textwidth}{!}{
    \begin{tabular}{lcccccccccccc}
    \hline
    Source & $T_{cold}$ & $T_{warm}$ & $T_{bol}$ &    $L_{bol}$  & $L_{bol}$/$L_{submm}$  & $M_{disk}$/$M_{env}$ &  Inc. & $\alpha_{IR}$ & $\beta_{mm}$ & $\alpha_{IR}$* & $T_{bol}$* \\
           & (K) &   (K)  &  (K)  & (10$^{-2}$\lsun) &      & (deg) &  & & & (K)\\
    \hline     
    \aztec  &   8.3$\pm$0.2 &       ...    &  10$\pm1$   &  1.0$\pm$0.3  & 1.7   &     0       &    ...    &  ...  & ...  &     ...     & ...      \\
    \aztecb &  10.9$\pm$0.3 & 1000$\pm$200 & 164$\pm$53  &  0.6$\pm$0.2  & 1.0   &     0       &    ...    &  0.78 & ...  &     ...     & ...      \\  
    \irasb  &  30.3$\pm$0.6 &  900$\pm$300 & 42$\pm5$    &   17$\pm$2    & 0.8   &     3.6     & 53$\pm$2  &  0.26 & 0.2  &  1.10/1.14  & 39/39    \\    
    \iras   &  33.1$\pm$0.9 &  700$\pm$300 & 68$\pm27$   &  110$\pm$48   & 2.6   &     0.4     & 66$\pm$14 &  0.60 & 0.1  &      ...    &   48     \\    
    \j      &  29.3$\pm$0.6 & 1150$\pm$40  & 90$\pm13$   & 15.3$\pm$0.7  & 18.9  &     0.3     & 66$\pm$3  & -0.73 & 0.8  & -0.19/-0.26 & 130/150  \\  
    \sz     &   144$\pm$6   & 4300$\pm$300 & 460$\pm$140 &    7$\pm$2    & 135   & $\rightarrow\infty$ & 58$\pm$3  &  0.76 & 0.1  &  0.72/0.63  & 810/3100 \\   
    \merin  &   270$\pm$20  &  750$\pm$70  & 529$\pm1$   &  7.8$\pm$0.1  & 155   &  $\rightarrow\infty$ & 68$\pm$1  &  0.16 & 0.4  &  0.22/0.71  & 560/640  \\    
    \hline     
    \end{tabular}
    }
    \label{tab:bol}
\end{table*}

\subsubsection{Chemical evolution}
\label{disc:chem}
Regarding the detected molecular lines in the present ALMA observations, \co\ and \cdo\ are present in all the observed sources. They are observed as part of the cloud emission, the envelope, the disk, and the outflow. Only \aztec, \iras\ and \irasb\ were associated with emission from other lines (see Tables \ref{tab:results}, \ref{tab:gauss_param} and Figure \ref{fig:figure1}). In particular, these three sources show emission from the deuterated species \nddp\ and DCN. Their hydrogenated counterparts are known to trace dense gas and low temperature regions. \nhhp\ is mainly associated with quiescent gas from envelopes and ISM filaments \citep[e.g.,][]{2013Tobin}, while HCN has also been observed at the cavity walls and head-shocks of outflows \citep[e.g.,][]{2017Busquet}. At very low temperatures, CO and other heavy and abundant molecules freeze-out onto dust grains. As temperatures increase due to the growing of a protostar, the CO molecules are released into the gas-phase again. 
It is worth to note that \nhhp\ is thought to be destroyed as CO is released during this process \citep[e.g.,][]{2003Roberts,2011Busquet,2015Stephens}. This is most important in HII regions or PDRs (Photo Dissociation Regions), but it is also a process seen when protostars evolve and increase their temperature \citep[e.g., L1157,][]{2015Kwon}. Therefore, \nhhp\ is expected mainly in cold (and therefore less evolved) dusty envelopes where CO is still depleted in the dust grain icy mantles, since nitrogen-bearing species are relatively unaffected by freeze-out \citep{1997Charnley}. 
Besides, larger deuterium fractionation (or D-enrichment) has been principally detected in pre-stellar and Class\,0 sources \citep{2014Ceccarelli}. Again, relatively larger abundances of deuterated species are expected due to the very cold temperatures on the order of 10\,K and the freeze-out of abundant neutral species like CO. DCN had been reported in low-velocity emission associated with outflows \citep[e.g.,][]{ 2012Codella}, while \nddp\ is found in dense envelopes of young protostars, mainly in their colder outskirts \citep[e.g.,][]{2013Tobin}.
We can use these molecules as indicators of stellar evolution in the first stages of star formation. The presence of DCN and \nddp\ in three of the sampled sources (\aztec, \iras\ and \irasb) place them at the earliest stages of protostellar evolution, while the non-detection of the nitrogen-bearing deuterated species in the rest of the sources suggests they are warmer and thereby more evolved.
Interpretation of the chemistry of disks and envelopes is complicated, even when large samples of objects in the same evolutionary phase are observed with high-angular resolution by sensitive molecular surveys  \citep[e.g.][]{Miotelloetal2019}, therefore our results should be taken with care.

For \iras\ the observations show the detection of other lines such as CH$_3$OH, SO and SiO. 
These molecules typically trace grain-mantle released into the gas phase, and their abundances in the shocked gas are enhanced with respect to the quiescent gas \citep{Jimenez-serraetal2005,Burkhardtetal2016}.
Some of these are known to trace shocks in outflows, and their presence is usually interpreted as a sign of youth in protostellar outflows (the presence of these molecules associated with \iras\ has already been discussed by \citealt{Jorgensenetal2013}). However, in this case, the presence of fresh gas enhancing the shock chemistry could be not only a matter of the protostellar youth, but also due to the ejections at different position angles of the outflow. The youngest ejections may encounter fresh gas (from a supply of non-previously shocked icy-coated grains) each time the system changes its orientation, thus producing a rich chemistry regardless of the protostellar age. This situation in \iras\ should be taken as a cautionary call when using shocks or chemistry as an outflow signpost of youth. The nature of the individual systems can modify the interpretation of the observations in at least one of every ten sources.   

\subsubsection{Outflow evolution}
Acknowledging the complications imposed by the inaccuracies of the measurements of the physical characteristics of the outflows, and despite realizing that they have different intrinsic natures (wind-driven vs jet-driven) or peculiarities (precession, pulsation), in this section we compare our findings with similar efforts in the literature. We also test some commonly referenced correlations used for evolutionary trends of outflows.

In order to compare the outflow physical characteristics we picked up two similar works on a sample of outflows from low-mass stars: \cite{ArceandSargent2006} and \cite{Dunhametal2014}. These two papers show a list of outflows whose originating sources are categorized in different evolutionary stages. We have split the sources into three main groups (Class\,0, Class\,I and Class\,II sources), and derived median values for the kinetic energy, the mass loss rate and the flux force (see Table \ref{tab:outflow_classes}). We left out three sources classified either as Class\,0 or Class\,I in \cite{Dunhametal2014}.

\begin{table}
    \centering
        \caption{Median outflow properties derived from some of the sources presented in the studies of \cite{ArceandSargent2006} and \cite{Dunhametal2014}.}
        \resizebox{0.5\textwidth}{!}{
    \begin{tabular}{lccccc}
    \hline
    Stage & Number of  & $E$ & $\dot{M}_{out}$ & $\dot{P}_{out}$ \\
           & outflows & (erg) &  \msunyr & \msun\,\kms yr$^{-1}$ \\
    \hline     
    Class 0  & 16 & $2.0\times10^{42}$  &  $3.1\times10^{-6}$ & $8.0\times10^{-6}$ \\
    Class I  & 12 & $3.6\times10^{42}$  &  $4.2\times10^{-6}$ & $1.4\times10^{-5}$ \\
    Class II & 3  & $2.0\times10^{41}$  &  $1.6\times10^{-6}$ & $3.0\times10^{-6}$ \\
    \hline     
    \end{tabular}
    }
    \label{tab:outflow_classes}
\end{table}

Molecular outflows have been presented in the literature as a possible evolutionary tracer for their corresponding protostellar system \citep[e.g.,][]{Curtisetal2010,ArceandSargent2006,Plunkettetal2013,Yildizetal2015,Arceetal2007}. 
Given their ubiquity, morphological and kinematic (velocity) extent, sufficient emission in order to be detected, and the fact that they do not need high-angular resolution observations to be analyzed, the outflows are indeed a promising observational probe.  
However, their measured dynamical times do not trace well the evolutionary stage, and molecular outflows from Class\,I and Class\,II sources can be found with similar dynamical times as outflows from Class\,0 protostars. Hence dynamical times of molecular outflows are not good age tracers and should be taken as lower limits for the age of their corresponding originating sources. Measurements of the energy and momentum of outflows resulted in the general consideration that powerful outflows are younger \citep[e.g.,][]{Bontempsetal1996,Saracenoetal1996}. This has been derived from the correlation of the outflow momentum flux (analogous of the rate of momentum, $\dot{P}_{out}$, reported in this work) with the bolometric luminosity and/or the envelope mass. As previously mentioned, one of the best age proxies for young stellar objects is the bolometric temperature \citep{Laddetal1998}, and therefore it would be good to test the correlation of the outflow parameters against $T_{bol}$ when seeking for an age estimator. Although this should be tested in a statistically significant sample, we will compare the three studied outflows under the view of previous statistical studies to see whether they fit in with general trends.

In this work, we have detected classical molecular outflows from three sources: \iras, \irasb, and \j\ (see Table \ref{tab:results}).  Based on the SOLA catalog SED fitting, these sources are classified as Class 0, Class 0, and late-Class I, respectively (see Table \ref{tab:sources}) but, in the analysis of section \ref{subsub:SED} based on $T_{bol}$ and the $L_{bol}$/$L_{submm}$ ratio, we classified them as late Class\,0, early Class\,0 and late Class\,0, respectively (note that all three may be Class\,0 protostars). We will use the latter arrangement in the following discussion. 

Regarding the bolometric temperature, although the measurements are somewhat uncertain due to the typical scarcity of experimental data, the results for the three outflows reflect that \j\ might be older than \iras, and both older than \irasb\ (90, 68 and 42\,K, respectively). Note however that if we use the ratio of disk mass to envelope mass (another typical protostellar age proxy, that becomes larger with time) the reverse order is obtained, with $M_{disk}$/$M_{env}$ ratios of 0.3, 0.4 and 3.6 for \j, \iras, and \irasb. 

One method of characterizing outflows is based on their emission structure in spatial-velocity space, specifically length, opening angle, and velocity extent. \citet{Curtisetal2010} found that the longest outflows in their sample were Class 0, compared with later classes; while reporting scatter, they indicate that the class-dependent length is significant.  They found average lobe lengths of (140$\pm$20) and (88$\pm$11) arcsec (which at their assumed distance of Perseus, d=250\,pc, corresponds to $3.5\times10^4$\ and $2\times10^4$\,AU) for Class 0 and I sources respectively.  According to Table \ref{tab:param_outflows} in the present work, we report the outflow lobe sizes (length, corrected for inclination) to be between $2\times10^3$\ and $2.5\times10^4$\,AU.  
\irasb\ shows the longest outflow (also likely extending beyond the FOV), consistent with the finding from \citet{Curtisetal2010} that the youngest Class\,0 objects drive longer outflows. The outflow of \j\ (also a late Class 0 or early Class\,I) is consistent with the Class\,I average outflow length from \citet{Curtisetal2010}. While for these two outflows the trend seems to qualitatively apply, on the other hand the shortest outflow in our sample  (\iras's) is also a Class\,0. Hence, the lobe size apparently is not a consistent evolutionary indicator for all sources in our sample, provided $T_{bol}$ is.  

Outflow opening angle ($\theta_{op}$) is also presented in Table \ref{tab:param_outflows}.  \citet{ArceandSargent2006} showed that, even while outflow morphology varies from source to source, a significant trend holds such that outflow opening angle appears to broaden with time.  They fit the outflow angle as a function of age for 17 sources with the relation: $$\log(\theta/deg)=(0.7\pm0.2)+(0.26\pm0.4)\log(t/yr)$$ 
The ages were determined from the $T_{bol}$-age relation from \citet{Laddetal1998}. Among the outflows in our sample, IRAS\,15398-3359 has a significantly smaller opening angle than IRAS\,16059-3857 and J160115-415235. The latter have opening angles more consistent with Class\,I sources in \citet{ArceandSargent2006}.  The trend of outflow opening angle is consistent with that of lobe size, previously discussed.  In both cases, it seems that IRAS 15398-3359 is the least evolved, while IRAS 16059-3857 and J160115-415235 have similar characteristics.

\citet{Curtisetal2010} also presents maximum outflow velocity as an evolutionary indicator; velocity is apparently correlated with length, so that the Class\,0 sources drive longer and faster outflows than Class\,I sources. The velocity extent is larger in \irasb\ than in \j\ (apparently decreasing with age). The velocity extent of \iras\ is the largest, although thought to be a late Class\,0. Although the velocity has not been corrected for inclination angle, either in Table \ref{tab:param_outflows}, or in \citet{Curtisetal2010}, the lack of a trend in our sample holds for speed, and not merely observed radial velocity. 

Momentum, energy, mass-loss rate, and momentum flux are kinematic properties of outflows that have also been linked to age in the literature.
Momentum and energy are thought to decrease from Class\,0 to I.  Momentum has been argued to be a more representative quantity because outflows are thought to be momentum-driven \citep{Shu1991,MassonandChernin1993,Shu2000,Leeetal2000}. \citet{Curtisetal2010} found average momentum for Class 0 and Class I sources to be $\langle$P$_0\rangle$=0.7\,\msun/\kms, $\langle$P$_I\rangle$=0.1\,\msun/\kms; and average energy to be $\langle$E$_0\rangle$=1.4$\times10^{44}$\,erg, $\langle$E$_I\rangle$=1.0$\times10^{43}$, respectively (omitting from the means SVS 13, which may be an anomalously strong Class I outflow).  

The outflows in our sample appear to follow a trend with age, as their energies are arranged in the same order as the $T_{bol}$. The momentum almost follows the same trend, with a reversal between \iras\ and \j.   

Momentum flux (or rate of injected momentum, $\dot{P}_{out}$) may be another age indicator, since it appears to correlate with the envelope mass in both Class\,0 and I sources \citep{Bontempsetal1996,Saracenoetal1996}. For \iras, \j\ and \irasb, the envelope masses are 2.3\,$M_J$, 20.2\,$M_J$ and 36.0\,$M_J$, respectively, and do not correlate with the measured momentum fluxes. However, the measured momentum fluxes inversely correlate with the bolometric temperatures (i.e. the lower the temperature, the larger the momentum flux), hence this could be another useful indicator for the age of outflows.

Our assessment of outflow characteristics, including morphology and energetics as evolutionary indicators, reveals that the nuances of the outflows themselves render these trends unreliable on a source-by-source comparison. In summary, if we assume that \irasb\ is the least evolved (``youngest''), and \j\ is the most evolved (``oldest''), with \iras\ intermediate, then most trends for evolution found in the literature do not consistently hold for all three sources.  Based on limited statistics in our sample, we do not negate the statistical significance of overall trends such as outflow opening angle increasing with time, or length decreasing with evolution. On the other hand, the age of our sample does apparently correlate with energy and momentum flux.  Nonetheless, we point out that specific characteristics of individual sources may render one (observational) trend more robust than another. Observational limitations may include, but are not limited to, uncertainties in distance determination or inclination of the outflow.  Additionally, a thorough investigation into the outflows known to be launched from sources undergoing episodic accretion \citep{Yildizetal2015}, may reveal additional revisions to the understanding of observational outflow trends.

\section{Summary}

In this paper we present the analysis and results of the millimeter continuum and  molecular gas emission associated with 7 prestellar and protostellar objects located in the Lupus molecular clouds.
Based on intermediate angular resolution dedicated observations and high angular resolution data from the ALMA archive, we have analyzed the dusty  disks/envelopes associated with each source and the molecular outflows observed for some of them. 
We present for the first time fully interferometrically mapped data of the outflows associated with \irasb\ and \j.
Based on our results we have attempted to delineate an evolutionary sequence for our source sample by testing different evolutionary indicators:

  \begin{itemize}
      
\item From the SEDs and the continuum emission we obtained the bolometric temperature $T_{bol}$, the ratios $L_{bol}$/$L_{submm}$ and $M_{disk}$/$M_{env}$, and the IR and millimetric spectral indices ($\alpha$ and $\beta$). We found that the first three indicators seem to be good at determining whether a source is inside one of the typical protostellar Classes, but they are not as effective in arranging (by age) sources in similar evolutionary stages. The spectral indices are not good diagnostic tools for our sample.

Based on these indicators we found that \aztec\ and \aztecb\ are the youngest sources (no disk detection) being probably in the pre-stellar phase, while \sz\ and \merin\ seem to be the most evolved (they are probably in the Class\,I) since the surroundings of the young stars appear have become more transparent.
The three remaining sources (\iras, \irasb\ and \j) show prominent \co\ emission associated with their outflows and appear to be in an intermediate evolutionary phase in between the two previous groups (they are probably in the Class\,0), having recently entered into an accretion phase. For these three sources we found distinct ordering results when using different age indicators. 

\item We detected \co\ and \cdo\ emission associated with all the sources in our sample. Three of them (\aztec,\iras\ and \irasb), chemically more complex, also present emission of deuterated species (DCN and \nddp) suggesting that they would be younger.
Only one source in our sample (\iras) presents CH$_3$OH and SiO emission, usually interpreted as a sign of youth in protostellar outflows. However, as the observations show, \iras\ (the most luminous protostar in our sample too) is possibly associated with a precessing outflow, which may produce shocks in the interactions between the ejections and the still fresh unperturbed gas in different directions. 

\item As we mentioned, the parameters obtained from the SEDs do not help us to discern between the evolutionary state of the three sources presenting clearly detected outflows. In an effort to find additional indicators, we have tested several evolutionary indicators obtained from the outflow emission such as the dynamical time, opening angle, maximum outflow velocity, momentum, energy and momentum flux. 
Although most of the evolutionary indicators do not consistently hold for the three sources, it is most likely that \irasb\ is the youngest, followed by \iras\ and \j.

   \end{itemize}
   
In table \ref{tab:summary} we summarize the evolutionary classification obtained for each source according to the different indicators.   

In addition to this, we have characterized the molecular outflows from their morphology and kinematics and contrasted them with the wide-angle and jet-driven models. 
We identified four pairs of bipolar elliptical structures in the outflow associated with \iras, which have different sizes and position angles. We propose that we are possibly detecting an episodic precessing outflow where every ellipse corresponds to a bipolar ejection. The PV diagram supports the hypothesis of an episodic outflow, showing high-velocity spur-like features quite evenly spaced in both lobes, reminiscent of a jet-driven wind model.
On the other hand, \irasb\ shows two lobes with very different morphologies. The redshifted lobe exhibits the typical kinematic features of an episodic wide-angle outflow, evidenced by the presence of more than one parabolic structure in the PV diagram. The elliptical structures in the redshifted lobe agree with the hypothesis of a wide-angle wind episodically ejected, rather than a jet-driven outflow. Otherwise, the terminal part of the blueshifted lobe shows a peculiar bubble-shaped structure which coincides with HH\,78. We propose this is caused because the outflow is emerging from the primordial cloud where the protostar forms.   
Finally, \j\ also shows some peculiarities. The blueshifted lobe suffers a change of direction and various quasi-parallel stripes of gas appear close to the cloud velocity, whereas the red-shifted lobe shows CO emission at position angles north of the main outflow lobe. We wonder if these features on both sides of the protostar position are related, or if it is possible that the outflow has driven several ejections with different orientations as in \iras. Data with higher angular resolution and more sensitivity will be needed to find an answer.

\begin{table*}
\centering
    \begin{tabular}{lccccc}
    \hline
    Source  & SED & $T_{bol}$ & Chemistry & Outflow & Overall \\
    \hline
    \aztec  &  pre-stellar    & early class 0 & pre-stellar/early class 0 & none & pre-stellar \\
    \aztecb &  pre-stellar(*) & early class 0(*) &   only CO envelope            & none & pre-stellar \\
    \irasb  & early class 0   & early class 0 & pre-stellar/early class 0 & early class 0 & early class 0 \\
    \iras   &  late class 0   & late class 0/early class I &        early class 0       &  class 0    & class 0  \\
    \j      &  late class 0   & late class 0/early class I & only CO disk  & late class 0 &  late class 0 \\
    \sz     &    class I      & class I & only CO disk & weak detection & class I \\
    \merin  &    class I      & class I & only CO disk & weak detection & class I \\
    \hline
    \end{tabular}
    \caption{Summary of the evolutionary classifications obtained for each source according to the different indicators. (*) We consider that only the FIR bump is associated with \aztecb.}
    \label{tab:summary}
\end{table*}


\begin{acknowledgements}
This paper makes use of the following ALMA data: ADS/JAO.ALMA 2016.1.01141.S, ADS/JAO.ALMA 2011.0.00777.S, ADS/JAO.ALMA 2013.1.00879.S, ADS/JAO.ALMA 2011.0.00628.S, ADS/JAO.ALMA 2013.1.00244.S, ADS/JAO.ALMA 2015.1.00306.S, ADS/JAO.ALMA 2013.1.00474.S, ADS/JAO.ALMA 2016.1.00571.S, ADS/JAO.ALMA 2016.1.00459.S, ADS/JAO.ALMA 2016.1.01239.S, ADS/JAO.ALMA 2016.1.01239.S, ADS/JAO.ALMA 2015.1.01510.S. ALMA is a partnership of ESO (representing its member states), NSF (USA) and NINS (Japan), together with NRC (Canada), MOST and ASIAA (Taiwan), and KASI (Republic of Korea), in cooperation with the Republic of Chile. The Joint ALMA Observatory is operated by ESO, AUI/NRAO and NAOJ. The National Radio Astronomy Observatory is a facility of the National Science Foundation operated under cooperative agreement by Associated Universities, Inc. 
The authors are grateful to the visiting program to JAO and ESO that facilitated this collaborative effort.
MMV is funded by post-graduate scholarships from the Argentine Consejo Nacional de Investigaciones Científicas y Técnicas (CONICET).
IdG-M is partially supported by MCIU-AEI (Spain) grant AYA2017-84390-C2-R (co-funded by FEDER).
\end{acknowledgements}

%
   \bibliographystyle{aa} 
   \bibliography{biblio_alpha} 
%

\begin{appendix} 
\section*{Appendix A: Summary of individual sources} \label{appendixA}
In this Appendix we briefly describe some respects of the bibliography of each source.

\subsection*{\aztec\ and \aztecb}
The sources \aztec\ and \aztecb\ are located in Lupus\,I and Lupus\,III clouds, respectively, and were classified in the SOLA catalog as prestellar cores. The peak intensity from ASTE/AzTEC 1.1\,mm continuum are I$_{peak}$= 455\,mJy beam$^{-1}$ and I$_{peak}$= 169\,mJy beam$^{-1}$, respectively. \aztecb\ is associated with the IR source J160859.7-390737 detected  in all the Spitzer bands. The cores where the sources are immersed were catalogued by \citet{Benedettinietal2018} from Herschel images. The authors calculate a column density N(H$_2$)= 29.61$\times$10$^{21}$\cmcub and N(H$_2$)= 7.59$\times$10$^{21}$\cmcub, for \aztec\ and \aztecb\ respectively, and classified both as prestellar.

\subsection*{\iras}
\iras\ is located in Lupus\,I and is embedded in the dense core B228 \citep{Barnardetal1927}.
B228 has a mass of 0.3-0.8\,\msun\ (\citealt{Reipurthetal1993}; \citealt{Shirleyetal2000}) and is located at a rest velocity of 5.1\,\kms\ (\citealt{Mardonesetal1997}; \citealt{vanKempenetal2009b}). 
The \iras\ protostar was first identified by \citet{HeyerandGraham1989}, who also discovered HH\,185, a Herbig-Haro object driven by this source, later on catalogued as Class 0 protostar by \cite{Evansetal2009}. 
The outflow driven by \iras\ was discovered by \citet{Tachiharaetal1996} and mapped in HCO$^+$(4-3) and various CO transitions by \citet{vanKempenetal2009b}, who determined a temperature of $\sim$100-200\,K for the outflowing gas, based on the line ratios in different transitions. 

From H$_2$CO and CCH ALMA observations, \citet{Oyaetal2014} detected the outflow extending northeast-southwest, with an inclination angle of 20\deg\ (almost along the line of sight). They estimated an upper limit to the outflow mass of 0.09\,\msun.
\citet{Bjerkelietal2016} detected the outflow in \co, \tco, HCO$^+$ and N$_2$H$^+$ using SMA (Submillimeter Array) data, and calculated the mass (2$\times$10$^{-4}$\,\msun), momentum (4$\times$10$^{-4}$\,\msun\kms), momentum rate (10$^{-6}$ \msun\kms\,yr$^{-1}$), mechanical luminosity (10$^{-3}$\,\lsun), kinetic energy (5$\times$10$^{40}$\,erg), and mass-loss rate (4$\times$10$^{-8}$\,\msunyr).
Using \cdo\ observations, \citet{Yenetal2017} identified some signatures of infall. 
Their kinematical models allowed them to estimate the protostellar mass of 0.01\,\msun, and the disk radius of 20\,AU.

\subsection*{\irasb}
\irasb (also named Lupus\,3\,MMS) is an embedded protostar in the Lupus\,III Molecular Cloud \citep{Comeron2008}. It was first discovered by \citet{Tachiharaetal2007} using H$^{13}$CO$^+$(1-0) and 1.2\,mm continuum observations. They point out the diffuse and faint cometary shape  IR emission pointing toward the source, suggesting that this emission is a jet coming from a embedded young protostar and physically associated with \irasb\ and the Herbig-Haro object HH\,78.
Lupus\,3\,MMS has been detected  in  the  mid-infrared  at  3.6–70\,\mic\ with Spitzer (\citealt{Tachiharaetal2007}; \citealt{Chapmanetal2007}; \citealt{Merinetal2008}) from which it has been classified as a Class\,0 protostar (\citealt{Tachiharaetal2007}; \citealt{Dunhametal2008}; \citealt{Evansetal2009}). 

\subsection*{\j}
The source SSTc2d\,J160115.6\,415235.3 is a deeply embedded protostar in the Lupus\,IV Molecular Cloud at a distance of 150\,pc  \citep{Comeron2008}, and is associated with 2MASS\,J16011549-4152351 located near the center of the globular filament GF\,17 \citep{Chapmanetal2007}.
Through the infrared detections, it was classified as a flat-spectrum source \citep{Merinetal2008}.
The source was also identified as a transition disk candidate by \cite{Ansdelletal2018} and the gas and dust masses of the surroundings of this young star have been determined in an ALMA disk survey carried out in Lupus previously (0.007 and $6\times10^{-5}$\msun for its envelope and disk respectively, \citealt{Ansdell2017PhD}).

\subsection*{\sz}
The source \sz\ \citep{Schwartz1977}, also known as Th\,28 \citep{The1962}, is associated with the Herbig-Haro object HH\,228, which consists of an east--west jet emanating from the driving source, and the emission of the Herbig--Haro objects HH\,228\,E, HH\,228\,W, HH\,228\,E2, HH\,228\,E3 and HH\,228\,E4 \citep{Krautter1986, GrahamandHeyer1988, Comeronandfernandez2011}. Proper motions of $>$260\kms have been measured for HH\,228\,W, E, E2, E3 and E4 \citep{WangandHenning2009,Comeronandfernandez2011}. 
The system of HH objects associated with \sz\ appears to nearly lie on the plane of the sky \citep{Comeronandfernandez2011}, which is also consistent with the faintness of the central source (probably due to the obscuration by a  circumstellar disk seen nearly edge-on, \citealt{Hughesetal1994}). Finally, the jet presents hints of rotation \citep{Coffeyetal2004,Coffeyetal2007} and a remarkably high mass--loss rate in the outflow ($\sim$ 1.2$\times$ 10$^{-8}$ \msunyr, \citealt{Coffeyetal2008}).

\subsection*{\merin}
\merin\ is associated with the 2MASS source J\,16245177-3956326 and was classified as a flat spectral source \citep{Spezzietal2011}. This object is not detected at optical wavelengths as expected for young objects still partially embedded, and appears point-like in all the 2MASS and Spitzer infrared images. The lack of detection in the HCO$^+$(3-2) line observations casted doubts on the existence of a residual dense envelope seen in other sources with flat infrared spectral indices \citep{vanKempeneetal2009c}. 

\newpage
\section*{Appendix B: Tables} \label{appendixB}
The tables in this appendix include information referred to the infrared data mentioned in Section \ref{sec:IR}

\begin{table}[h!]
    \centering
    \resizebox{0.48\textwidth}{!}{
    \begin{tabular}{ccccc}
        \hline
        \multicolumn{5}{c}{{\bf \aztec}}\\
        \hline
         Band & Wavelength &  FWHM   &  Flux & Reference \\
              &  (\mic)    &(\arcsec)& (mJy) &           \\
         \hline
         PACS70   &  70.0 &  5.0 & 0.81$\pm$10  & 1  \\
         PACS160  & 160.0 & 12.0 &   48$\pm$50  & 1  \\
         SPIRE250 & 250.0 & 17.6 &  750$\pm$200 & 1  \\
         SPIRE350 & 350.0 & 23.9 & 2500$\pm$300 & 1  \\
         SPIRE500 & 500.0 & 35.2 & 3600$\pm$200 & 1  \\
        \hline
        \multicolumn{5}{c}{{\bf \aztecb}}\\
        \hline
         Band & Wavelength & FWHM & Flux & Reference \\
              &  (\mic)    &(\arcsec)& (mJy) &           \\
         \hline
         WISE-W1       & 3.4    & 6.1  & 2.15$\pm$0.05 & 2 \\
         Spitzer-IRAC1 & 3.6    & 1.66 & 3.26$\pm$0.18 & 3 \\
         Spitzer-IRAC2 & 4.5    & 1.72 & 4.36$\pm$0.22 & 3 \\
         WISE-W2       & 4.6    & 6.4  & 5.23$\pm$0.11 & 2 \\
         Spitzer-IRAC3 & 5.8    & 1.88 & 4.63$\pm$0.23 & 3 \\
         Spitzer-IRAC4 & 8.0    & 1.98 & 2.94$\pm$0.15 & 3 \\
         WISE-W3       & 12.0   & 6.5  & 0.73$\pm$0.35 & 2 \\
         Spitzer-MIPS1 & 23.68  & 6.0  & 0.77$\pm$0.52 & 3 \\
         PACS70   &  70.0 &  5.0 &  1.1$\pm$10  & 1  \\
         PACS160  & 160.0 & 12.0 &  220$\pm$80  & 1  \\
         SPIRE250 & 250.0 & 17.6 & 1000$\pm$200 & 1  \\
         SPIRE350 & 350.0 & 23.9 & 1400$\pm$200 & 1  \\
         SPIRE500 & 500.0 & 35.2 & 1800$\pm$200 & 1  \\
         \hline
         \multicolumn{5}{c}{{\bf \iras}}\\
         \hline
         Band & Wavelength & FWHM & Flux & Reference \\
              &  (\mic)    &(\arcsec)& (mJy) &           \\
         \hline
         Spitzer-IRAC1 & 3.6    & 1.66 &  6.06$\pm$0.50 & 3 \\
         Spitzer-IRAC2 & 4.5    & 1.72 &  32.5$\pm$2.7  & 3 \\
         Spitzer-IRAC3 & 5.8    & 1.88 &  61.1$\pm$4.9  & 3 \\
         Spitzer-IRAC4 & 8.0    & 1.98 &  21.5$\pm$3.9  & 3 \\
         Spitzer-MIPS1 & 23.68  & 6.0  & 107.0$\pm$31.9 & 4 \\
         PACS70     &  70.0 &  5.0 & 16000$\pm$30  & 1  \\
         PACS160    & 160.0 & 12.0 & 53000$\pm$80  & 1  \\
         SPIRE250   & 250.0 & 17.6 & 44000$\pm$200 & 1  \\
         SPIRE350   & 350.0 & 23.9 & 26000$\pm$400 & 1  \\
         SPIRE500   & 500.0 & 35.2 & 14000$\pm$200 & 1  \\
         ALMA-Band6 & 1340  &  7.0 & 35.0$\pm$2.6  & 8 \\
         \hline
         \multicolumn{5}{c}{{\bf \irasb}}\\
         \hline
         Band & Wavelength & FWHM & Flux & Reference \\
              &  (\mic)    &(\arcsec)& (mJy) &           \\
         \hline
         WISE-W1       & 3.4    & 6.1  & 0.30$\pm$0.01 & 2 \\
         Spitzer-IRAC1 & 3.6    & 1.66 & 0.26$\pm$0.03 & 3 \\
         Spitzer-IRAC2 & 4.5    & 1.72 & 1.00$\pm$0.06 & 3 \\
         WISE-W2       & 4.6    & 6.4  & 2.00$\pm$0.05 & 2 \\
         Spitzer-IRAC3 & 5.8    & 1.88 & 0.99$\pm$0.07 & 3 \\
         Spitzer-IRAC4 & 8.0    & 1.98 & 0.55$\pm$0.05 & 3 \\
         WISE-W4       & 22.0   & 12.0  & 25.9$\pm$1.7 & 2 \\
         Spitzer-MIPS1 & 23.68  & 6.0  & 32.4$\pm$3.04 & 3,4 \\
         PACS70        & 70.0   & 5.0  & 2800$\pm$20   & 1  \\
         Spitzer-MIPS2 & 71.42  & 18.0 & 2610$\pm$252  & 3,4 \\
         Spitzer-MIPS3 & 155.9  &  --  & 8708$\pm$1742 &  4 \\
         PACS160  & 160.0 & 12.0 & 9000$\pm$60  & 1  \\
         SPIRE250 & 250.0 & 17.6 & 8600$\pm$100 & 1  \\
         SPIRE350 & 350.0 & 23.9 & 5600$\pm$200 & 1  \\
         SPIRE500 & 500.0 & 35.2 & 3600$\pm$100 & 1  \\
         ALMA-Band6 & 1340  &  7.0 & 230.0$\pm$3.3  & 8 \\
         \hline
    \end{tabular}
    }
    \caption{(1):\citet{Benedettinietal2018}. (2):\citet{Cutrietal2013}. (3):\citet{Evansetal2014}. (4):\citet{Merinetal2008}. (5):\citet{DENISConsortium2005}. (6):\citet{GaiaCollaboration2018}. (7):\citet{Yamamuraetal2010}. (8): this work. ALMA fluxes consider only statistical errors (10\% calibration error is not included).}
    \label{tab:IR}
\end{table}       

\renewcommand{\thetable}{\arabic{table} (Cont.)}
\addtocounter{table}{-1}
         
\begin{table}
    \centering
    \resizebox{0.48\textwidth}{!}{
    \begin{tabular}{ccccc}     
         \hline
         \multicolumn{5}{c}{{\bf \j}}\\
         \hline
         Band & Wavelength & FWHM & Flux & Reference \\
              &  (\mic)    &(\arcsec)& (mJy) &           \\
         \hline
         2MASS-J       & 1.235  & 1.22 & 0.42$\pm$0.06 & 2 \\
         2MASS-H       & 1.662  & 1.63 & 1.67$\pm$0.10 & 2 \\
         DENIS-K       & 2.15   & 2.15 & 5.97$\pm$0.66 & 5 \\
         WISE-W1       & 3.4    & 6.1  & 5.06$\pm$0.11 & 2 \\
         Spitzer-IRAC1 & 3.6    & 1.66 & 8.36$\pm$0.45 & 3 \\
         Spitzer-IRAC2 & 4.5    & 1.72 & 9.92$\pm$0.49 & 3 \\
         WISE-W2       & 4.6    & 6.4  & 9.55$\pm$0.18 & 2 \\
         Spitzer-IRAC3 & 5.8    & 1.88 & 8.98$\pm$0.44 & 3 \\
         Spitzer-IRAC4 & 8.0    & 1.98 & 7.70$\pm$0.37 & 3 \\
         WISE-W3       & 12.0   & 6.5  & 5.35$\pm$0.25 & 2 \\
         WISE-W4       & 22.0   & 12.0 & 70.59$\pm$1.89 & 2 \\
         Spitzer-MIPS1 & 23.68  & 6.0  & 75.90$\pm$7.03 & 3,4 \\
         Spitzer-MIPS2 & 71.42  & 18.0 & 1220$\pm$123 & 3,4 \\
         Spitzer-MIPS3 & 155.9  &  --  & 5386$\pm$1077 &  4 \\
         ALMA-Band6 & 1340  &  7.0 & 81.9$\pm$2.7  & 8 \\
         \hline         
                  \multicolumn{5}{c}{{\bf \sz}}\\
         \hline
         Band & Wavelength & FWHM & Flux & Reference \\
              &  (\mic)    &(\arcsec)& (mJy) &           \\
         \hline
         APASS-DR9-B   & 0.438  &  --  &  0.63$\pm$0.01 & -- \\  
         GAIA-GBP-DR2  & 0.532  &  --  &  0.93$\pm$0.01 & 6 \\
         APASS-DR9-V   & 0.545  &  --  &  0.94$\pm$0.01 & -- \\
         GAIA-G-DR2    & 0.673  &  --  &  0.96$\pm$0.01 & 6 \\
         GAIA-GRP-DR2  & 0.797  &  --  &  1.76$\pm$0.03 & 6 \\
         DENIS-I       & 0.82   & 0.82 &  1.52$\pm$0.07 & 5 \\
         2MASS-J       & 1.235  & 1.22 &  2.39$\pm$0.09 & 2 \\
         2MASS-H       & 1.662  & 1.63 &  3.54$\pm$0.12 & 2 \\
         DENIS-K       & 2.15   & 2.15 &  6.37$\pm$0.65 & 5 \\
         WISE-W1       & 3.4    & 6.1  & 11.27$\pm$0.24 & 2 \\
         WISE-W2       & 4.6    & 6.4  & 22.96$\pm$0.38 & 2 \\
         WISE-W3       & 12.0   & 6.5  & 105.3$\pm$1.4  & 2 \\
         WISE-W4       & 22.0   & 12.0 & 371.2$\pm$8.2  & 2 \\
         Spitzer-MIPS1 & 23.68  & 6.0  & 347.0$\pm$32.6 & 3,4 \\
         PACS70        & 70     & 5.2  &   330$\pm$20   & 1 \\
         Spitzer-MIPS2 & 71.42  & 18.0 & 257.0$\pm$30.2 & 3,4 \\
         PACS160       & 160    &  12  &   230$\pm$80   & 4 \\
         ALMA-Band6 & 1340  &  7.0 & 4.7$\pm$2.6  & 8 \\         
         \hline         
         \multicolumn{5}{c}{{\bf \merin}}\\
         \hline
         Band & Wavelength & FWHM & Flux & Reference \\
              &  (\mic)    &(\arcsec)& (mJy) &           \\
         \hline
         2MASS-J       & 1.235 & 1.22 &   0.52$\pm$0.06 & 2 \\
         DENIS-J       & 1.25  & 1.25 &   0.56$\pm$0.11 & 5 \\
         2MASS-H       & 1.662 & 1.63 &   2.06$\pm$0.11 & 2 \\
         DENIS-K       & 2.15  & 2.15 &  10.24$\pm$1.04 & 5 \\
         WISE-W1       & 3.4   & 6.1  &  25.35$\pm$0.51 & 2 \\
         WISE-W2       & 4.6   & 6.4  &  52.12$\pm$0.91 & 2 \\
         AKARI-9       & 9.0   & --   &  137.4$\pm$5.9  & 7 \\
         WISE-W3       & 12.0  & 6.5  & 122.66$\pm$1.47 & 2 \\
         AKARI-18      & 18.0  & --   &  224.6$\pm$5.37 & 7 \\
         WISE-W4       & 22.0  & 12.0 & 235.27$\pm$4.33 & 2 \\
         ALMA-Band6 & 1340  &  7.0 & 31.9$\pm$11.2  & 8 \\         
         \hline         
    \end{tabular}
    }
    \caption{(1):\citet{Benedettinietal2018}. (2):\citet{Cutrietal2013}. (3):\citet{Evansetal2014}. (4):\citet{Merinetal2008}. (5):\citet{DENISConsortium2005}. (6):\citet{GaiaCollaboration2018}. (7):\citet{Yamamuraetal2010}. (8): this work. ALMA fluxes consider only statistical errors (10\% calibration error is not included).}
    \label{tab:IR2}
\end{table}

\renewcommand{\thetable}{\arabic{table}}
\section*{Appendix C: Peculiarities of the individual outflows} \label{appendixC}
In this Appendix we describe some peculiar features of the outflows associated with \iras\ and \irasb, and intend to explain them.

\subsection*{\iras}
The source \iras\ has been very well studied (e.g. \citealt{Jorgensenetal2013}, \citealt{Oyaetal2014}, \citealt{Bjerkelietal_2_2016},  \citealt{Yenetal2017}).
However we have detected molecular emission which is not related with the source nor with its previously-identified outflow.

\begin{figure*}
    \centering
    \includegraphics[width=0.8\textwidth]{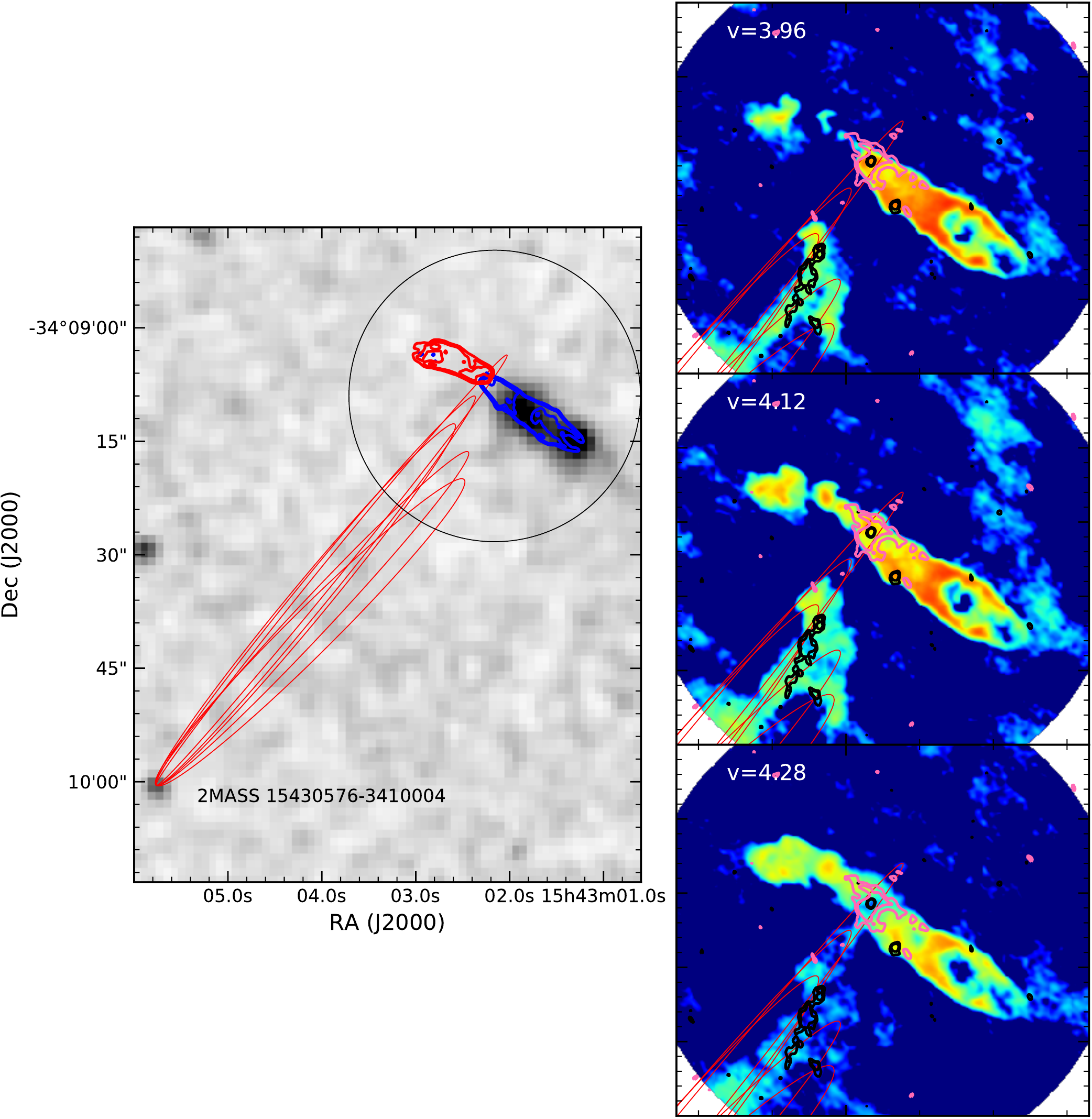}
    \caption{\textit{Left}: 2MASS K$_s$ band image (grey-scale) and \co(2-1) emission associated with the blue and red lobes of \iras\ (blue and red contours, respectively). The source 2MASS\,15430576-3410004, located in the left bottom corner in the image, could be driving the \co(2-1) emission detected southeast of \iras. The red ellipses indicate possible ejection interacting with the gas in the vicinity of \iras. The black circle indicates the region zoomed in the right panels. \textit{Right}: Velocity channel maps of the \co(2-1) emission (color scale) near the cloud velocity. SO(6$_5$, 5$_4$) and \cdo(2-1) moment 0 maps are shown in black and pink contours, respectively.} 
    \label{fig:IRAS15398_2MASS}
\end{figure*}

In the right panels of Figure \ref{fig:IRAS15398_2MASS} we show the \co(J=2--1) emission in three channels close to the cloud velocity. We also show the \cdo(J=2--1) and SO(J$_N$=6$_5$, 5$_4$) moment 0 emission (integrated from 3.24 to 4.52\,\kms). 
At these velocities a structure can be seen almost perpendicular to the main outflow associated with \iras. This structure is also revealed by the SO(J$_N$=6$_5$, 5$_4$) and SiO(4--3) emission (the latter is not shown here), indicating the possible presence of shocked gas.
The latter two molecules are known as good tracers of physical shocks in the interstellar and circumstellar material, and are frequently detected associated with bipolar outflows of Class\,0 and I protostars (e.g., \citealt{Tafallaetal2010}, \citealt{Podioetal2015}, \citealt{Sakaietal2016}).
The \cdo(J=2-1) emission, which shows the disk associated with the central source and traces parts of the internal \iras\ outflow walls, also presents an elongation in the direction of this filament.

One possibility to explain this feature is the presence of another source located southeast of \iras, throwing an outflow that impacts and excites the dense environment of \iras. In the panels of Figure \ref{fig:IRAS15398_2MASS} the red ellipses indicate possible ejections coming from the southeast. The variations in the position of these shocks may be due to the source being precessing and therefore outflowing in different directions. These bullets could be even crossing the \iras\ outflow, being responsible for the lower CO emission intensity at the source position in the 4.28\,\kms\ velocity channel (bottom left panel in Figure\,\ref{fig:IRAS15398_2MASS}), if the gas in these southern ejections had a lower temperature. Moreover, the crossing of the southern outflow may explain the \cdo\ elongation perpendicular to the \iras\ outflow. Although there may have been some contamination of the outflow originating from \iras\ our main analysis of that outflow is not affected.

We have also identified a possible source responsible for launching this second outflow: 2MASS\,15430576-3410004, located about 69\arcsec south east of \iras\ (exactly at \radec\ 15:43:05.76, --34:10:00.4). The ellipses describing the tips of the observed shocks can all have the same origin at the position of this source. 

\subsection*{\irasb}
The blue lobe of the outflow associated with \irasb\ in the vicinity of the protostar ($<$3500\,AU) is V-shaped, but at greater distances it has two very particular features: on the one hand the shreds of CO emission south of the outflow cone
, and on the other hand the bubble-shaped structures at the terminal part of the outflow. The bubble-shaped structures spatially coincide with some red-shifted emission (see upper left panel of Figure\,\ref{fig:IRAS16_moments}) and with the Herbig--Haro object HH\,78 (\radec = 16:09:12.8, --39:05:02). This HH object, can be observed in the Spitzer bands as a small ($\sim$ 4\arcsec) faint nebulosity, mostly at 3.6 and 4.5 \mic.

\citet*{Comeronandfernandez2011} proposed that HH\,78 originates from a jet emerging from the original molecular cloud, since deep infrared observations of the region \citep{Nakajimaetal2003} showed that the visible portion of HH\,78 can be traced as a well collimated jet up to the position of \irasb\ \citep{Tachiharaetal2007}, which is almost surely at the origin of this jet. The near-infrared part of the jet is bright in the K$_S$ band, probably because of shock-excited H$_2$ emission coincident with the  molecular outflow detected by \citet{Tachiharaetal2007}. 
This suggests that the jet would originate from a location deeply embedded in the cloud that hosts \irasb\ and moves towards the side of the cloud that faces our direction, showing itself as the visible counterpart of HH\,78 as it emerges.
This scenario is in line with what is revealed by the present observations. The data suggests that the CO bubble-shaped structure seen at the western tip of the outflow is showing the outflow escaping from the molecular cloud in which the source is immersed. 

\begin{figure*}
    \centering
    \includegraphics[width=1.0\textwidth]{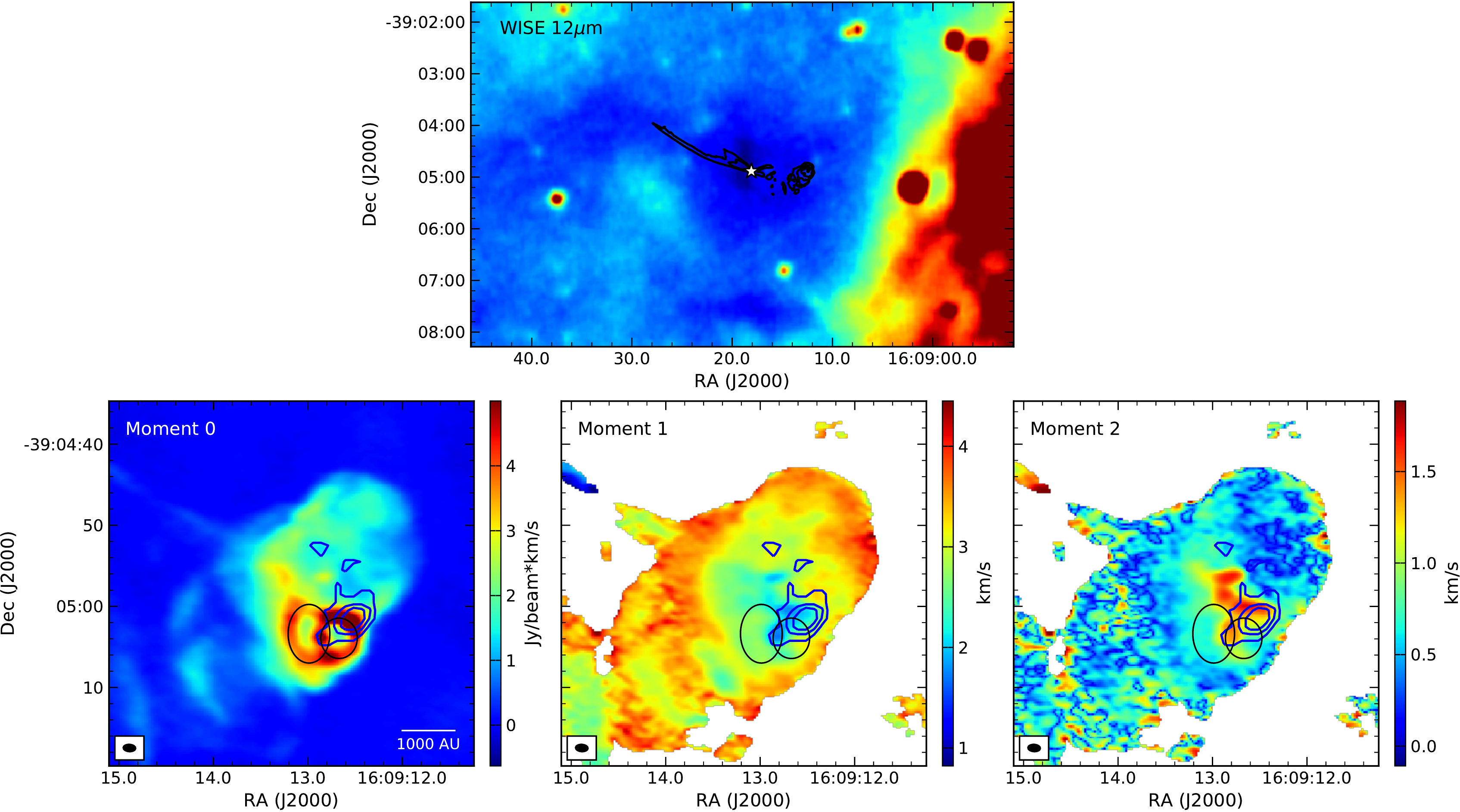}
    \caption{{\bf \irasb.} \textit{Up:} 12\,\mic\ WISE emission. The star symbol indicates the position of the compact source associated with \irasb. The black contours shows the outflow emission. \textit{Bottom:} Moment maps of the shell-shape structures detected in the blue lobe. Black ellipses show the structures detected from the moment\,0 map. Blue contours show the dss2 emission associated with HH\,78.}
    \label{fig:IRAS16_bubble_mom}
\end{figure*}

In the upper panel of Figure \ref{fig:IRAS16_bubble_mom} (WISE 12\,\mic\ emission), it can be seen that the outflow associated with IRAS 16059-3857 is immersed in a dark region or molecular pocket, at the center of which is forming the protostar. 
We see that the blue lobe tip coincides with the border of that dark region so we think that the structures detected there could be produced when the outflow reaches the limit of the dense pocket of gas and dust.
The lower panels of the figure, show two elliptical structures clearly revealed by the moment 0 map. These structures spatially coincide with the DSS2 optical emission (shown in blue contours), produced at the position of HH\,78.
The two ellipses match the regions of highest velocity and dispersion, as observed in the moment 1 and 2 maps.
The complex arc structures detected in these images could be due to the interaction of the molecular outflow with the edge of the dense pocket in which the system is embedded.

\begin{figure*}
    \centering
    \includegraphics[width=1.0\textwidth]{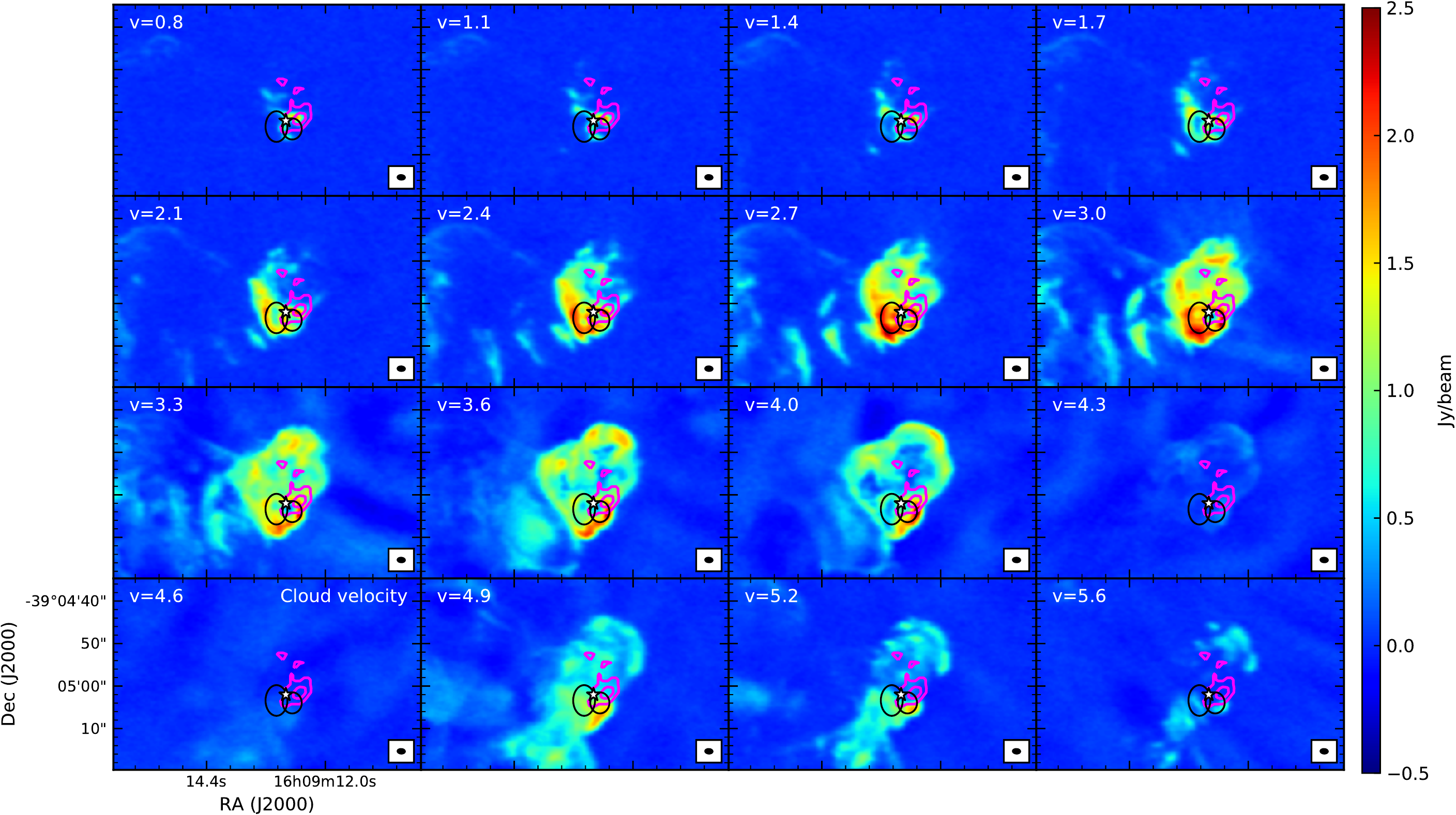}
    \caption{{\bf \irasb:} Velocity channel maps showing the structures in the blue lobe. Shell-like structures are detected coinciding with the position of the Herbig Haro object HH\,78 (dss2 emission shown in magenta contours). The ellipses are the same as those in Figure\,\ref{fig:IRAS16_bubble_mom}.}
    \label{fig:IRAS16_bubble}
\end{figure*}

In Figure\,\ref{fig:IRAS16_bubble} we display the same ellipses over the velocity channel maps from 0.8 to 5.6\,\kms, covering the blueshifted channels, the cloud velocity (4.6\,\kms) and also part of the redshifted emission. The edges of the ellipses are detected in the extent of this velocity range, along with a very complex structure comprising arcs and filaments.

We suggest that the high velocity gas emission (both at blue- and red-shifted velocities) at the position of HH\,78 is possibly due to physical interactions between different ejections from the protostellar system or between the jet and the walls of the molecular cloud.

\section*{Appendix D: Complete velocity cubes} \label{appendixD}

In this Appendix we include the complete velocity cubes with 0.32\,\kms\ width channels, for \iras\ and \irasb, and 0.17\,\kms\ width channels for \j.

\begin{figure*}
    \centering
    \includegraphics[width=1.0\textwidth]{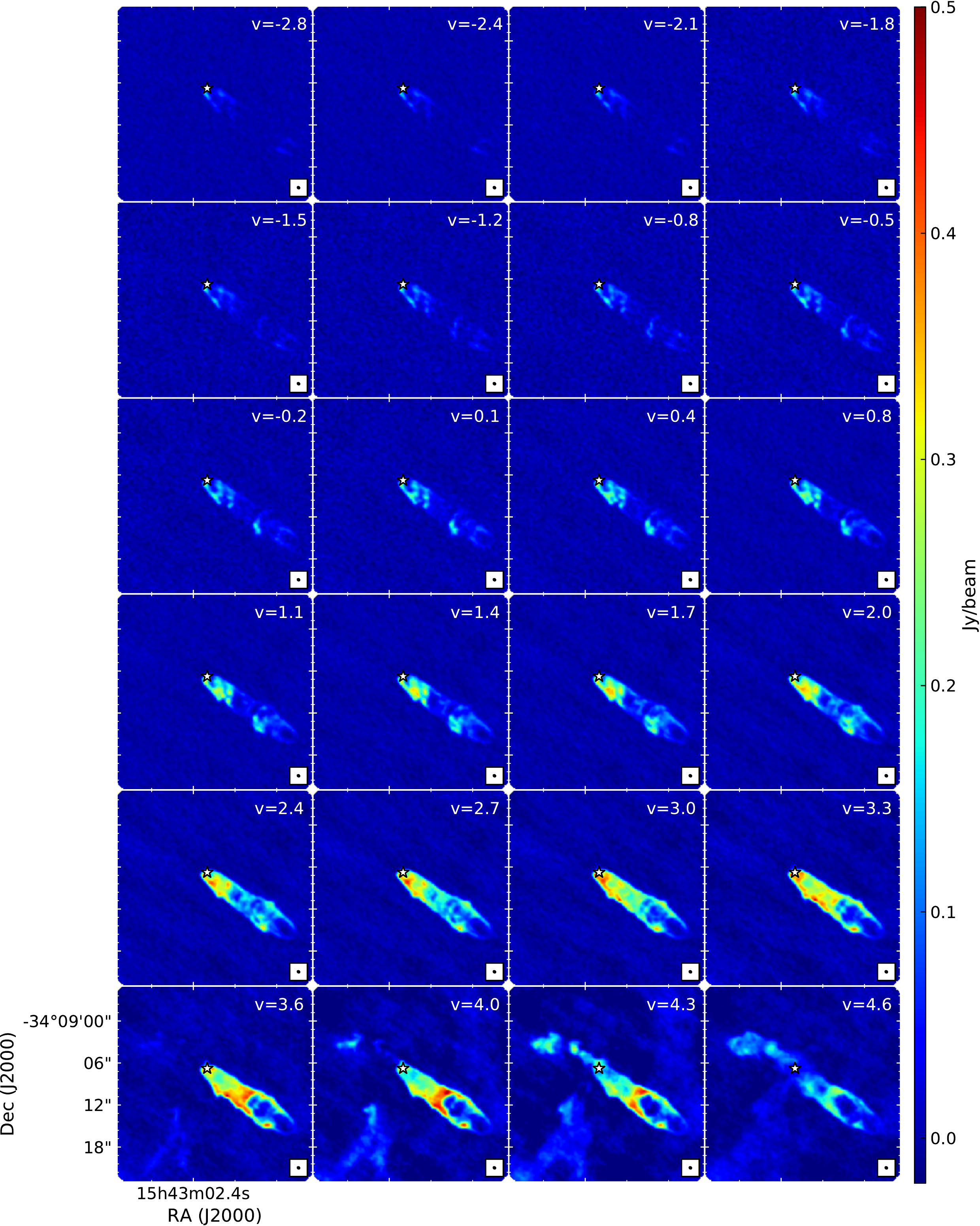}
    \caption{{\bf \iras:} Blue lobs channel maps.}
    \label{fig:IRAS15-chan-blue}
\end{figure*}{}

\begin{figure*}
    \centering
    \includegraphics[width=1.0\textwidth]{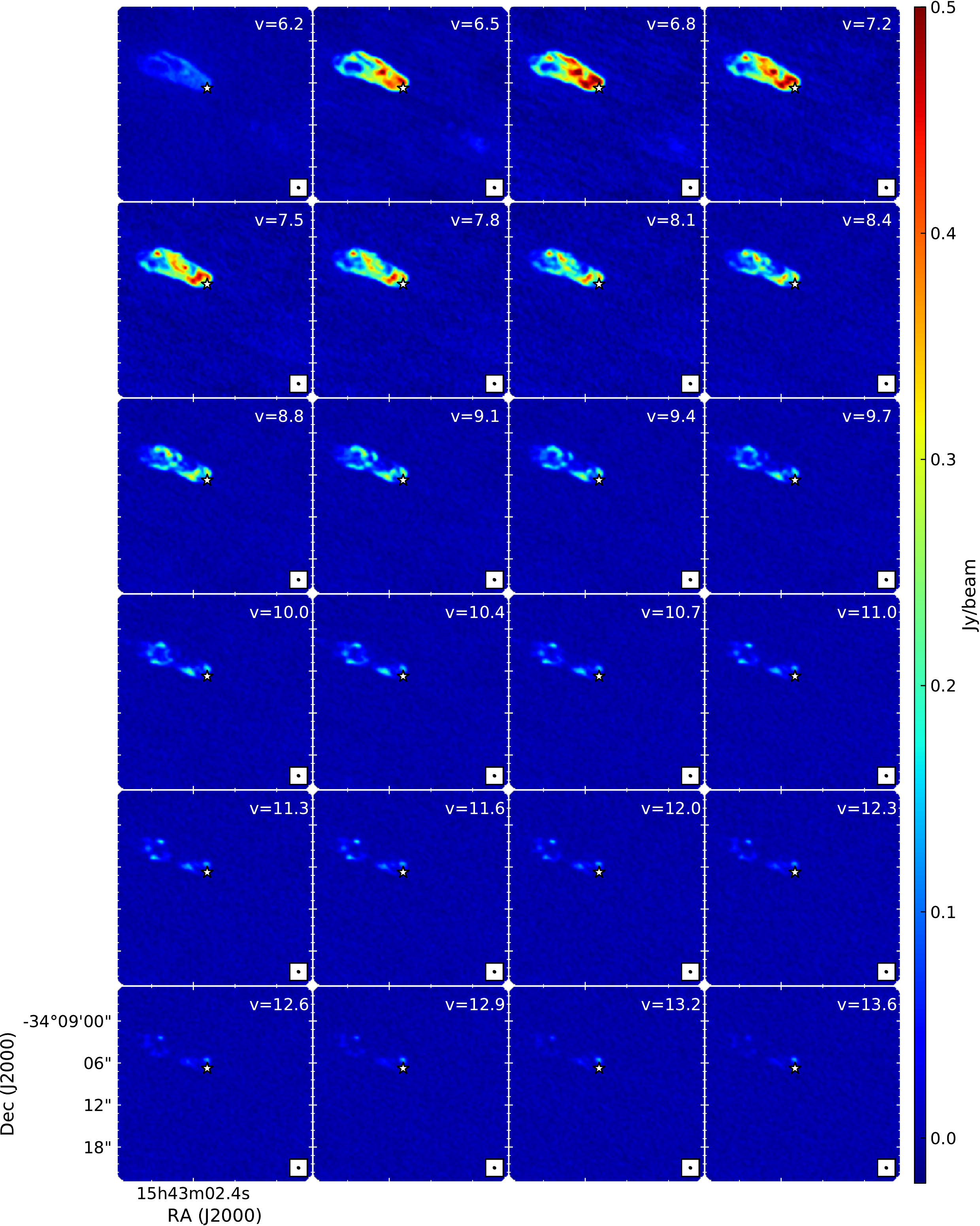}
    \caption{{\bf \iras:} Red lobe channel maps.}
    \label{fig:IRAS15-chan-red}
\end{figure*}{}

\begin{figure*}
    \centering
    \includegraphics[width=1.0\textwidth]{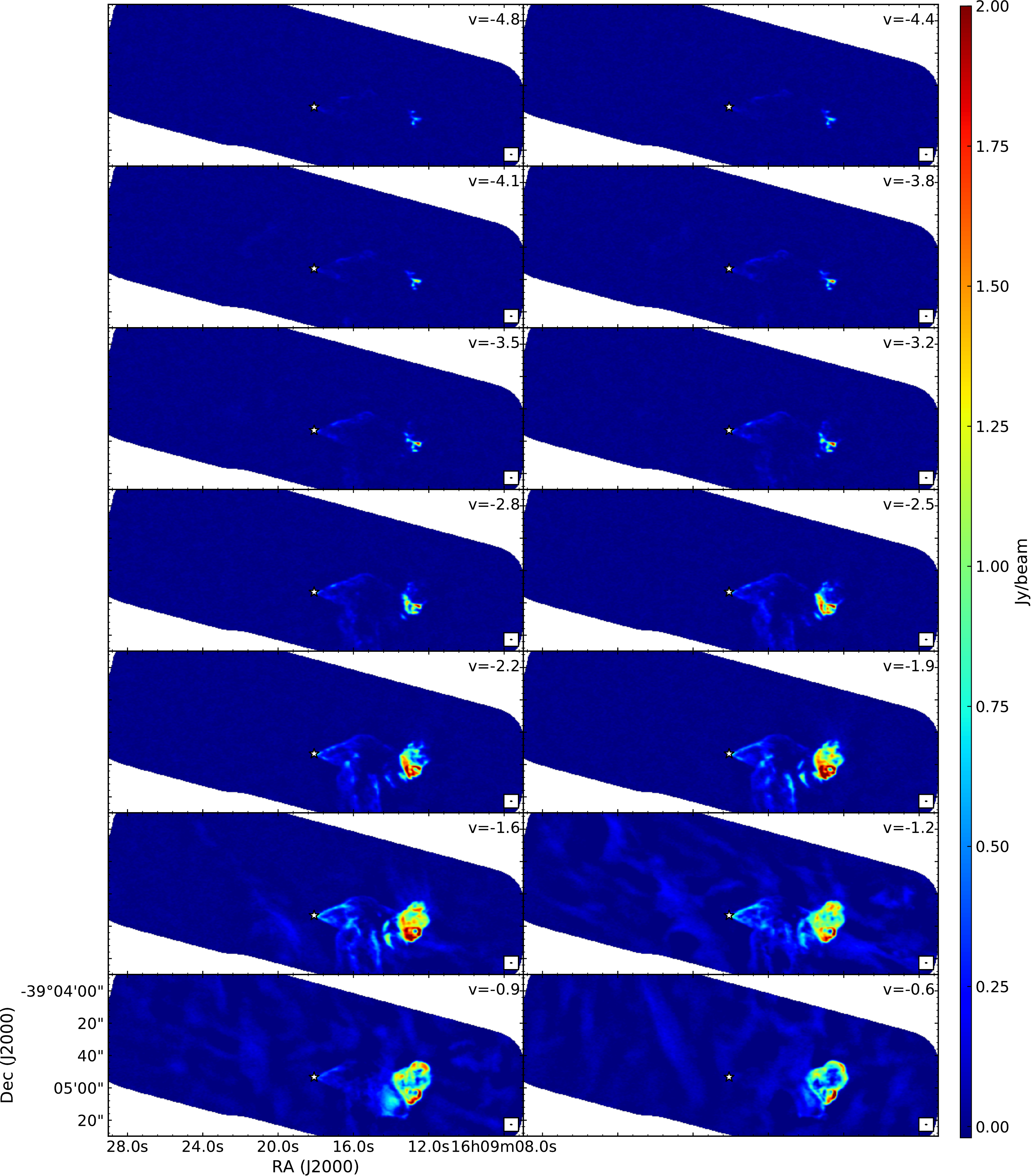}
    \caption{{\bf \irasb:} Blue lobs channel maps.}
    \label{fig:IRAS16-chan-blue}
\end{figure*}{}

\begin{figure*}
    \centering
    \includegraphics[width=1.0\textwidth]{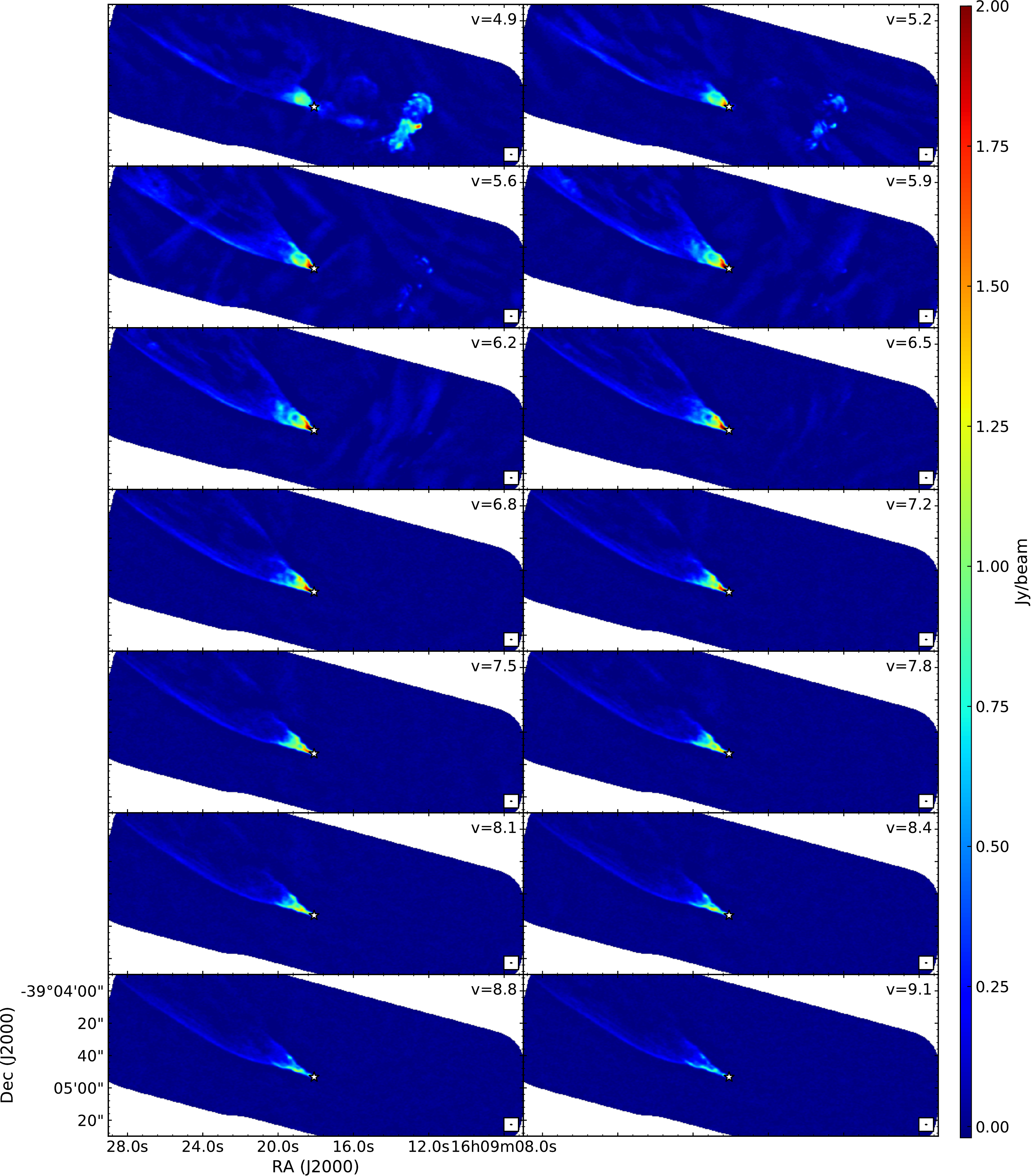}
    \caption{{\bf \irasb:} Red lobe channel maps.}
    \label{fig:IRAS16-chan-red}
\end{figure*}{}
\begin{figure*}
    \centering
    \includegraphics[width=1.0\textwidth]{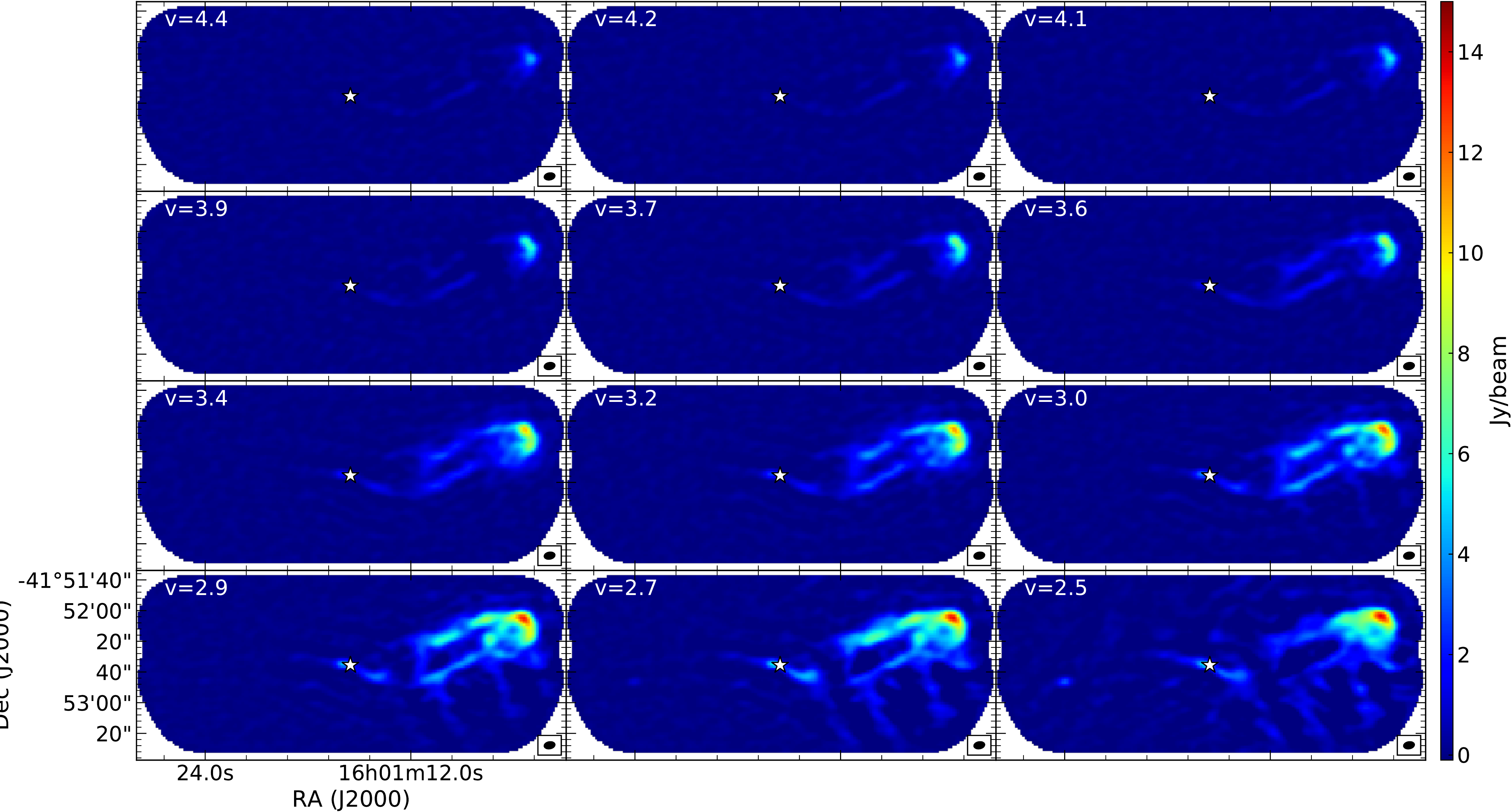}
    \includegraphics[width=1.0\textwidth]{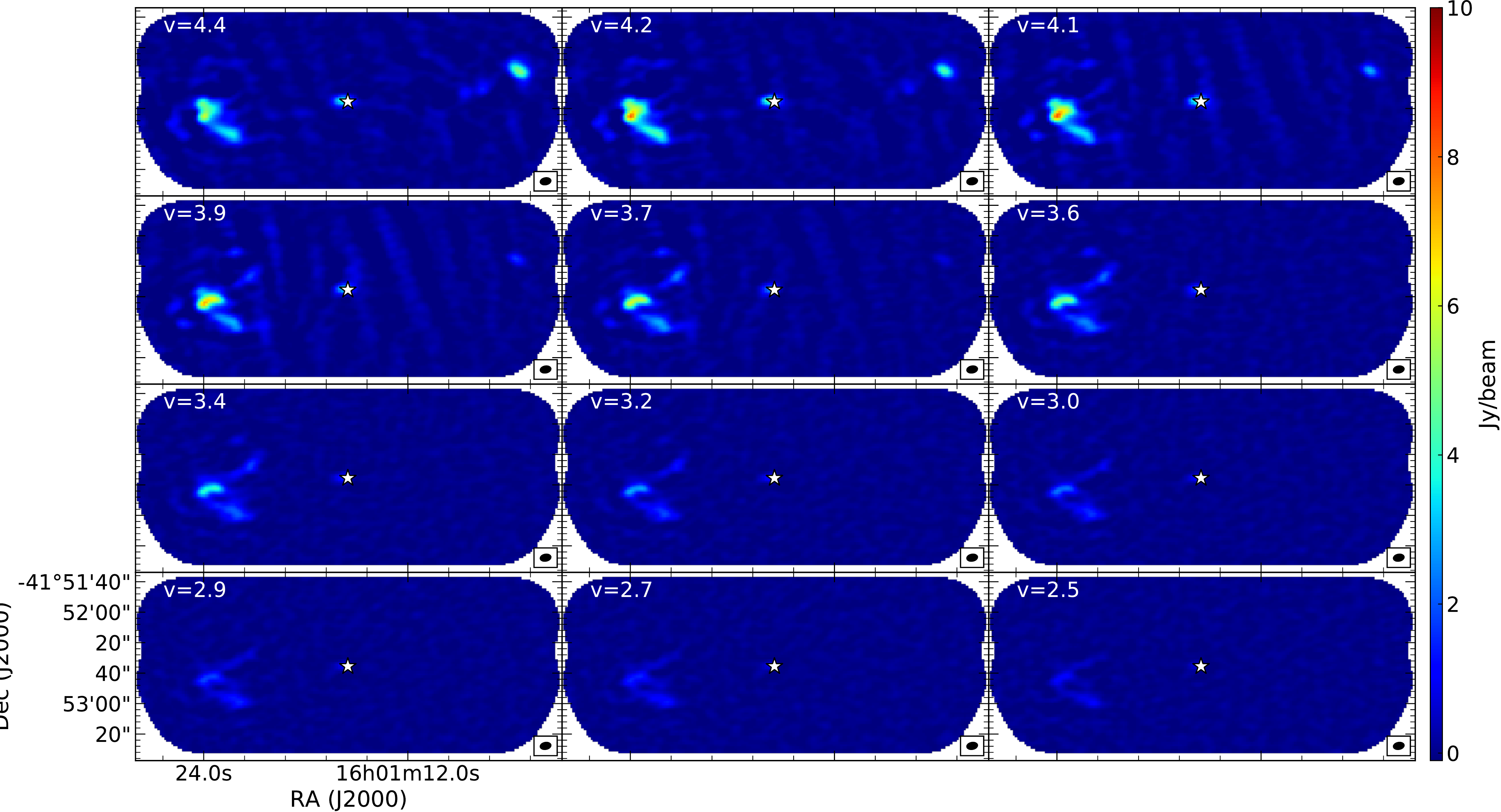}    
    \caption{{\bf \j:} Blue and red lobes channel maps.}
    \label{fig:J160115_chan}
\end{figure*}{}

\end{appendix}

\end{document}